\documentclass[nofootinbib,aps,prb,twocolumn,superscriptaddress,amssymb,floatfix,longbibliography]{revtex4-2}

\newcommand{\GOEaffiliation}{\affiliation{Institut für Theoretische Physik, Georg-August-Universität Göttingen, Friedrich-Hund-Platz 1, 37077 Göttingen, Germany}}

\usepackage{xcolor}
\usepackage{graphicx}
\usepackage{hyperref}
\hypersetup{
    colorlinks=true,
    linkcolor=blue,
    citecolor=blue,
    filecolor=blue,      
    urlcolor=blue,
    pdftitle={Hierarchy of timescales in a disordered spin-1/2 XX ladder},
}
\usepackage[all]{hypcap}
\usepackage{mathtools}
\usepackage{physics}
\usepackage{bbm}
\usepackage{orcidlink}

\usepackage{lmodern}
\usepackage{microtype}

\usepackage[utf8]{inputenc}
\usepackage[T1]{fontenc}

\usepackage[english]{babel}
\makeatletter
\adddialect\l@en\l@english
\makeatother

\newcommand{\wmat}{\boldsymbol{\mathcal{W}}}
\newcommand{\E}[1]{\mathbb{E}_{\wmat}\qty[#1]}
\newcommand{\Ec}[1]{\mathbb{E}_{\mathrm{c}}\qty[#1]}
\DeclareMathOperator*{\argmin}{arg\,min}
\newcommand{\reflabel}[2]{\hyperref[#1]{{\ref*{#1}}#2}}

\begin{document}

\title{Hierarchy of timescales in a disordered spin-\texorpdfstring{$1/2$}{1/2} XX ladder}

\author{Kadir Çeven\orcidlink{0000-0002-1770-1255}}
\email{kadir.ceven@uni-goettingen.de}
\GOEaffiliation

\author{Lukas Peinemann\orcidlink{0009-0006-7228-1288}}

\GOEaffiliation

\author{Fabian Heidrich-Meisner\orcidlink{0000-0002-3463-1121}}
\email{heidrich-meisner@uni-goettingen.de}
\GOEaffiliation

\date{\today}

\begin{abstract}
Understanding the timescales associated with relaxation to equilibrium in closed quantum many-body systems is one of the central focuses in the study of their nonequilibrium dynamics. At late times, these relaxation processes exhibit universal behavior, emerging from the inherent randomness of chaotic Hamiltonians. In this work, we investigate a disordered spin-$1/2$ XX ladder---an experimentally realizable model known for its diffusive dynamics---to explore the connection between transport properties and spectral measures derived solely from the energy levels of the system via these relaxation timescales. We begin by analyzing the spectral form factor, which yields the time when the system begins to follow the random matrix theory (RMT) statistics, known as the RMT time. We then determine the Thouless times---the average times for a local excitation to diffuse across the entire finite system---through the linear-response theory for both spin and energy transport. Our numerical results confirm that the RMT time scales quadratically with system size and upper bounds the Thouless times. Interestingly, we also find that, unlike other nonintegrable models, spin diffusion proceeds faster than energy diffusion. 
\end{abstract}

\maketitle

\section{\label{sec:intro}Introduction}

In the study of nonequilibrium dynamics in closed quantum systems, identifying the relevant timescales is a long-standing problem \cite{dalessioQuantumChaosEigenstate2016}. Efforts to obtain rigorous bounds on such timescales were reported in Refs.~\cite{shortQuantumEquilibrationFinite2012,  goldsteinTimeScalesApproach2013, malabarbaQuantumSystemsEquilibrate2014, goldsteinExtremelyQuickThermalization2015, torres-herreraRelaxationThermalizationIsolated2015, reimannTypicalFastThermalization2016}. A defining aspect in the nonequilibrium dynamics of closed quantum many-body systems \cite{gogolinEquilibrationThermalisationEmergence2016} is that relevant relaxation timescales can depend on system size \cite{dalessioQuantumChaosEigenstate2016,friedmanSpectralStatisticsManyBody2019}. Most notably, the time to diffuse through a finite system scales quadratically with linear system size and is called the Thouless time \cite{edwardsNumericalStudiesLocalization1972, thoulessElectronsDisorderedSystems1974, thoulessMaximumMetallicResistance1977}. Generic nonintegrable many-body systems are believed to be described by quantum chaos and the eigenstate thermalization hypothesis (ETH) \cite{deutschQuantumStatisticalMechanics1991, srednickiChaosQuantumThermalization1994, srednickiApproachThermalEquilibrium1999, rigolThermalizationItsMechanism2008,  dalessioQuantumChaosEigenstate2016}. In this context, the question of how physical timescales are reflected in,  e.g.,~spectral properties \cite{bertiniExactSpectralForm2018, chanSolutionMinimalModel2018,chanSpectralStatisticsSpatially2018,gharibyanOnsetRandomMatrix2018,kosManyBodyQuantumChaos2018,friedmanSpectralStatisticsManyBody2019,suntajsQuantumChaosChallenges2020, liaoManyBodyLevelStatistics2020,royRandomMatrixSpectral2020,sierantThoulessTimeAnalysis2020, bertiniRandomMatrixSpectral2021,moudgalyaSpectralStatisticsConstrained2021,suntajsSpectralPropertiesThreedimensional2021, chanSpectralLyapunovExponents2021,suntajsErgodicityBreakingTransition2022, chanManybodyQuantumChaos2022,roySpectralFormFactor2022,kliczkowskiFadingErgodicity2024,martinez-azconaDecomposingSpectralForm2025,kumarLeadingLeadingorderSpectral2025}, survival probabilities \cite{tavoraInevitablePowerlawBehavior2016,tavoraPowerlawDecayExponents2017,torres-herreraDynamicalManifestationsQuantum2017, torres-herreraGenericDynamicalFeatures2018,schiulazThoulessRelaxationTime2019,lerma-hernandezDynamicalSignaturesQuantum2019,schiulazSelfaveragingManybodyQuantum2020, lezamaEquilibrationTimeManybody2021}, the ETH itself \cite{khatamiFluctuationDissipationTheoremIsolated2013, Serbyn2017,brenesLowfrequencyBehaviorOffdiagonal2020, wangEigenstateThermalizationHypothesis2022}, or in autocorrelations of local observables \cite{wangEigenstateThermalizationHypothesis2022, bartschEstimationEquilibrationTime2024, wangEstimateEquilibrationTimes2025, maceiraThermalizationDynamicsClosed2025} has attracted significant attention. Setting these different approaches into a comprehensive context \cite{bouverot-dupuisRandomMatrixUniversality2025} and establishing connections to experiments \cite{dongMeasuringSpectralForm2025,dasProposalManybodyQuantum2025,karchProbingQuantumManybody2025} are important goals of ongoing research.

To contribute to  this objective, we first draw upon \emph{quantum chaos}  \cite{bohigasCharacterizationChaoticQuantum1984,berrySemiclassicalTheorySpectral1985, haakeQuantumSignaturesChaos1991, stockmannQuantumChaosIntroduction1999, mehtaRandomMatrices2004}---which is closely connected to the ETH through its relationship with random matrix theory (RMT) \cite{dalessioQuantumChaosEigenstate2016}. One way to study the onset of random matrix behavior in a microscopic model is the utilization of the spectral form factor (SFF)---a widely-used measure to analyze the universal RMT dynamics by examining the spectral fluctuations \cite{berrySemiclassicalTheorySpectral1985, haakeQuantumSignaturesChaos1991, guhrRandommatrixTheoriesQuantum1998, stockmannQuantumChaosIntroduction1999, mullerSemiclassicalFoundationUniversality2004, mullerPeriodicorbitTheoryUniversality2005} in quantum many-body physics (see Ref.~\cite{sierantManybodyLocalizationAge2025} for a review). This measure does not require the selection of an initial state or an observable as it is solely based on the correlations in the energy spectrum. The onset of this RMT dynamics defines the first timescale that is of particular interest to us---the RMT time. Notably, it has been theoretically established that this timescale exhibits different scaling behaviors with system size \cite{sierantThoulessTimeAnalysis2020, gharibyanOnsetRandomMatrix2018, friedmanSpectralStatisticsManyBody2019, royRandomMatrixSpectral2020, kumarLeadingLeadingorderSpectral2025, kumarManybodyQuantumChaos2024, chanSpectralStatisticsSpatially2018}. While the RMT time scales quadratically with the system size for a wide range of nonintegrable many-body models \cite{sierantThoulessTimeAnalysis2020, kosManyBodyQuantumChaos2018, gharibyanOnsetRandomMatrix2018, friedmanSpectralStatisticsManyBody2019, suntajsQuantumChaosChallenges2020, royRandomMatrixSpectral2020, roySpectralFormFactor2022, kumarLeadingLeadingorderSpectral2025, kumarManybodyQuantumChaos2024}, other scaling behaviors also exist, including other power laws \cite{ colmenarezSubdiffusiveThoulessTime2022}, logarithmic \cite{kosManyBodyQuantumChaos2018, gharibyanOnsetRandomMatrix2018, kumarManybodyQuantumChaos2024,  kumarLeadingLeadingorderSpectral2025}, or no scaling with system size at all  \cite{roySpectralFormFactor2022, kumarLeadingLeadingorderSpectral2025}. The broad spectrum of scaling properties can be attributed to the complex and varied mechanisms that govern thermalization processes of the corresponding systems \cite{kumarLeadingLeadingorderSpectral2025, kumarManybodyQuantumChaos2024, roySpectralFormFactor2022, royRandomMatrixSpectral2020, gharibyanOnsetRandomMatrix2018, suntajsQuantumChaosChallenges2020}. The RMT time is commonly argued to be identical to the slowest physical relaxation time of the system \cite{sierantManybodyLocalizationAge2025}.

Very often, the RMT time is related to diffusion, and hence an $L^2$ scaling is expected and indeed observed \cite{sierantThoulessTimeAnalysis2020, gharibyanOnsetRandomMatrix2018, friedmanSpectralStatisticsManyBody2019, suntajsQuantumChaosChallenges2020, royRandomMatrixSpectral2020} ($L$ is the linear system size). For instance, in Ref.~\cite{prelovsekSlowDiffusionThouless2023}, RMT properties are utilized to obtain the diffusion constant from level curvature.  In our work we aim at a quantitative comparison between the RMT time and the Thouless time, the relevant timescale for diffusion. Transport in interacting one-dimensional models is a long-studied problem (for a review, see, e.g.,~Refs.~\cite{zotosTransportConservationLaws1997, zotosEvidenceIdealInsulating1996, bertiniFinitetemperatureTransportOnedimensional2021, hessHeatConductionLowdimensional2007}). At finite temperatures, generic nonintegrable systems are considered to exhibit diffusive dynamics \cite{bertiniFinitetemperatureTransportOnedimensional2021} while anomalous transport is found in integrable \cite{Bulchandani2021,Vidmar2016, bertiniFinitetemperatureTransportOnedimensional2021, Ljubotina2023,jepsenSpinTransportTunable2020,Scheie2021,Wei2022} or weakly ergodicity-breaking systems (see, e.g.,~Refs.~\cite{Ljubotina2023, Feldmeier2020}). A straightforward way to compute diffusion constants is the utilization of linear-response theory \cite{bertiniFinitetemperatureTransportOnedimensional2021}. How to actually accomplish this in an efficient way in numerical simulations is the topic of ongoing method development \cite{znidaricMagnetizationTransportSpin2013,steinigewegScalingDiffusionConstants2014, klossTimedependentVariationalPrinciple2018, rakovszkyDissipationassistedOperatorEvolution2022, uskovQuantumDynamicsOne2024, artiacoEfficientLargeScaleManyBody2024, wangDiffusionConstantsRecursion2024, fullgrafLanczosPascalApproachCorrelation2025, longFinitetemperatureDynamicalCorrelations2003, prelovsekSlowDiffusionThouless2023, pawlowskiThoulessApproachTransport2025}. While experiments with optical lattices investigated transport in the far-from-equilibrium regime \cite{schneiderFermionicTransportOutofequilibrium2012, ronzheimerExpansionDynamicsInteracting2013,Hild2014}, recently, direct measurements of diffusion constants in nonintegrable models \cite{wienandEmergenceFluctuatingHydrodynamics2024} and classifications of transport in integrable models have become possible \cite{jepsenSpinTransportTunable2020,Wei2022}. In earlier experiments with quantum magnets, both thermal and spin transport of spin chain and ladders systems were studied \cite{hessHeatTransportCupratebased2019} (see also Ref.~\cite{Scheie2021}).

Crucially, transport phenomena also yield a  timescale called the Thouless time---the average time for a local perturbation to spread through a  finite system \cite{edwardsNumericalStudiesLocalization1972, thoulessElectronsDisorderedSystems1974, thoulessMaximumMetallicResistance1977}. Its scaling behavior significantly depends on the type of transport: Under diffusive dynamics it scales quadratically with the linear system size, whereas under ballistic dynamics it scales linearly with the system size \cite{hopjanCriticalDynamicsShortRange2025}. (Super)Subdiffusive behavior yields $L^\gamma$ with $\gamma>2$ ($1<\gamma<2$) \cite{Capizzi2025}.

In the presence of more than one transport channel, there are several Thouless times, e.g.,~the energy-Thouless time associated with energy transport or the particle-Thouless time associated with particle transport (see, e.g.,~Ref.~\cite{mierzejewskiMultipleRelaxationTimes2022}). A priori, it is not clear how these Thouless times for energy, spin and particle diffusion relate to each other. In the best-studied examples of integrable one-dimensional spin-$1/2$ systems, energy transport is always faster than spin transport, even qualitatively \cite{bulchandaniSuperdiffusiveTransportEnergy2020,bertiniFinitetemperatureTransportOnedimensional2021}.

\begin{figure*}
    \centering
    \includegraphics[width=\linewidth]{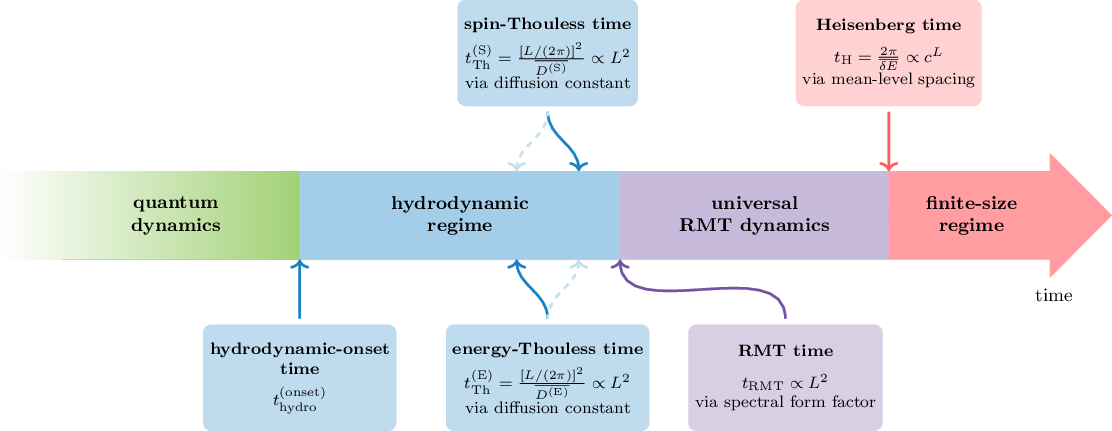}
    \caption{Sequence of regimes for a diffusive nonintegrable many-body quantum spin system with a finite system size $L$. The diffusion of a conserved quantity for such a system is governed by the $L^2$-scaling timescale called Thouless time $t_\mathrm{Th}$, which represents the average time for a transport quantity to spread through the finite system. Often, energy diffusion is expected to occur more rapidly than spin diffusion, i.e.,~$t_\mathrm{Th}^{(\mathrm{E})} < t_\mathrm{Th}^{(\mathrm{S})}$. The RMT time $t_\mathrm{RMT}$ indicates when the system begins to behave like a random matrix, following universal RMT statistics. This time is expected to be identical (or related) to the longest physical timescale. On top of all of these $L^2$-scaling relaxation timescales, the Heisenberg time $t_\mathrm{H}$, determined by the mean-level spacing $\overline{\delta E}$, is the longest timescale and exhibits exponential scaling with respect to the system size.}
    \label{fig:hierarchy}
\end{figure*}

The expected hierarchy of relaxation timescales for a diffusive nonintegrable many-body quantum system is illustrated in Fig.~\ref{fig:hierarchy}. The sequence begins with the onset of the hydrodynamic regime, marked by the hydrodynamic-onset time, at which point the system reaches local equilibrium \cite{chen-linTheoryDiffusiveFluctuations2019, delacretazBoundThermalizationDiffusive2025}. Perturbations in the density associated with a conserved quantity reach the boundaries of the finite system at their respective Thouless times. After all local perturbations have spread through the finite-dimensional system, the dynamics of the system become featureless and similar to that of random matrices at the RMT time \cite{sierantThoulessTimeAnalysis2020, kosManyBodyQuantumChaos2018, gharibyanOnsetRandomMatrix2018, friedmanSpectralStatisticsManyBody2019, suntajsQuantumChaosChallenges2020, royRandomMatrixSpectral2020, roySpectralFormFactor2022, kumarLeadingLeadingorderSpectral2025, kumarManybodyQuantumChaos2024, chanSpectralStatisticsSpatially2018, colmenarezSubdiffusiveThoulessTime2022}. As a result, the RMT time is expected to be identical to the slowest physical relaxation timescale \cite{sierantManybodyLocalizationAge2025}.  Eventually, beyond the Heisenberg time, the dynamics of the system are controlled by the inverse level spacing of the finite-sized system due to the energy-time uncertainty relation, as the system becomes aware of the discrete nature of its energy spectrum \cite{Kiendl2017}.

\begin{figure}
    \centering
    \includegraphics[width=\linewidth]{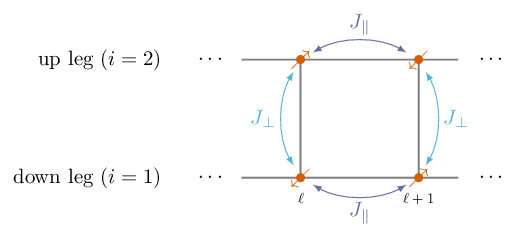}
    \caption{Sketch of a spin-$1/2$ XX model on a two-leg ladder lattice with periodic boundary conditions along the legs. The double-sided arrows along the legs and rungs represent the exchange coupling constants with strengths $J_\parallel$ and $J_\perp$, respectively.
    }
    \label{fig:model_diagram}
\end{figure}

To investigate the hierarchy of these relaxation times, we choose a disordered spin-$1/2$ XX ladder as a study case, as shown in Fig.~\ref{fig:model_diagram}.  We induce disorder to overcome the non-self-averaging behavior of the SFF \cite{matsoukas-roubeasUnitarityBreakingSelfaveraging2023}. Transport in the clean case of this model is well studied, both theoretically  \cite{znidaricMagnetizationTransportSpin2013,vidmarSuddenExpansionMott2013, karraschRealtimeRealspaceSpin2014, steinigewegScalingDiffusionConstants2014, karraschSpinThermalConductivity2015, klossTimedependentVariationalPrinciple2018, rakovszkyDissipationassistedOperatorEvolution2022, wangDiffusionConstantsRecursion2024} and experimentally \cite{wienandEmergenceFluctuatingHydrodynamics2024}. Theoretical and experimental results yield comparable results for the diffusion constant \cite{wienandEmergenceFluctuatingHydrodynamics2024}. However, these studies are focused exclusively on spin transport, whereas in our investigation, we consider both transport channels in the disordered case. Authors of previous studies of transport looked at a larger family of related spin-ladder models that include XXZ-type interactions on the rungs \cite{znidaricMagnetizationTransportSpin2013}, also including disorder that respects $\mathbb{Z}_2$ symmetry associated with exchanging the legs \cite{Iadecola2019}.

Our computations reveal that RMT and Thouless times both scale linearly with $L^2$ and therefore, the former appears to be controlled by diffusive hydrodynamics. The RMT time is larger than both Thouless times for energy and spin transport, which we shall discuss further.

Additionally, our results show, for both disordered and clean cases, that the two-leg ladder model possesses a slower energy diffusion than spin diffusion, i.e.,~the energy-Thouless time is longer than the spin-Thouless time. Other well-studied nonintegrable models exhibit the opposite behavior \cite{luitzErgodicSideManybody2017, richterMagnetizationEnergyDynamics2019, herbrychSpinEnergyDiffusion2025}. As an example, we consider
a disordered spin-$1/2$ XXZ chain with next-nearest-neighbor couplings \cite{zotosEvidenceIdealInsulating1996, narozhnyTransportXXZModel1998, rigolBreakdownThermalizationFinite2009, rigolQuantumQuenchesThermalization2009, steinigewegDecayCurrentsStrong2011, znidaricMagnetizationTransportSpin2013}. 
While the RMT time in this model also scales quadratically with system size, here we find the more conventional scenario with a slower spin than energy diffusion. We provide an intuitive explanation for the existence of faster-than-energy spin diffusion in the XX ladder.

This paper is organized as follows: In Sec.~\ref{sec:model}, we begin by introducing the model. Sec.~\ref{sec:relax_times} describes the methodology used to extract the relaxation timescales from the relevant measures. The results of these measures are then presented in Secs.~\ref{sec:sff_results} and \ref{sec:diff_const_results}, where we discuss the outcomes for varying system sizes of a disordered spin-$1/2$ XX ladder. In Sec.~\ref{sec:relax_time_results}, we compare the relaxation timescales obtained from these measures for both a disordered spin-$1/2$ XX ladder and, for comparison, a disordered spin-$1/2$ XXZ chain with next-nearest-neighbor couplings. Finally, our conclusions are summarized in Sec.~\ref{sec:conclusions}, and additional technical details regarding the computations are provided in the Appendixes.

\section{\label{sec:model}Model}

We consider a system of spin-$1/2$ objects interacting with their nearest-neighbor spins only on the XY plane of the spin space on a two-leg ladder lattice with periodic boundary conditions along the legs, i.e.,~a spin-$1/2$ XX ladder, as shown in Fig.~\ref{fig:model_diagram}. To make it disordered, we apply random magnetic fields along the z-direction in spin space, which yields a disorder realization $\wmat$. For the rest of the paper, we mostly avoid using $\wmat$ for ease of complex notation. Setting $\hbar = 1$ throughout the paper, its Hamiltonian can be written as a sum of three different operators as follows:
\begin{equation}\label{eq:XX_ladder_ham}
    H{(\wmat)} = H_\parallel + H_\perp + H_{\mathrm{dis}}{(\wmat)}\,.
\end{equation}
The first operator $H_\parallel$ is for the spin couplings along the legs, given by
\begin{equation}
    H_\parallel = \frac{J_\parallel}{2} \sum_{\ell=1}^L \sum_{i=1}^2 \qty(S^+_{\ell, i} S^-_{\ell+1, i} + S^-_{\ell, i} S^+_{\ell+1, i})\,,
\end{equation}
where $J_\parallel$ is the leg-coupling strength, $L$ is the number of the rungs on the lattice, which defines the number of spins $N=2L$, and $S^\pm_{\ell, i}=S^x_{\ell, i} \pm i S^y_{\ell, i}$ are the spin-$1/2$ ladder operators at $\ell$ rung and $i$ leg. The second part is responsible for the spin couplings on the rungs, given by
\begin{equation}
    H_\perp = \frac{J_\perp}{2} \sum_{\ell=1}^L \qty(S^+_{\ell, 1} S^-_{\ell, 2} + S^-_{\ell, 1} S^+_{\ell, 2})\,,
\end{equation}
where $J_\perp$ is the rung-coupling strength. The last part is for the disorder terms, defined as
\begin{equation}\label{eq:XX_ladder_ham_disorder_part}
    H_{\mathrm{dis}}{(\wmat)} = \sum_{\ell=1}^L \sum_{i=1}^2 w_{\ell, i} S^z_{\ell, i}\,,
\end{equation}
where $w_{\ell, i}$ is the strength of the random magnetic field applied to the spin located at the lattice site $(\ell, i)$, which is drawn from the uniform distribution on the interval $[-W,W]$ with the disorder strength $W>0$.

In a clean spin-$1/2$ XX ladder (where $W=0$), the coupling ratio $J_\perp/J_\parallel$ determines the behavior of the system. When $J_\perp/J_\parallel=0$, the system consists of two XX chains, which can be mapped to free spinless fermions through the Jordan-Wigner transformation. In the opposite limit where $J_\perp/J_\parallel \rightarrow \infty$, the system consists of  $L$ uncoupled dimers, where each dimer acts as an independent two-spin-$1/2$ system. For any intermediate coupling ratio $J_\perp/J_\parallel$, the Jordan-Wigner transformation converts the system from spin-1/2 objects into interacting spinless fermions. The model can equivalently be written as a system of hardcore bosons (see, e.g.,~Refs.~\cite{vidmarSuddenExpansionMott2013,steinigewegScalingDiffusionConstants2014}).

Moreover, depending on the parameters, the model displays several symmetries. In a system without any disorder $(W=0)$, the model has the symmetries of, for example, the total magnetization, the spin inversion, lattice translations, and reflection along the legs. Whenever $W\neq 0$, the translation and reflection symmetries are broken because of the introduced random local magnetic fields acting on  each spin. 

In the disordered case, to choose the parameter sets located in the delocalized regime throughout our paper, we search the $W$-$J_\perp$-$J_\parallel$ parameter space using the adjacent gap ratio $\overline{r}$ \cite{oganesyanLocalizationInteractingFermions2007, santosOnsetQuantumChaos2010, palManybodyLocalizationPhase2010, atasDistributionRatioConsecutive2013, dalessioQuantumChaosEigenstate2016} and the entanglement entropy $\overline{S}_\mathrm{vNE}$ \cite{songBipartiteFluctuationsProbe2012, bauerAreaLawsManybody2013, kjallManyBodyLocalizationDisordered2014, luitzManybodyLocalizationEdge2015, nandkishoreManyBodyLocalizationThermalization2015, limManybodyLocalizationTransition2016} (see  Appendix~\ref{appen:sec:crossover-diagram}). Based on these measures, we find that the system is sufficiently deep in the delocalized regime up to at least disorder strength of $W/J_\parallel \approx 3.0$ when the coupling ratio is set to $J_\perp/J_\parallel = 1.0$ for the accessible system sizes.

\section{\label{sec:relax_times}Relaxation timescales}

In this section, we introduce how we extract the relaxation timescales for our quantum many-body system, utilizing measures from different areas of physics. First, we use the SFF to obtain the RMT time. Secondly, we compute  diffusion constants to determine the corresponding timescales, referred to as the Thouless times. For each measure mentioned in this section, we describe the  numerical methods and technical aspects.

\subsection{\label{sec:RMT_time}RMT time}

The RMT time, denoted as $t_\mathrm{RMT}$, is a fundamental timescale in the context of quantum chaos, and a widely employed method for extracting this timescale is through the analysis of the SFF. However, despite its popularity, the SFF has a known limitation: It is not self-averaging \cite{prangeSpectralFormFactor1997}, necessitating the construction of a Hamiltonian ensemble and additional averaging procedures. In our paper, we construct such an ensemble using the disorder realizations.

\subsubsection{SFF}

\begin{figure}
    \centering
    \includegraphics[width=\linewidth]{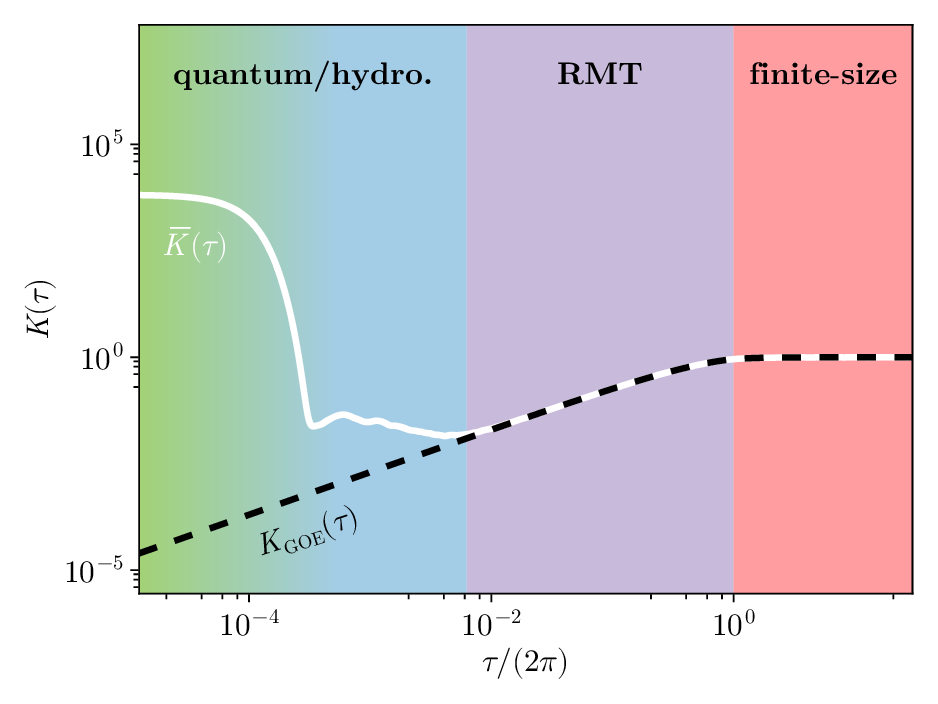}
    \caption{Comparison between a typical ensemble-averaged SFF of a time-reversal symmetric Hamiltonian (white solid line) $\overline{K}{(\tau)}$ and the SFF of the Gaussian orthogonal ensemble (black dashed line) $K_\mathrm{GOE}{(\tau)}$ as a function of normalized time $\tau$ at disorder strength $W/J_\parallel = 1.5$ and coupling ratio $J_\perp/J_\parallel = 1.0$. $\overline{K}{(\tau)}$ exhibits four regimes: quantum, hydrodynamic, universal RMT, and finite size; in contrast, $K_\mathrm{GOE}{(\tau)}$ displays only the RMT and finite-size regimes \cite{haakeQuantumSignaturesChaos1991, stockmannQuantumChaosIntroduction1999, mehtaRandomMatrices2004}.}
    \label{fig:typical_sff}
\end{figure}

As shown in Fig.~\ref{fig:typical_sff}, a typical SFF of a generic model generally exhibits four distinct regimes over time: quantum, hydrodynamic, universal RMT, and finite size. Initially, in the nonuniversal quantum regime, it starts at a finite value and smoothly decreases. Then this is followed by the hydrodynamic regime. Upon completion of these two initial regimes, the SFF transitions into the RMT regime. This regime is characterized by universal behavior, as it begins to follow the SFF of an eigenvalue distribution from RMT (see Ref.~\cite{bertiniExactSpectralForm2018} and references therein). The time at which this transition occurs is referred to as the normalized RMT time $\tau_\mathrm{RMT}$, marking the onset of universal RMT behavior. Finally, at the end of this ramp, it saturates to a constant value at the normalized Heisenberg time $\tau_\mathrm{H}$. It is inversely proportional to the mean-level spacing and is located at $2\pi$ on the time axis due to the spectral unfolding procedure.

We define the disorder-averaged SFF of a Hamiltonian, $\overline{K}{(\tau)}$, as follows:
\begin{equation}\label{eq:sff}
    \overline{K}{(\tau)} = \qty(\overline{Z})^{-1} \, \E{\abs{\sum_{\alpha} \rho_\mathrm{f}{\qty(\varepsilon_\alpha; \eta)} e^{-i \varepsilon_\alpha \tau}}^2}\,,
\end{equation}
which corresponds to the Fourier transform of the two-point correlation function of energy eigenvalues. In Eq.~\eqref{eq:sff}, $\E{\,\cdots}$ denotes the disorder average, $\varepsilon_\alpha$ is the unfolded energy eigenvalue of the disorder realization $\wmat$, $\tau$ is the normalized time with respect to the Heisenberg time $t_\mathrm{H}$, $\overline{Z} \coloneqq \E{ \sum_{\alpha} \abs{\rho_\mathrm{f}{\qty(\varepsilon_\alpha; \eta)}}^2 }$ is the disorder-averaged partition function, and $\rho_\mathrm{f}{\qty(\varepsilon_\alpha; \eta)}$ is the filter function to smoothen out the spectral edge effects appearing in $\overline{K}{(\tau)}$ with a tuning factor $\eta$ \cite{gharibyanOnsetRandomMatrix2018}.

\subsubsection{Numerical extraction of \texorpdfstring{$t_\mathrm{RMT}$}{RMT time}}

A straightforward approach to extract $\tau_\mathrm{RMT}$ from $\overline{K}{(\tau)}$ involves comparing it with the SFF of a RMT eigenvalue distribution, such as the Gaussian orthogonal ensemble (GOE). This comparison leads to the following criterion function \cite{suntajsQuantumChaosChallenges2020}:
\begin{equation}\label{eq:SFF_criterion}
    \Delta K{(\tau)} = \left\lvert \log_{10}\frac{\overline{K}_\mathrm{RA}{(\tau)}}{K_\mathrm{GOE}{(\tau)}} \right\rvert\,,
\end{equation}
where $\overline{K}_\mathrm{RA}{(\tau)}$ is the running-and-disorder-averaged SFF of our model, and $K_\mathrm{GOE}(\tau)$ is the SFF of the GOE (see Appendix~\ref{app:GOE} for further details). We calculate the running average of $\overline{K}{(\tau)}$, denoted as $\overline{K}_{\mathrm{RA}}{(\tau)}$, in addition to the disorder average because it reduces  fluctuations in the numerical data. 

Pragmatically, to extract $\tau_\mathrm{RMT}$  using Eq.~\eqref{eq:SFF_criterion}, we introduce a threshold value $\epsilon_{\Delta K}$ for $\Delta K{(\tau)}$. We then determine $\tau_\mathrm{RMT}$ as the time at which
\begin{equation}\label{eq:tau_RMT}
  \Delta K{(\tau_\mathrm{RMT})} = \epsilon_{\Delta K}\,.
\end{equation}

For the comparison with other relaxation timescales, the RMT time must be expressed in real units. This can be achieved by multiplying it by the real Heisenberg time $t_\mathrm{H}$ as follows:
\begin{equation}\label{eq:t_RMT}
    t_\mathrm{RMT} = \frac{\tau_\mathrm{RMT}}{2\pi} t_\mathrm{H} \propto L^2\,.
\end{equation}
Due the unfolding, we empirically determine the Heisenberg time using the mean-level spacing. Its calculation is presented in detail in Appendix~\ref{app:real_t_Hs}.

For the computation of these SFFs, we use the canonical ensemble, i.e.,~the zero-magnetization symmetry sector ($S^z = 0$), where the energy eigenvalues are computed via the exact diagonalization for $M=5000$ independent disorder realizations.

To eliminate the variations in the density of states $\rho{(E)}$ due to the ensemble of the various disorder realizations appearing in the SFF computations, we unfold the energy spectrum of each realization $\wmat$, which also makes comparing the numerical results with RMT easier and thus highlights the universal RMT behavior \cite{brodyRandommatrixPhysicsSpectrum1981, guhrRandommatrixTheoriesQuantum1998, gomezMisleadingSignaturesQuantum2002, abul-magdUnfoldingSpectrumChaotic2014}. As a filter function, we use a Gaussian distribution with a given filter width factor $\eta$. However, for some disorder realizations, even after unfolding the energy spectrum, $\tau_\mathrm{H}$ may shift from $2\pi$ because of the spectral edge effects \cite{liSpectralFormFactor2024}. To prevent this shifting, we renormalize $\tau_\mathrm{H}$ by applying the filter function to the mean-level spacing of the unfolded spectrum. We explain this shift-fixing procedure in Appendix~\ref{app:fix_shift_in_tau_H} and further details regarding the unfolding and filtering procedures can be found in Appendixes~\ref{app:unfolding} and \ref{app:filtering}, respectively.

The choices for $\epsilon_{\Delta K}$ and $\eta$ quantitatively affect the result for $t_\mathrm{RMT}$. However, the $L^2$ scaling of $t_\mathrm{RMT}$ observed in our model remains robust if $\epsilon_{\Delta K}$ is chosen to be sufficiently large to surpass the fluctuations that occur just prior to the onset of RMT behavior in $\overline{K}_\mathrm{RA}{(t)}$, and $\eta$ falls within a reasonable range, typically between $0.2$ and $1.0$. As part of our analysis, we estimate the RMT time $t_\mathrm{RMT}$ for the models by varying the threshold value $\epsilon_{\Delta K}$ over a range of $0.035$ to $0.2$, while keeping the parameter $\eta$ fixed at $0.5$. Generally, we search for the range of pairs $(\eta, \epsilon_{\Delta K})$ that yield the same $L$ dependence of $t_\mathrm{RMT}$, from which we then extract $t_\mathrm{RMT}$ up to this systematic uncertainty. Thus the exact prefactor of $t_\mathrm{RMT}$ is only fixed within these limitations.

In our results, we obtain $\overline{K}_\mathrm{RA}{(\tau)}$ in Eq.~\eqref{eq:SFF_criterion} by time-averaging $\overline{K}{(\tau)}$ with a running average window in $\log_{10}$ scale $\delta{(\log_{10}\tau)}$ of $0.025$ while the discrete time step in $\log_{10}$ scale $\Delta{(\log_{10}\tau)}$ is $0.001$.

\subsection{\label{sec:thouless_times}Spin- and energy-Thouless times}

Our main interest is in a direct comparison of Thouless times, extracted from transport, with the RMT time. We recall that the Thouless time $t_\mathrm{Th}$ is the average time for a local excitation to spread through a finite system with length $L$
\begin{equation}\label{eq:t_Th}
    t^{(\mu)}_\mathrm{Th} = \frac{[L/(2\pi)]^2}{D^{(\mu)}} \propto L^2\,,
\end{equation}
here assuming diffusion. The symbol $\mu$  indicates the transport channel, such as spin, particle, charge, or energy, and $D^{(\mu)}$ is the diffusion constant for the corresponding transport channel in the thermodynamic limit. In Eq.~\eqref{eq:t_Th}, we divide $L$ by $2\pi$ to obtain the smallest wavenumber $k_0 = 2\pi / L$. This wavenumber also appears in the autocorrelation function of the local densities $c^{(\mu)}{(k,t)}$. For late times, this autocorrelation function decays exponentially at wavenumber $k=k_0$ \cite{steinigewegSpinTransportChain2011}, with a decay rate given by
\begin{equation}
    c^{(\mu)}({k_0, t )} \propto e^{-D^{(\mu)} k_0^2 t},
\end{equation}
where the exponent $D^{(\mu)} k_0^2$ corresponds to the Thouless energy,\footnote{Since our model includes disorder, we replace the diffusion constant $D^{(\mu)}$ in the denominator in Eq.~\eqref{eq:t_Th} with the disorder-averaged diffusion constant $\overline{D^{(\mu)}}$ for the rest of this paper.} which is the inverse of the Thouless time $t_\mathrm{Th}$.

\subsubsection{Transport quantities}

As we mentioned previously, to extract the diffusion constants and therefore $t^{(\mu)}_\mathrm{Th}$ for our model, we employ the linear-response theory, which relates the current autocorrelation function to the diffusion constant through the Green-Kubo relations and the Einstein relation. 
Note that we exclusively work at infinite temperature in our analysis of transport coefficients.

To set up the formalism, we first need to choose a conserved  quantity $Q^{(\mu)}$. In a disordered spin-$1/2$ XX ladder, the obvious transport quantities are
\begin{equation}\label{eq:conserved_quantities}
    \begin{cases}
        Q^{(\mathrm{S})} = \sum_{\ell, i} S^z_{\ell, i} =: S^z\\
        Q^{(\mathrm{E})} = H
    \end{cases}\,,
\end{equation}
where $Q^{(\mathrm{S})}$ is the total magnetization in the z direction of the spins, and $Q^{(\mathrm{E})}$ is the total energy.

As a second step, we utilize the continuity equation to relate the local densities of transport quantities to the corresponding local current operators. The continuity equation is given by
\begin{eqnarray}
    \dot{q}^{(\mu)}_\ell &=& i \comm{H}{q^{(\mu)}_\ell}\,,\nonumber\\
    &=& j^{(\mu)}_{\ell - 1} - j^{(\mu)}_\ell\,,
\end{eqnarray}
where $q^{(\mu)}_{\ell}$ is the local density of transport quantity at the $\ell$ rung, so that $Q^{(\mu)} = \sum_{\ell} q_{\ell}^{(\mu)}$. From this equation, we can derive expressions for the local current operator $j^{(\mu)}_\ell$, which are essential for characterizing the transport properties of the system. As expected, these current operators may differ depending on the choice of local density.

\subsubsection{Local densities of transport quantities and local current operators}

For instance, as the first transport quantity given in Eq.~\eqref{eq:conserved_quantities}, the total magnetization can be decomposed into a sum of local magnetization densities as $Q^\mathrm{(S)} = \sum_\ell{q^\mathrm{(S)}_\ell}$, where the local magnetization at each rung is defined as $q^\mathrm{(S)}_\ell \coloneqq \sum_i S^z_{\ell, i} =  S^z_{\ell, 1} +  S^z_{\ell, 2}$, yielding the local spin-current operator below
\begin{equation}\label{eq:loc_spin_current}
    j_\ell^\mathrm{(S)} = J_\parallel \sum_{i=1}^2 j_{(\ell, \ell+1), (i, i)}\,,
\end{equation}
where  $j_{(\ell, \ell^\prime), (i, i^\prime)}$ is defined as
\begin{equation}
    j_{(\ell, \ell^\prime), (i, i^\prime)} \coloneqq \frac{i}{2}\qty(S^+_{\ell, i} S^-_{\ell^\prime, i^\prime} - S^-_{\ell, i} S^+_{\ell^\prime, i^\prime})\,.
\end{equation}

Similarly, the local energy current $j_\ell^\mathrm{(E)}$ can be obtained by selecting the total energy as the transport quantity, i.e.,~$Q^\mathrm{(E)} = \sum_{\ell} q^\mathrm{(E)}_\ell$, where $q^\mathrm{(E)}_\ell$ is the local energy density. For our model, we express $q^\mathrm{(E)}_\ell$ as the sum of three terms
\begin{equation}
    q^\mathrm{(E)}_{\ell} \coloneqq q^\mathrm{(E)}_{\parallel, \ell} + q^\mathrm{(E)}_{_\perp, \ell} + q^\mathrm{(E)}_{\mathrm{dis}, \ell}\,,
\end{equation}
which represent the parallel, perpendicular, and disorder contributions to the local energy, respectively. Specifically, the parallel term is given by
\begin{equation}\label{eq:XX_ladder_local_energy_parallel}
    q^\mathrm{(E)}_{\parallel, \ell} = \frac{J_\parallel}{2} \sum_{i=1}^2 \qty(S^+_{\ell, i} S^-_{\ell+1, i} + S^-_{\ell, i} S^+_{\ell+1, i})\,,
\end{equation}
the perpendicular term by
\begin{equation}\label{eq:XX_ladder_local_energy_perp}
    q^\mathrm{(E)}_{\perp, \ell} = \frac{J_\perp}{4} \sum_{n=\ell}^{\ell+1} \qty(S^+_{n, 1} S^-_{n, 2} + S^-_{n, 1} S^+_{n, 2})\,,
\end{equation}
and the disorder term by
\begin{equation}\label{eq:XX_ladder_local_energy_disorder}
    q^\mathrm{(E)}_{\mathrm{dis}, \ell} = \frac{1}{2} \sum_{n=\ell}^{\ell+1} \sum_{i=1}^2 w_{n, i} S^z_{n, i}\,.
\end{equation}
With these choices for the local energy density in Eqs.~\eqref{eq:XX_ladder_local_energy_parallel}--\eqref{eq:XX_ladder_local_energy_disorder}, the local energy current $j_\ell^\mathrm{(E)}$ amounts to\footnote{Note that the local energy density is not uniquely defined.}
\begin{equation}\label{eq:loc_energy_current}
    j_\ell^\mathrm{(E)} = j_{\parallel,\ell}^\mathrm{(E)} + j_{\perp,\ell}^\mathrm{(E)},
\end{equation}
where its leg part $j_{\parallel,\ell}^\mathrm{(E)}$ and its rung part $j_{\perp,\ell}^\mathrm{(E)}$ are given by
\begin{eqnarray}
    j_{\parallel,\ell}^\mathrm{(E)} & = & J^2_\parallel \sum_{i=1}^2 j_{(\ell+1, \ell-1), (i, i)} S^z_{\ell, i}\\
    && + J_\parallel \sum_{i = 1}^2 \frac{w_{\ell, i} + w_{\ell+1, i}}{2} j_{(\ell, \ell+1),(i, i)}\,,\nonumber
\end{eqnarray}
and
\begin{eqnarray}
    j_{\perp,\ell}^\mathrm{(E)} &=& \frac{J_\parallel J_\perp}{2} \sum_{i=1}^2 j_{(\ell+1, \ell), (i, 3-i)} S^z_{\ell, i}\\
    && - \frac{J_\parallel J_\perp}{2} \sum_{i=1}^2 j_{(\ell-1, \ell), (i, 3-i)} S^z_{\ell, i}\,.\nonumber
\end{eqnarray}
Since the calculation of the current autocorrelation function requires the total current operator $J^{(\mu)}$, we sum the local current operators in Eqs.~\eqref{eq:loc_spin_current} and \eqref{eq:loc_energy_current} over the rungs, i.e.,
\begin{equation}\label{eq:tot_current}
    J^{(\mu)} = \sum_{\ell} j_\ell^{(\mu)}\,.
\end{equation}

\subsubsection{Current autocorrelation functions}

With Eq.~\eqref{eq:tot_current}, the autocorrelation function of the total current operator for the disorder realization $\wmat$, denoted by $C^{(\mu)}{(t; \beta)}$, can be defined as
\begin{equation}\label{eq:autocorrelation_C_J}
    C^{(\mu)}{(t; \beta)} = \Re{\frac{\expval{J^{(\mu)}{(t)} \, J^{(\mu)}{(0)} }_\beta}{N}}\,,
\end{equation}
where $\expval{\cdots}_\beta = \Trace\!\qty{e^{-\beta H}\,\cdots}/\Trace\!\qty{e^{-\beta H}}$ is the thermal ensemble average of an operator at the inverse temperature $\beta$. Due to the disorder, we need to average this current autocorrelation function over the disorder realizations,\footnote{Averaging over diffusion constants instead of current autocorrelation functions also gives similar results for a large system size.} which yields the disorder-averaged current autocorrelation function
\begin{equation}\label{eq:disorder-averaged_C_J}
    \overline{C^{(\mu)}}{(t; \beta)} = \E{C^{(\mu)}{(t; \beta)} }\,.
\end{equation}

\subsubsection{Time-dependent disorder-averaged diffusion constants}

\begin{figure}
    \centering
    \includegraphics[width=\linewidth]{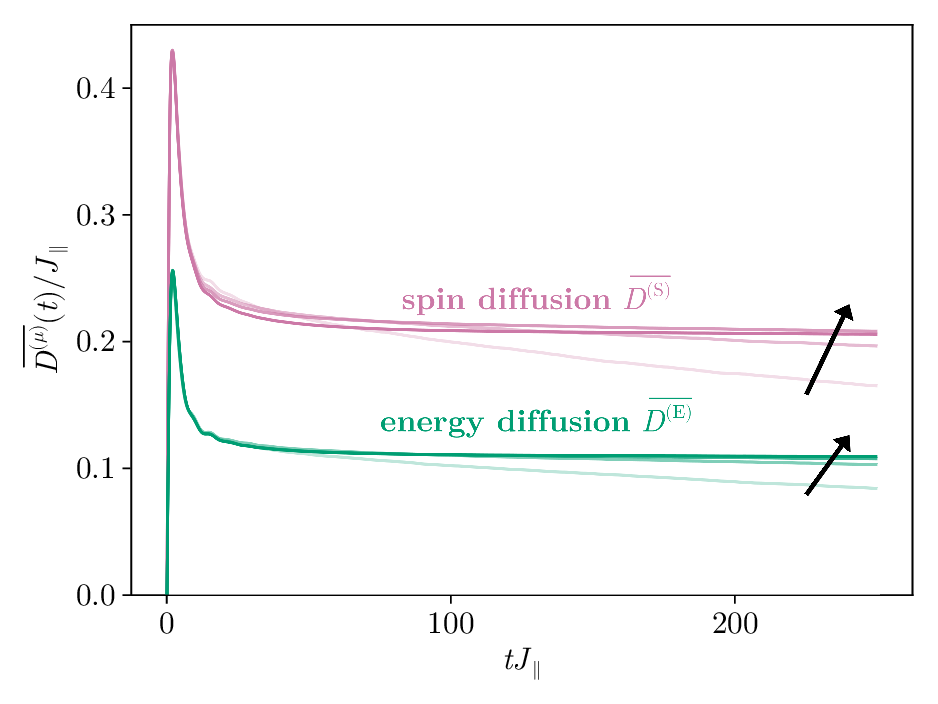}
    \caption{Comparison between disorder-averaged time-dependent diffusion constants of a disordered spin-$1/2$ XX ladder as a function of time at disorder strength $W/J_\parallel = 1.5$ and coupling ratio $J_\perp/J_\parallel = 1.0$ for various numbers of rungs $L=7,8,9,10$. The spin diffusion (purple lines) occurs at a faster rate than the energy diffusion (green lines). The opacity of each line denotes the result for a chosen $L$---lower opacity corresponds to a smaller system, while higher opacity indicates a larger one. The arrows specify how the time-dependent diffusion constant converges with an increase in $L$. }
    \label{fig:typical_diff_const}
\end{figure}

Furthermore, as the (normalized) disorder strength $W/J_\parallel$ or the coupling ratio $J_\perp/J_\parallel$ increases beyond a certain threshold, the spin- and energy-diffusion constants tend to converge to the same result that is independent of $J_\perp$.

From the average in Eq.~\eqref{eq:disorder-averaged_C_J}, we obtain the time-dependent disorder-averaged diffusion constant $\overline{D^{(\mu)}}{(t;\beta)}$, shown in Fig.~\ref{fig:typical_diff_const}, as \cite{steinigewegDensityDynamicsTranslationally2009}
\begin{equation}
    \overline{D^{(\mu)}}{(t;\beta)} = \qty[\overline{\chi_\mu}{(\beta)}]^{-1} \int_0^t \mathrm{d} t^\prime \, \overline{C^{(\mu)}}{(t^\prime;\beta)}\,,
\end{equation}
where $\overline{\chi_\mu}{(\beta)}$ is the disorder-averaged static susceptibility of the corresponding conserved quantity at the inverse temperature $\beta$. In the thermodynamic and infinite-time limits, $\overline{D^{(\mu)}}{(t;\beta)}$ is expected to give the actual disorder-averaged diffusion constant $\overline{D^{(\mu)}}{(\beta)}$, i.e.,
\begin{equation}\label{eq:diff_const_limits}
    \overline{D^{(\mu)}}{(\beta)} = \lim_{t\rightarrow \infty} \lim_{N\rightarrow \infty} \overline{D^{(\mu)}}{(t;\beta)}\,.
\end{equation}

\subsubsection{\label{sec:static_suscep}Disorder-averaged static susceptibilities}

In the infinite-temperature and thermodynamic limits, the disorder-averaged static spin susceptibility per spin of our model $\overline{\chi_\mathrm{S}}{(\beta=0)}$ turns out to be disorder independent and becomes
\begin{equation}\label{eq:spin_suscep}
    \overline{\chi_\mathrm{S}}{(\beta=0)} = \frac{1}{4}\,.
\end{equation}
The disorder-averaged static energy susceptibility per spin $\overline{\chi_\mathrm{E}}{(\beta=0)}$ is disorder dependent and amounts to
\begin{equation}\label{eq:energy_suscep}
    \overline{\chi_\mathrm{E}}{(\beta=0)} = \frac{2J^2_\parallel + J^2_\perp}{16} + \frac{W^2}{12}\,.
\end{equation}
This result is obtained by taking the disorder average analytically.

\subsubsection{Numerical extraction of \texorpdfstring{$t_\mathrm{Th}$}{Thouless time}}

Following the computation of the disorder-averaged diffusion constant in the thermodynamic and infinite-time limits, denoted as $\overline{D^{(\mu)}}{(\beta)}$, as outlined in Eq.~\eqref{eq:diff_const_limits}, we proceed to extract the Thouless time $t^{(\mu)}_\mathrm{Th}$. To achieve this, we simply substitute the obtained value of $\overline{D^{(\mu)}}{(\beta)}$ into the formula for $t^{(\mu)}_\mathrm{Th}$ given in Eq.~\eqref{eq:t_Th}. For all graphical and quantitative analyses of $t^{(\mu)}_\mathrm{Th}$ as a function of the linear system size $L$, we use the estimated large-system-size values of the diffusion constant. Corrections must be sufficient small such that $t^{(\mu)}_\mathrm{Th} \propto L^2$ is the leading $L$ dependence.

We also note that, for some large $W/J_\parallel$ or $J_\perp/J_\parallel$, the spin- and energy-diffusion constants decrease and become nearly identical, which can be understood by looking at the scales in Eqs.~\eqref{eq:loc_spin_current}, \eqref{eq:loc_energy_current}, \eqref{eq:spin_suscep} and \eqref{eq:energy_suscep}. This yields identical Thouless times for $W \gg J_\parallel$.

For the computation of these diffusion constants, we use the grand-canonical ensemble of the spin models, corresponding to all of the total magnetization symmetry sectors ($\expval{S^z} = 0$), where the time evolution is carried out via the Lanczos method with dynamical quantum typicality (DQT) \cite{hamsFastAlgorithmFinding2000,bartschDynamicalTypicalityQuantum2009, elsayedRegressionRelationPure2013, steinigewegSpinCurrentAutocorrelationsSingle2014, steinigewegPushingLimitsEigenstate2014, steinigewegSpinEnergyCurrents2015, steinigewegTypicalityApproachOptical2016} (see Appendix~\ref{app:dqt} for the details). For the Lanczos algorithm, we discretize time with a time step of $\Delta t = 0.01/J_\parallel$ and carry out a maximum of $30$ Lanczos iterations. To compute the disorder-averaged diffusion constants, we employ an ensemble of $M=1000$ independent disorder realizations.

We estimate $\overline{D^{(\mu)}}$ by taking an average of $\overline{D^{(\mu)}}{(t)}$ in the time interval $200 \leq t J_\parallel \leq 250$ for the largest two systems sizes we have ($N=18, 20$). In contrast with the clean case \cite{steinigewegScalingDiffusionConstants2014}, we find that taking late-time results for the estimate is more convenient because a saturation in $\overline{D^{(\mu)}}{(t)}$ emerges late in the disordered case.

Further details related to the conserved quantities, currents, diffusion constants, and static susceptibilities are discussed in Appendix~\ref{app:diff_const}.

\section{\label{sec:results}Results}

\subsection{\label{sec:sff_results}SFF and RMT time}

\begin{figure}
    \centering
    \includegraphics[width=\linewidth]{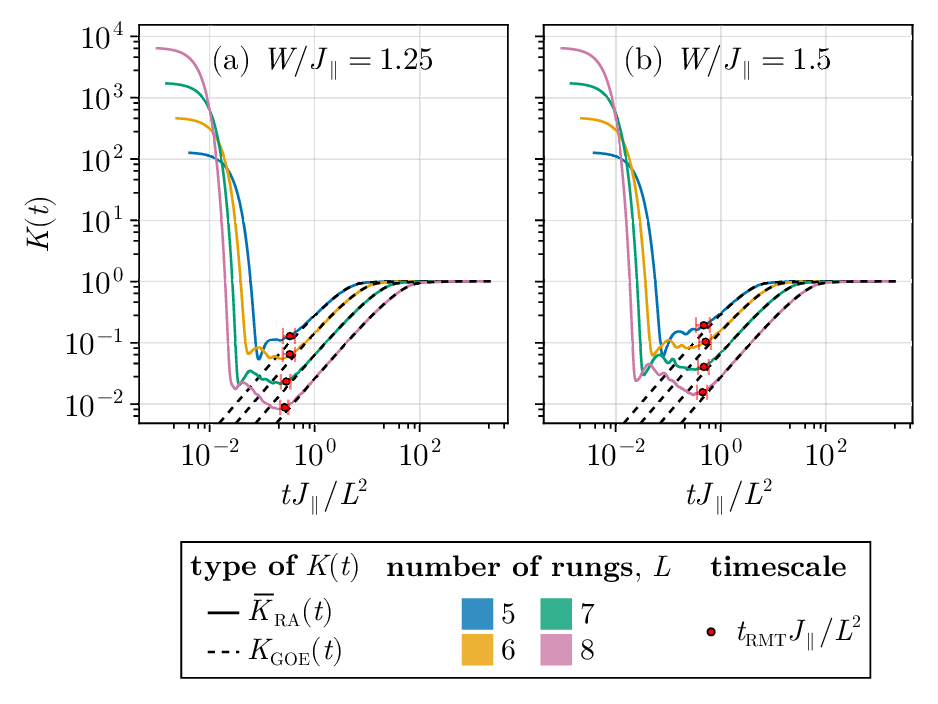}
    \caption{Running-and-disorder-averaged
SFF $\overline{K}_\mathrm{RA}{(t)}$ of a spin-$1/2$ XX ladder as a function of time normalized by $L^2$ at disorder strengths (a) $W/J_\parallel = 1.25$ and (b) $W/J_\parallel = 1.5$ and coupling ratio $J_\perp/J_\parallel = 1.0$. The (real) time is obtained from the normalized time $\tau$ via $t = \frac{\tau}{2\pi} t_\mathrm{H}$, where $t_\mathrm{H}$ denotes the Heisenberg time. The color of each line corresponds to a specific number of rungs, $L$. Solid lines represent $\overline{K}_\mathrm{RA}{(t)}$, while black dashed lines denote the SFF of the GOE $K_\mathrm{GOE}{(t)}$. Each SFF is computed with a discrete time step in $\log_{10}$ scale $\Delta(\log_{10}\tau)$ of $0.001$ using a running average window in $\log_{10}$ scale $\delta(\log_{10}\tau)$ of $0.025$ and a Gaussian filter width factor $\eta$ of $0.5$. All SFFs are averaged over $M = 5000$ disorder realizations. The RMT time $t_\mathrm{RMT}$ (red dots with error bars) is determined for each $L$ by varying the threshold value for the criterion function $\Delta K{(t)}$, $\epsilon_{\Delta K}$, within the range $[0.035, 0.2]$.}
    \label{fig:disorder-averaged_SFF_t_for_Ls}
\end{figure}

For a disordered spin-$1/2$ XX ladder, we first present the running-and-disorder-averaged SFF $\overline{K}_\mathrm{RA}{(t)}$ in Fig.~\ref{fig:disorder-averaged_SFF_t_for_Ls} as a function of time normalized by the squared number of rungs $L^2$ at the coupling ratio $J_\perp/J_\parallel = 1.0$ and the disorder strengths $W/J_\parallel = 1.25$ in Fig.~\reflabel{fig:disorder-averaged_SFF_t_for_Ls}{(a)} and $W/J_\parallel = 1.5$ in Fig.~\reflabel{fig:disorder-averaged_SFF_t_for_Ls}{(b)}.

Due to the choice of a large number of disorder realizations in the SFF and the exponential growth of the Hilbert-space dimension of our model with increasing the system size $N$, we compute $\overline{K}_\mathrm{RA}{(t)}$ for four different numbers of rungs $L \in \{5,6,7,8\}$, where we indicate these numbers of rungs by different colors. For all system sizes we consider, we can clearly distinguish the expected three regimes in the SFF in Fig.~\ref{fig:disorder-averaged_SFF_t_for_Ls}. More importantly, all values of $t_\mathrm{RMT}$, indicated by the red dots with the error bars, align linearly along the SFF axis within the error range determined by the threshold value $\epsilon_{\Delta{K}}$ and the Gaussian filter width factor $\eta$. This implies the existence of an $L^2$ scaling in the RMT time, i.e.,~$t_\mathrm{RMT}{(L)} \propto L^2$. We notice that we could not identify $L^2$-scaling of RMT times for small disorder strengths such as  $W/J_\parallel \lesssim 1.0$ even for large system sizes. 

\subsection{\label{sec:diff_const_results} Thouless times from diffusion constants}

\begin{figure}
    \centering
    \includegraphics[width=\linewidth]{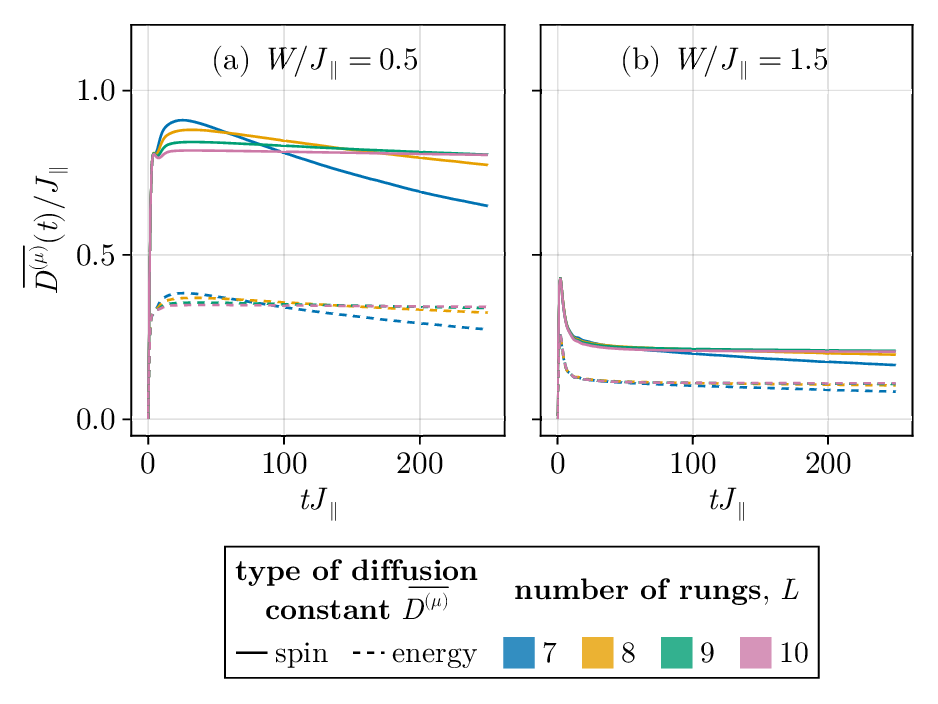}
    \caption{Time-dependent disorder-averaged diffusion constants $\overline{D^{(\mu)}}{(t)}$ of a spin-$1/2$ XX ladder as a function of time $t$ for disorder strengths (a) $W/J_\parallel = 0.5$ and (b) $W/J_\parallel = 1.5$ and coupling ratio $J_\perp/J_\parallel = 1.0$. The color of each line corresponds to a specific number of rungs $L$. Solid lines represent the spin-diffusion constant $\overline{D^{(\mathrm{S})}}{(t)}$, while dashed lines denote the energy-diffusion constant $\overline{D^{(\mathrm{E})}}{(t)}$. Each diffusion constant is averaged over $M = 1000$ disorder realizations. The time evolution is performed using the Lanczos algorithm with a discrete time step of $\Delta t = 0.01 / J_\parallel$ and a maximum of 30 Lanczos iterations.}
    \label{fig:disorder-averaged_D_t_for_Ls}
\end{figure}

We plot the time-dependent disorder-averaged diffusion constant of spin transport (solid lines) and energy transport (dashed lines), indicated by $\overline{D^{(\mathrm{S})}}{(t)}$ and $\overline{D^{(\mathrm{E})}}{(t)}$, as a function of time at  the coupling ratio $J_\perp/J_\parallel = 1.0$ and the disorder strengths $W/J_\parallel = 0.5$ in Fig.~\reflabel{fig:disorder-averaged_D_t_for_Ls}{(a)} and $W/J_\parallel = 1.5$ in Fig.~\reflabel{fig:disorder-averaged_D_t_for_Ls}{(b)}. The line colors represent four different numbers of rungs, with $L \in \{ 7, 8, 9, 10 \}$.

At weak disorder strengths, exemplified by $W/J_\parallel=0.5$ in Fig.~\reflabel{fig:disorder-averaged_D_t_for_Ls}{(a)}, the disorder-averaged diffusion constant $\overline{D^{(\mu)}}{(t)}$ exhibits an initial increase, driven by the positive current autocorrelation function $\overline{C^{(\mu)}}{(t)} > 0$ at short times. However, in contrast with the clean limit \cite{steinigewegScalingDiffusionConstants2014}, this growth is transient, and $\overline{C^{(\mu)}}{(t)}$ eventually undergoes a first sign change at a certain time (followed by oscillations and more sign changes). For stronger disorder strengths, such as $W/J_\parallel = 1.5$ in Fig.~\reflabel{fig:disorder-averaged_D_t_for_Ls}{(b)}, the sign change in $\overline{C^{(\mu)}}{(t)}$ occurs at an earlier time, marking a more rapid departure from the initial growth regime.

The slow approach toward a constant long-time value in Fig.~\ref{fig:disorder-averaged_D_t_for_Ls} is caused by a transient, presumably subdiffusive behavior. For sufficiently large system sizes, we expect this transient behavior to give way to standard diffusion as seen in other disordered models \cite{karahaliosFinitetemperatureTransportDisordered2009, barisicConductivityDisorderedOnedimensional2010, barisicDynamicalConductivityIts2016, steinigewegTypicalityApproachOptical2016, prelovsekSlowDiffusionThouless2023, herbrychSpinEnergyDiffusion2025}, leading to the formation of a plateau in the time-dependent diffusion constant at late times and thus providing an approximate value for the diffusion constant for our case. We also note that increasing the disorder strength accelerates the formation of the late-time plateau for small system sizes, as shown in Fig.~\ref{fig:disorder-averaged_D_t_for_Ls}.

Moreover, we find the energy-diffusion constant to be smaller than the spin-diffusion constant, indicating that spin diffusion occurs more rapidly than energy diffusion. As shown in Fig.~\ref{fig:disorder-averaged_D_t_for_Ls}, increasing the disorder strength $W/J_\parallel$ leads to a decrease in the difference between the diffusion constants. We also observe a similar reduction in the difference when the coupling ratio $J_\perp/J_\parallel$ is increased. Beyond a certain threshold value for these parameters, the diffusion constants converge to the same value. This is also evident from the large limiting cases of $W/J_\parallel$ and $J_\perp/J_\parallel$ in Eqs.~\eqref{eq:loc_spin_current}, \eqref{eq:loc_energy_current}, \eqref{eq:spin_suscep}, and \eqref{eq:energy_suscep}, where the equivalence of the diffusion constants can be inferred without the need for explicit computation of $\overline{D^{(\mathrm{S})}}{(t)}$ and $\overline{D^{(\mathrm{E})}}${(t)} as we do in Fig.~\ref{fig:disorder-averaged_D_t_for_Ls}.

\subsection{\label{sec:relax_time_results}Hierarchy of relaxation timescales}

\subsubsection{Disordered spin-\texorpdfstring{$1/2$}{1/2} XX ladder}

\begin{figure}
    \centering
    \includegraphics[width=\linewidth]{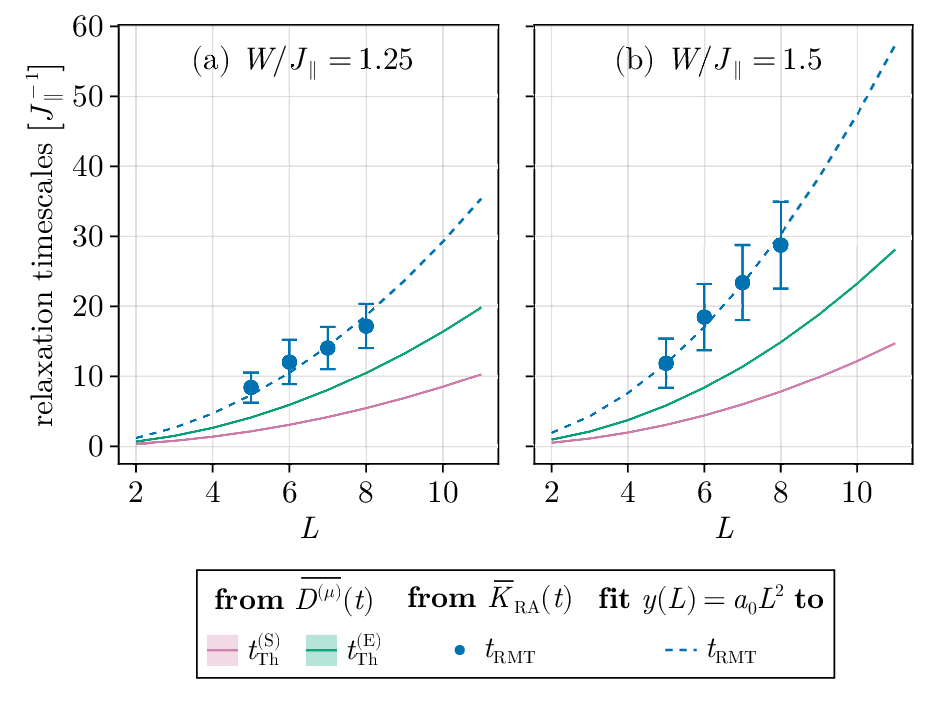}
    \caption{Relaxation timescales of a spin-$1/2$ XX ladder as a function of the number of the rungs, $L$, at disorder strengths (a) $W/J_\parallel = 1.25$ and (b) $W/J_\parallel = 1.5$ and coupling ratio $J_\perp/J_\parallel = 1.0$. The blue circles with error bars denote the RMT times $t_\mathrm{RMT}$ derived from the SFF, and these error bars represent the systematic uncertainties in extracting $t_\mathrm{RMT}$. The blue dashed lines correspond to the fits of the function $y(L)=a_0 L^2$ to the RMT time $t_\mathrm{RMT}$. The spin-Thouless time $t^{(\mathrm{S})}_\mathrm{Th}$ (purple lines with shaded bands) and the energy-Thouless time $t^{(\mathrm{E})}_\mathrm{Th}$ (green lines with shaded bands) are obtained via the combined mean and standard deviation of the disorder-averaged diffusion constant $\overline{D^{(\mu)}}{(t)}$ computed over the interval $200 \leq t J_\parallel \leq 250$ for $L = 9, 10$. The uncertainty of the diffusion constants is within the linewidth. The number of disorder realizations is $M=5000$ for $t_\mathrm{RMT}$ and $M=1000$ for $t_\mathrm{Th}$. The other parameters used to obtain this figure can be found in Figs.~\ref{fig:disorder-averaged_SFF_t_for_Ls} and \ref{fig:disorder-averaged_D_t_for_Ls}.}
    \label{fig:relax_ts_vs_L_for_XX_ladder}
\end{figure}

Building on our previous results from Figs.~\ref{fig:disorder-averaged_SFF_t_for_Ls} and \ref{fig:disorder-averaged_D_t_for_Ls}, we finally show the relaxation timescales as a function of the number of rungs $L$ for a coupling ratio of $J_\perp/J_\parallel = 1.0$ and two different disorder strengths: $W/J_\parallel = 1.25$ in Fig.~\reflabel{fig:relax_ts_vs_L_for_XX_ladder}{(a)} and $W/J_\parallel = 1.5$ in Fig.~\reflabel{fig:relax_ts_vs_L_for_XX_ladder}{(b)}. The blue circles with error bars represent the RMT times $t_\mathrm{RMT}$ from Eq.~\eqref{eq:t_RMT}, which are obtained using the Heisenberg time $t_\mathrm{H}$. To estimate $t_\mathrm{RMT}$ at larger values of $L$, we also fit the function $y(L)=a_0 L^2$ to the data for $t_\mathrm{RMT}$ (dashed lines) for the set $L \in \{ 5, 6, 7, 8 \}$. We do not extract the exponent independently; therefore, our procedure only demonstrates a consistent $L^2$ scaling. Likewise, the solid lines with shaded bands indicate the Thouless times $t^{(\mu)}_\mathrm{Th}$ defined in Eq.~\eqref{eq:t_Th}. The energy-Thouless time $t^{(\mathrm{E})}_\mathrm{Th}$ is represented by the green color, while the spin-Thouless time $t^{(\mathrm{S})}_\mathrm{Th}$ is denoted by the purple color. As mentioned in Sec.~\ref{sec:thouless_times}, the computation of $t^{(\mu)}_\mathrm{Th}$ involves estimating the time-independent disorder-averaged diffusion constants $\overline{D^{(\mu)}}$, by considering a portion of the late-time behavior at sufficiently large values of $L$, which yields their combined mean value (solid lines) and standard deviation (shaded bands). The width of those bands is smaller than the linewidth.

As expected, Fig.~\ref{fig:relax_ts_vs_L_for_XX_ladder} reveals that for both transport channels,
\begin{equation} 
    t^{(\mu)}_\mathrm{Th} < t_\mathrm{RMT}\,,
\end{equation}
that is, the Thouless times are upper bounded by the RMT time.

As was previously noted in Fig.~\ref{fig:disorder-averaged_D_t_for_Ls} and evident from Fig.~\ref{fig:relax_ts_vs_L_for_XX_ladder}, spin diffusion processes faster than energy diffusion in the spin-$1/2$ XX ladder. This is  reflected in 
\begin{equation}
    t^{(\mathrm{S})}_\mathrm{Th} < t^{(\mathrm{E})}_\mathrm{Th} \,.
\end{equation}
We also note that $t^{(\mathrm{S})}_\mathrm{Th}< t^{(\mathrm{E})}_\mathrm{Th}$ is seen at all disorder strengths investigated, i.e.,~$W/J_\parallel \lesssim 2$ including the clean model. This hierarchy of Thouless times shown in Fig.~\ref{fig:relax_ts_vs_L_for_XX_ladder} for a spin‑$1/2$ XX ladder is different from other examples of nonintegrable models studied previously \cite{luitzErgodicSideManybody2017, richterMagnetizationEnergyDynamics2019, herbrychSpinEnergyDiffusion2025}. 

\subsubsection{Disordered spin-\texorpdfstring{$1/2$}{1/2} XXZ chain with next-nearest-neighbor couplings}

\begin{figure}
    \centering
    \includegraphics[width=\linewidth]{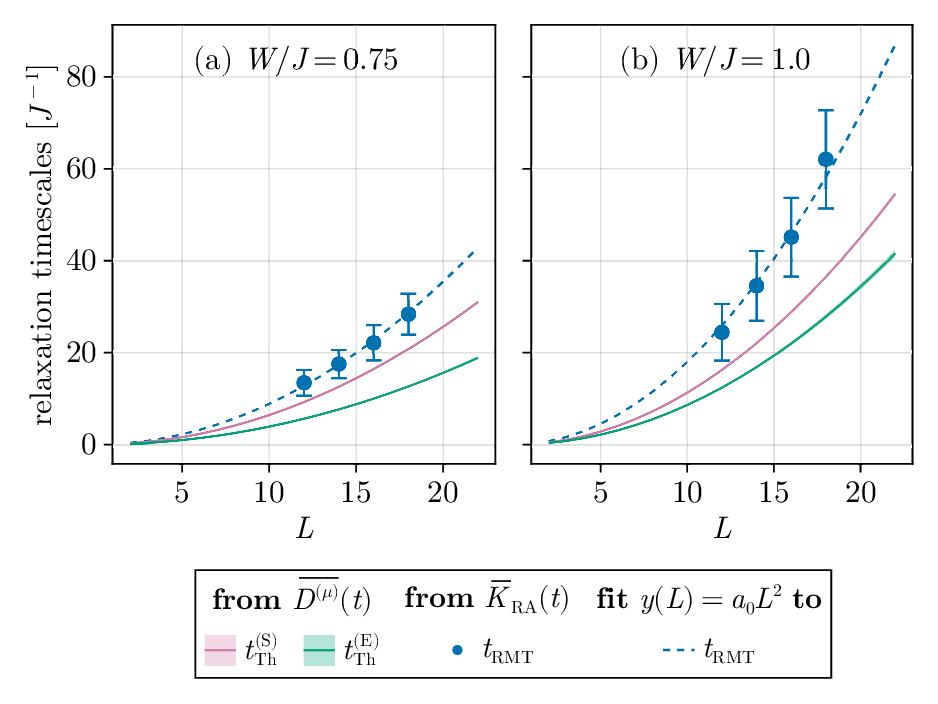}
    \caption{Relaxation timescales of a spin-$1/2$ XXZ chain with next-nearest-neighbor couplings as a function of system size $L$ at disorder strengths (a) $W/J = 0.75$ and (b) $W/J = 1.0$, nearest-neighboring anisotropy $\Delta/J = 1.0$, and next-nearest-neighboring anisotropy $\Delta_2/J = 0.3$. The blue circles with error bars denote the RMT times $t_\mathrm{RMT}$ derived from the SFF, and these error bars represent the systematic uncertainties in extracting $t_\mathrm{RMT}$. The blue dashed lines correspond to  the fits of the function $y(L)=a_0 L^2$ to the RMT time $t_\mathrm{RMT}$. The spin-Thouless time $t^{(\mathrm{S})}_\mathrm{Th}$ (purple lines with shaded bands) and the energy-Thouless time $t^{(\mathrm{E})}_\mathrm{Th}$ (green lines with shaded bands)  are obtained via the combined mean and standard deviation of the disorder-averaged diffusion constant $\overline{D^{(\mu)}}{(t)}$ computed over the interval $200 \leq t J \leq 250$ for $L = 18, 20$. The uncertainty of the diffusion constants is within the linewidth. The number of disorder realizations is $M=5000$ for $t_\mathrm{RMT}$ and $M=1000$ for $t_\mathrm{Th}$. The other parameters used to obtain this figure can be found in Figs.~\ref{fig:disorder-averaged_SFF_t_for_Ls} and \ref{fig:disorder-averaged_D_t_for_Ls}.}
    \label{fig:relax_ts_vs_L_for_XXZ_chain}
\end{figure}

In fact, slower energy diffusion relative to spin diffusion is indeed an exception because there are other nonintegrable models that do not exhibit this unusual hierarchy. To check that, we consider a disordered spin-$1/2$ XXZ chain with next-nearest-neighbor couplings. In the literature, it has been established that this model obeys ETH in its clean limit ($W=0$) \cite{rigolQuantumQuenchesThermalization2009, rigolBreakdownThermalizationFinite2009}, while its transport properties were studied in Refs.~\cite{znidaricMagnetizationTransportSpin2013, steinigewegDecayCurrentsStrong2011, zotosEvidenceIdealInsulating1996, narozhnyTransportXXZModel1998}.

The Hamiltonian of this system can be expressed as a sum of four distinct operators:
\begin{equation}
    H{(\wmat)} = H_0 + H_{\Delta} + H_{\Delta_2} + H_\mathrm{dis}{(\wmat)}\,.
\end{equation}

The first component $H_0$ describes the spin couplings along the chain and is given by
\begin{equation}
    H_0 = \frac{J}{2} \sum_{\ell=1}^L \qty(S^+_{\ell} S^-_{\ell+1} + S^-_{\ell} S^+_{\ell+1})\,,
\end{equation}
where $J$ represents the coupling strength between neighboring spins, and $L$ denotes the length of the chain. The second component $H_\Delta$ accounts for the longitudinal coupling between nearest-neighbor spins and is expressed as
\begin{equation}\label{eq:XXZ_ham_Delta_part}
    H_\Delta = \Delta \sum_{\ell=1}^L S^z_{\ell} S^z_{\ell+1}\,,
\end{equation}
with $\Delta$ representing the nearest-neighbor anisotropy. The third component $H_{\Delta_2}$ is similar to $H_\Delta$ but describes the coupling between next-nearest-neighbor spins, given by
\begin{equation}
    H_{\Delta_2} = \Delta_2 \sum_{\ell=1}^L S^z_{\ell} S^z_{\ell+2}\,,
\end{equation}
where $\Delta_2$ represents the next-nearest-neighboring anisotropy. The last part is for the disorder terms, as in Eq.~\eqref{eq:XX_ladder_ham_disorder_part}, defined as
\begin{equation}
    H_{\mathrm{dis}}{(\wmat)} = \sum_{\ell=1}^L w_{\ell} S^z_{\ell}\,.
\end{equation}

A notable observation can be made from Fig.~\ref{fig:relax_ts_vs_L_for_XXZ_chain}, where the additional model exhibits a reversal in the order of diffusion constants. Despite this reversal, the Thouless times $t^{(\mu)}_\mathrm{Th}$ remain smaller than the RMT time $t_\mathrm{RMT}$ and have an $L^2$ scaling, 
as expected for diffusive models.

\subsection{Discussion}

Let us discuss the key results: (i) the fact that the RMT time is larger than all Thouless times and (ii) that in the spin-1/2 XX ladder, energy transport is slower than spin transport. Our results are consistent with the usual picture that the $L^2$ dependence of the RMT time is inherited from diffusive transport. The fact that the RMT is larger than the largest Thouless time may have several reasons. First of all, the SFF measures the entire spectrum and therefore, it sees contributions from the temperature dependence of the diffusion constant, while we computed the diffusion constants at infinite temperature from the Kubo formalism. At sufficiently large systems, the contributions to $t_{\mathrm{RMT}}$ should be dominated by the bulk of the spectrum, though. Technically, there is no unique way of extracting the RMT time from the SFF as we detailed before, and hence, $t_{\mathrm{RMT}}$ is only determined up to a prefactor. The difference $t_{\mathrm{RMT}} - t_{\mathrm{Th}}^{(\mathrm{E})}$ is, however, larger than this uncertainty according to our analysis. Conceptually, the prefactor between $t_{\mathrm{RMT}}$ and the largest Thouless time may simply not be one, as one could check by a calculation of the diffusion constant from the spectrum with twisted boundary conditions \cite{prelovsekSlowDiffusionThouless2023}, which is left for future work. Finally, it cannot be completely ruled out  that another unidentified relaxation timescale is larger than the Thouless times and closer $t_{\mathrm{RMT}}$, even though this appears unlikely in this model.

We now turn to the observation that
\begin{equation}
    t_{\mathrm{Th}}^{(\mathrm{S})} < t_{\mathrm{Th}}^{(\mathrm{E})}\,
\end{equation}
for the spin ladder. To understand this behavior, it is instructive to consider the clean limit. Special to XX models, there are no diagonal terms at all in the computational basis (i.e.,~the joint eigenbasis of all $S^z_{\ell}$ operators). Since we compute the diffusion constants at infinite temperature, we can use any basis. In the computational basis, it is obvious that a spin excitation can move between different rungs (as the center of the local energy density) without changing the local energy since the energy expectation value vanishes in all basis states. Consequently, we expect spin to be able to  move faster than energy at infinite temperature in the absence of diagonal terms.

To substantiate this picture, we also carried out simulations for a spin-$1/2$ XXZ ladder with $\Delta>0$ (results not shown here). At sufficiently large values of $\Delta$,  we find $D^{(\mathrm{E})} > D^{(\mathrm{S})}$, consistent with the results of Ref.~\cite{richterMagnetizationEnergyDynamics2019} obtained for Heisenberg ladders. Whether a finite nonzero value of the exchange anisotropy $\Delta$ is required to reverse the order of the diffusion constants is left for future work.
 
In principle, disorder introduces diagonal terms, coupling the modification of the local spin moment to the local energy density again.
Our results suggest that, at a sufficiently large disorder strength $W$, the diffusion constants converge to the same value because the local energy density picks up a term proportional to the local spin density, hence tying energy transport to spin transport. An interesting open question is the connection of our results to the existence of ballistic and localized subspaces in certain spin-ladder models \cite{znidaricMagnetizationTransportSpin2013,Iadecola2019}, which we leave for future work.

\section{\label{sec:conclusions}Conclusions}

We investigated the nonequilibrium dynamics of a disordered spin-$1/2$ XX ladder by utilizing a pair of diagnostic measures from quantum chaos and transport. Specifically, we employed the SFF to extract the timescale associated with the onset of RMT, i.e.,~the RMT time. Additionally, we used the infinite-temperature diffusion constant  to characterize the different transport properties of the system and to determine the Thouless times associated with spin and energy diffusion.

We found that the RMT time scales quadratically with linear system size, suggesting that it is controlled by diffusive transport. Moreover, we found that the RMT time is larger than the Thouless times. We pointed out several possible reasons for this quantitative difference.

Among the two transport channels in the spin-$1/2$ XX ladder, the  spin-Thouless time is smaller than the energy-Thouless time. We argue that this is due to the absence of any diagonal terms in the computational basis, implying that spin can move without affecting the local energy density. We expect that most spin models should exhibit the more generic behavior, i.e.,~$D^{(\mathrm{E})} > D^{(\mathrm{S})}$. We verified this for a nonintegrable spin-$1/2$ XXZ chain and for spin-1/2 XXZ ladders, where the RMT time again scales with $L^2$ yet $t_\mathrm{Th}^{(\mathrm{E})} < t_\mathrm{Th}^{(\mathrm{S})}$.

Possible extensions of our work may concern the temperature dependence of the diffusion constant and the search for further examples where the 
RMT time is not controlled by transport (see, e.g.,~Refs.~\cite{kosManyBodyQuantumChaos2018, bertiniExactSpectralForm2018, chanSpectralStatisticsSpatially2018, chanSolutionMinimalModel2018, friedmanSpectralStatisticsManyBody2019, bertiniRandomMatrixSpectral2021, fritzschEigenstateThermalizationDualunitary2021}).

\begin{acknowledgments}
We thank Sarang Gopalakrishnan, Miroslav Hopjan, Andreas M. Läuchli, Tomaž Prosen, Marcos Rigol, Rafał \'{S}wi\c{e}tek, and Lev Vidmar for helpful discussions. Furthermore, we are grateful to Marko \v{Z}nidari\v{c} for comments on a previous version of the manuscript and for bringing Ref.~\cite{Iadecola2019}
to our attention.
This work was funded by the Deutsche Forschungsgemeinschaft (DFG, German Research Foundation)---499180199,  493420525 via FOR 5522 and a large-equipment grant (GOEGrid cluster).
\end{acknowledgments}

\section*{Data \texorpdfstring{\&}{&} code availability}

The data that support the findings of this article are partially
openly available \cite{zenodo}.

\appendix

\section{\label{appen:sec:crossover-diagram}Disorder-induced delocalized-localized regime crossover diagram}

\begin{figure}
    \centering
    \includegraphics[width=\linewidth]{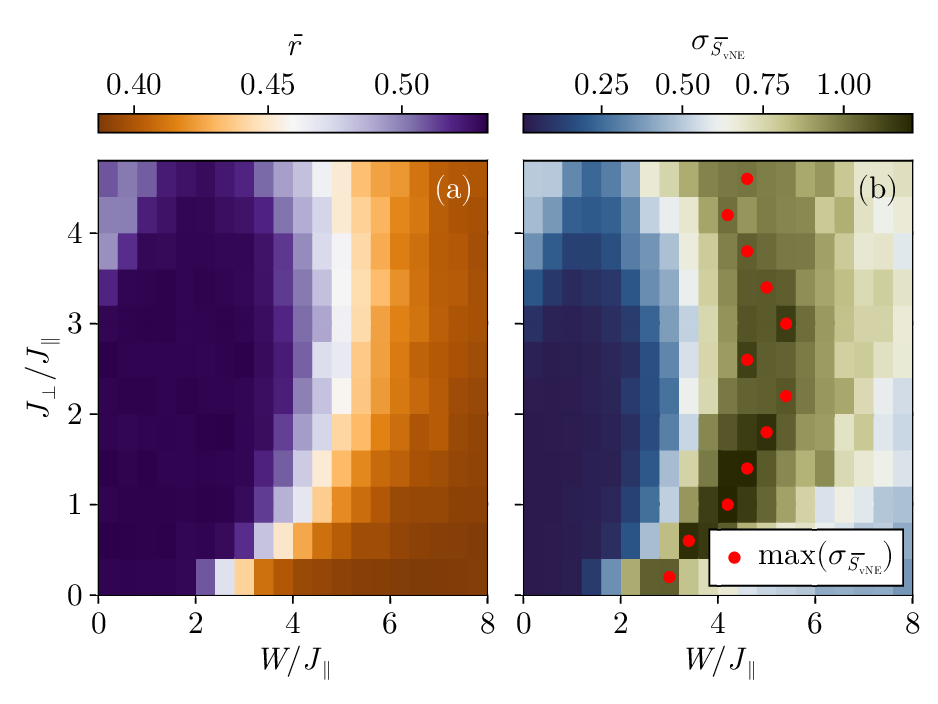}
    \caption{Disorder-induced delocalization-localization crossover diagrams of a spin-$1/2$ XX ladder as function of the disorder strength $W/J_\parallel$ and the coupling ratio $J_\perp/J_\parallel$ for $L=9$ for a dimensionless normalized energy in the center of the spectrum. The diagrams are based on two different measures: (a) the state-to-state and sample-to-sample average of the adjacent gap ratio $\overline{r}$ and (b) the state-to-state and sample-to-sample average of the standard deviation of the entanglement entropy $\sigma_{\overline{S}_\mathrm{vNE}}$. Each colored rectangle represents an average over two steps: first, a subset of eigenstates around the center of the energy spectrum and then $M = 90$ disorder realizations. In (b), the red dots indicate the maximum value of $\sigma_{\overline{S}_\mathrm{vNE}}{(W/J_\parallel)}$ at a fixed value of $J_\perp/J_\parallel$. }
    \label{fig:crossover-diagrams}
\end{figure}

Inducing disorder can lead to localization  \cite{sierantManybodyLocalizationAge2025}. The localization regime can be avoided by choosing an appropriate parameter set with respect to  the finite-size crossover from the delocalized regime to the localized one. For the disordered spin-$1/2$ XX ladder, we use two measures: the adjacent gap ratio $\overline{r}$ \cite{oganesyanLocalizationInteractingFermions2007, santosOnsetQuantumChaos2010, palManybodyLocalizationPhase2010, atasDistributionRatioConsecutive2013, dalessioQuantumChaosEigenstate2016} and the maximal standard deviation of the entanglement entropy $\sigma_{\overline{S}_\mathrm{vNE}}$ \cite{songBipartiteFluctuationsProbe2012, bauerAreaLawsManybody2013, kjallManyBodyLocalizationDisordered2014, luitzManybodyLocalizationEdge2015, nandkishoreManyBodyLocalizationThermalization2015, limManybodyLocalizationTransition2016}. In Fig.~\ref{fig:crossover-diagrams}, we display the delocalization-localization crossover diagram using these heuristics for $L=9$ and $M=90$.

\subsection{\label{app:gap_ratio}Adjacent gap ratio}

The adjacent gap ratio allows us to check whether the spectral level statistics of the system follow the predictions from the RMT. These predictions purely originate from the adjacent energy gap $\omega_{\alpha}$, defined as
\begin{equation}\label{eq:adjacent_energy_gap}
    \omega_{\alpha} \coloneqq E_\alpha - E_{\alpha-1}\,,
\end{equation}
where $\alpha$ is the index for the larger eigenvalue of the adjacent energy pair $\{ E_{\alpha-1}, E_\alpha \}$. We then introduce the dimensionless adjacent gap ratio for a single disorder realization, $r_\alpha$ as follows
\begin{equation}\label{eq:dimensionless_gap_ratio}
    r_\alpha = \frac{\min\{\omega_{\alpha}, \omega_{\alpha+1}\}}{\max\{\omega_{\alpha}, \omega_{\alpha+1}\}}\,.
\end{equation}
 The motivation for utilizing the ratios of energy gaps rather than the energy gaps themselves results from the necessity to unfold the spectrum when using the latter. 

Due to the potential edge effects in the energy spectrum, which arise from the very low density of states at extreme energy eigenvalues, we only use some portion of the energy eigenvalues from the spectral center for the adjacent gap ratio. We introduce the dimensionless normalized energy $\tilde{E}_\alpha$, given by
\begin{equation}
    \tilde{E}_\alpha = \frac{E_\alpha - E_\mathrm{min}}{E_\mathrm{max} - E_\mathrm{min}}\,,
\end{equation}
which reduces the total energy interval to $[0,1]$. 
For the ladder model from Eq.~\eqref{eq:XX_ladder_ham}, we only consider the $10\%$ closest eigenstates to the center of the energy spectrum in the $S^z=0$ subspace.

Based on Eqs.~\eqref{eq:adjacent_energy_gap} and \eqref{eq:dimensionless_gap_ratio}, the state-to-state and sample-to-sample average of adjacent gap ratio $\overline{r}$ amounts to
\begin{equation}
    \overline{r} = \E{\Ec{r_\alpha}}\,,
\end{equation}
where we first carry out the central-portion and disorder averages in succession denoted by $\mathbb{E}_{\mathrm{c}}$ and $\mathbb{E}_{\wmat}$, respectively.

In the localized regime, the systems follow a Poisson distribution of uncorrelated level spacings, signaling the absence of level repulsion. As a consequence, the averaged adjacent gap ratio in the localized regime is $\overline{r}_\mathrm{Pois} \approx 0.3863$ \cite{dalessioQuantumChaosEigenstate2016}. In the delocalized regime, the systems exhibit the spectral features of random matrices. For time-reversal symmetric systems, the adjacent-gap distribution equals a Wigner-Dyson distribution for the GOE, reflecting energy-level repulsion. For a large random matrix of dimension  $1000 \times 1000$, this results in $\overline{r}_\mathrm{GOE} \approx 0.5307$  \cite{atasDistributionRatioConsecutive2013,dalessioQuantumChaosEigenstate2016}.

In Fig.~\reflabel{fig:crossover-diagrams}{(a)}, we plot the state-to-state and sample-to-sample average of the adjacent gap ratio as a function of both the disorder strength $W/J_\parallel$ and the coupling ratio $J_\perp/J_\parallel$. As expected, there are two distinct regions, distinguished by their averaged adjacent gap ratios, which are consistent with the theoretical values from GOE and Poisson statistics. Specifically, these regions correspond to a delocalized regime (dark purple) and a localized regime (dark orange).

\subsection{Standard deviation of the entanglement entropy}

Our second measure for identifying the crossover between delocalized and localized regimes is the von Neumann entanglement entropy $S_\mathrm{vNE}$ or more specifically, its maximal standard deviation $\sigma_{S_\mathrm{vNE}}$. To compute this quantity, we consider a bipartition of the system across the bond between rungs $L/2$ and $L/2 + 1$, which divides the system into two subsystems $A$ and $B$. Then the entanglement entropy $S_\mathrm{vNE}$ of a many-body pure state $\ket{\psi}$ for this bipartition can be written as
\begin{equation}\label{eq:S_vNE_single_state_realization}
    S_\mathrm{vNE}{\qty(\ket{\psi})} = - \Trace_{A}\!\qty{\rho_A{\qty(\ket{\psi})} \ln[\rho_A{\qty(\ket{\psi})}]}\,,
\end{equation}
where $\rho_A{\qty(\ket{\psi})} = \Tr_B(\dyad{\psi})$ is the reduced density matrix of the subsystem $A$ with the many-body pure state $\ket{\psi}$. 

With Eq.~\eqref{eq:S_vNE_single_state_realization}, the state-to-state and sample-to-sample average of entanglement entropy, denoted by $\overline{S}_\mathrm{vNE}$, can be defined as the average of the von Neumann entanglement entropy over a subset of energy eigenstates and disorder realizations. This average can be expressed as
\begin{equation}
    \overline{S}_\mathrm{vNE} = \E{\Ec{S_\mathrm{vNE}{(\ket{E})}}}\,.
\end{equation}
In a similar manner to the approach given in Appendix~\ref{app:gap_ratio}, we focus on the 10\% of eigenstates closest to the center of the energy spectrum in the 
$S^z=0$ subspace.

In the delocalized regime, since the reduced density matrix of the subsystem $A$, $\rho_A$, is expected to satisfy the ETH for an energy eigenstate located in the middle of spectrum, the von Neumann entropy exhibits a volume law. In the localized regime and on finite systems, the energy eigenstates follow an area-law scaling \cite{bauerAreaLawsManybody2013, kjallManyBodyLocalizationDisordered2014}. The crossover region can be also predicted by examining the standard deviation of $\overline{S}_\mathrm{vNE}$, defined as
\begin{equation}
    \sigma_{\overline{S}_\mathrm{vNE}} = \sqrt{\E{\Ec{S^2_\mathrm{vNE}{(\ket{E})}}} - \overline{S}_\mathrm{vNE}^2}\,.
\end{equation}
In the delocalized regime and the localized regime, $\sigma_{\overline{S}_\mathrm{vNE}}$ converges toward zero as the system size increases.  At the onset of crossover regime, $\sigma_{\overline{S}_\mathrm{vNE}}$ exhibits a peak whose amplitude increases and whose width decreases with increasing system size.

In Fig.~\reflabel{fig:crossover-diagrams}{(b)}, we show the state-to-state and sample-to-sample average of the standard devitation of the von Neumann entanglement entropy $\sigma_{\overline{S}_\mathrm{vNE}}$ as a function of both the disorder strength $W/J_\parallel$ and the coupling ratio $J_\perp/J_\parallel$. We observe the same behavior as for the gap ratio shown Fig.~\reflabel{fig:crossover-diagrams}{(a)}. For our study, we choose $W/J_\parallel < 2$, which puts us safely away from the finite-size localized regime.

\section{SFF}

\subsection{\label{app:unfolding}Unfolding the energy spectrum}

We begin the spectral unfolding procedure by defining the cumulative spectral function of an energy for a given disorder realization as
\begin{equation}\label{eq:cum_spec_func}
    G{\qty(E)} \coloneqq \sum_{\alpha=1}^{\mathcal{D}} \Theta{\qty(E - E_\alpha)}\,,
\end{equation}
where $\Theta{(\,\cdots)}$ denotes the Heaviside step function, and $\mathcal{D}$ is the Hilbert-space dimension. We then smoothen out this function based on the polynomial regression with the degree $n$, resulting in an approximating function $\hat{G}^{(n)}(E)$. This  function is used to define the $n$ th-degree unfolded energy eigenvalues $\varepsilon^{(n)}_\alpha$, given by
\begin{equation}
    \varepsilon^{(n)}_\alpha = \hat{G}^{(n)}{(E_\alpha)}\,.
\end{equation}
For certain energy spectra, the sensitivity of unfolded energies to different polynomial degrees can be a problem. To mitigate this issue, we employ the following criterion function \cite{abul-magdUnfoldingSpectrumChaotic2014}:
\begin{equation}\label{eq:Lambda_criterion_func}
    \Lambda^{(n)} = \frac{4}{\mathcal{D}} \sum_{\alpha=1}^\mathcal{D} \qty[\frac{\varepsilon^{(n)}_\alpha - G{\qty(E_\alpha)}}{\varepsilon^{(n)}_\alpha + G{\qty(E_\alpha)}}]^2\,.
\end{equation}
This criterion function measures how the unfolded energies for a chosen polynomial degree deviate from the cumulative spectral function defined in Eq.~\eqref{eq:cum_spec_func}. By searching in a certain interval for the polynomial degrees, the optimum polynomial degree for unfolding $n_0{\qty( \wmat )}$ can be found depending on the criterion function in Eq.~\eqref{eq:Lambda_criterion_func}. In this paper, we limit the polynomial degrees to be in the interval $[2,18]$, so that
\begin{equation}\label{eq:opt_poly_n}
    n_0 \coloneqq \argmin_{2 \leq n \leq 18}{\Lambda^{(n)}}\,.
\end{equation}
Based on the optimal polynomial degree given in Eq.~\eqref{eq:opt_poly_n}, the (optimal) unfolded energy eigenvalue for a disorder realization, denoted by $\varepsilon_\alpha$, can be expressed as
\begin{equation}
    \varepsilon_\alpha = \varepsilon^{(n_0)}_\alpha\,.
\end{equation}

\subsection{\label{app:filtering}Gaussian filtering}

In addition to the spectral unfolding procedure covered in Appendix~\ref{app:unfolding}, we also apply a Gaussian filter function to mitigate the edge effects arising in $\overline{K}{(\tau)}$, as defined in Eq.~\eqref{eq:sff}, as a result of the unfolding procedure itself. We define this filter function for an unfolded energy eigenvalue in a disorder realization $\rho_\mathrm{f}{\qty(\varepsilon_\alpha)}$ as follows:
\begin{equation}
    \rho_\mathrm{f}{\qty(\varepsilon_\alpha)} = \exp[- \frac{1}{2}\qty(\frac{\varepsilon_\alpha - \overline{\varepsilon}}{\eta \sigma_\varepsilon})^2]\,,
\end{equation}
where $\overline{\varepsilon}$ and $\sigma_\varepsilon$ are the mean and the standard deviation of a given unfolded energy spectrum, and $\eta$ is a factor that controls the width of the Gaussian filter, thereby tuning its filtering strength.

\subsection{\label{app:fix_shift_in_tau_H}Fixing the shift in the normalized Heisenberg time}

To fix the shifting problem of the normalized Heisenberg time discussed in Sec.~\ref{sec:RMT_time}, we reweight each adjacent unfolded energy gap by utilizing the filter function given in Appendix~\ref{app:filtering}, yielding the weighted-mean-level spacing
\begin{equation}\label{eq:weighted-mean_level_spacing} 
\overline{\tilde{\delta \varepsilon}}_\mathrm{mean} \coloneqq \E{\tilde{N}_\omega^{-1} \sum_{\alpha} \omega_{\alpha} \, \rho_\mathrm{f}{\qty(\varepsilon_\alpha)} \, \rho_\mathrm{f}{\qty(\varepsilon_{\alpha-1})}}\,,
\end{equation}
where $\tilde{N}_\omega \coloneqq \sum_\alpha \rho_\mathrm{f}{\qty(\varepsilon_\alpha)} \, \rho_\mathrm{f}{\qty(\varepsilon_{\alpha-1})}$ is the weighted number of adjacent energy gaps with respect to the filter $\rho_\mathrm{f}$. We then rescale the normalized times, denoted by $\tau$, with respect to $\overline{\tilde{\delta \varepsilon}}_\mathrm{mean}$, i.e.,
\begin{equation}
    \tau \leftarrow \overline{\tilde{\delta \varepsilon}}_\mathrm{mean} \times  \tau\,.
\end{equation}

\subsection{\label{app:GOE}Gaussian orthogonal ensemble}

The analytical results for the SFF vary depending on the specific ensemble from the RMT being considered, but they can still be obtained analytically. For instance, in the case of a chaotic Hamiltonian following the time-reversal symmetry, the random matrix distributions correspond to the GOE. The SFF for the GOE is given by \cite{mehtaRandomMatrices2004}
\begin{equation}
    K_\mathrm{GOE}{(\tau)} = \begin{cases}
         \frac{\tau}{\pi} - \frac{\tau}{2\pi}\ln(\frac{\tau}{\pi} + 1) & \qif* \tau \leq 2\pi\\
         2 -\frac{\tau}{2\pi} \ln(\frac{\frac{\tau}{\pi} + 1}{\frac{\tau}{\pi} - 1}) & \qif* \tau > 2\pi
    \end{cases}\,.
\end{equation}

\subsection{\label{app:real_t_Hs}Heisenberg time}

To revert to real units on the time axis, we first calculate the Heisenberg time, defined as
\begin{equation}
t_\mathrm{H} = \frac{2\pi}{\overline{\delta E}}\,,
\end{equation}
where $\overline{\delta E}$ is the mean-level spacing, given by
\begin{equation}
\overline{\delta E} = \E{N_\omega^{-1}\sum_\alpha \omega_{\alpha}}\,,
\end{equation}
where $N_\omega$ is the number of adjacent gaps. The time $t$ measured in real units is related to $\tau $ by
\begin{equation}
    t = \frac{\tau}{2\pi} t_\mathrm{H} = \frac{\tau}{\overline{\delta E}}\,.
\end{equation}

\section{\label{app:diff_const}Diffusion constant}

\subsection{Initial current autocorrelation functions}

With the local densities of transport quantities and local current operators defined as described in Sec.~\ref{sec:thouless_times}, we can proceed to establish the current autocorrelation function $C^{(\mu)}{(t; {\beta})}$ within the framework of linear-response theory, as expressed in Eq.~\eqref{eq:autocorrelation_C_J}. In the limit of infinite temperature, where $\beta \rightarrow 0$, the current autocorrelation function simplifies to
\begin{equation}\label{eq:C_func_infty_T_canonical}
    C^{(\mu)}{(t; \beta=0)} = \Re{\frac{\Trace\!\qty{ J^{(\mu)}{(t)} \, J^{(\mu)}{(0)}}}{N \mathcal{D}}}\,,
\end{equation}
where $\mathcal{D}$ is the Hilbert-space dimension and $N$ is the total number of spins. This simplified expression provides a foundation for further analysis of the transport properties of the system in the high-temperature limit.

In this limit, the current autocorrelation functions for spin and energy currents can be evaluated at $t=0$. Specifically, the spin-current autocorrelation function at $t=0$ is given by
\begin{equation}
    C^\mathrm{(S)}{(t=0, \beta=0)} = \frac{J_\parallel^2}{8}\,,
\end{equation}
while the energy-current autocorrelation function at $t=0$ is found to be
\begin{widetext}
\begin{equation}
    C^\mathrm{(E)}{(t=0; \beta=0)} = \frac{J_\parallel^2}{64} \qty[ 2J_\parallel^2 + J_\perp^2 + \frac{2}{N} \sum_{\ell, i} \qty(w_{\ell, i} + w_{\ell+1, i})^2 ]\,.
\end{equation}
\end{widetext}
These results serve as a useful checkmark for confirming that the computation of transport measures is properly implemented in the high-temperature regime.

\subsection{Static susceptibility per spin}

The disorder-averaged infinite-temperature static susceptibilities, as presented in Eqs.~\eqref{eq:spin_suscep} and \eqref{eq:energy_suscep}, can be defined as
\begin{equation}
    \overline{\chi_\mu}{(\beta=0)} = \E{\chi_{\mu}{(\beta=0)}}\,,
\end{equation}
where $\chi_{\mu}{(\beta)}$ represents the static susceptibility per spin for a specific disorder realization at a given inverse temperature. This quantity is calculated using the following expression
\begin{equation}
    \chi_\mu{(\beta)} = \frac{\expval{\qty[Q^{(\mu)}]^2}_\beta - \expval{Q^{(\mu)}}_\beta^2}{N}\,.
\end{equation}

To obtain the disorder-averaged infinite-temperature static susceptibilities given in Eqs.~\eqref{eq:spin_suscep} and \eqref{eq:energy_suscep}, we first rewrite their conserved quantities as a linear combinations of Pauli strings for a single disorder realization, where every parameter of the Hamiltonian is stored symbolically, and then trace them out to find their expectation values. For the disorder average, we symbolically integrate them over each $w_{\ell, i}$ with respect to their distributions, which eventually yields $\overline{\chi_\mu}{(\beta=0)}$.

\subsection{\label{app:dqt}DQT}

The concept of DQT leverages the fact that, in a high-dimensional Hilbert space, a single pure state can replicate the characteristics of an entire statistical ensemble. This thus enables an approximate computation of the expectation value of current autocorrelation functions \cite{elsayedRegressionRelationPure2013, steinigewegSpinCurrentAutocorrelationsSingle2014, steinigewegPushingLimitsEigenstate2014, steinigewegSpinEnergyCurrents2015, steinigewegTypicalityApproachOptical2016}.

The autocorrelation function of total currents, typically defined as in Eq.~\eqref{eq:autocorrelation_C_J}, can be reexpressed using the  DQT framework. This still yields an exact expression for the current autocorrelation function up to the order of the inverse of the effective Hilbert space dimension $d_\mathrm{eff}{(\beta)} = \Trace\!\qty{e^{-\beta ( H - E_0 )}}$, given by
\begin{widetext}
\begin{equation}\label{eq:C_dqt_with_O}
    C^{(\mu)}{(t; \beta)} = \Re{\frac{\mel{\psi_\beta}{J^{(\mu)}{(t)} \, J^{(\mu)}{(0)}}{\psi_\beta}}{N\braket{\psi_\beta}}} + \order{\frac{1}{ \sqrt{d_\mathrm{eff}{(\beta)}}}}\,,
\end{equation}
\end{widetext}
where $\ket{\psi_\beta} \coloneqq e^{-\beta H/2} \ket{\psi_\mathrm{Haar}}$ is the thermal random state constructed by the Haar-random state for a given disorder realization realization $\wmat$, $\ket{\psi_\mathrm{Haar}}$. Notably, $d_\mathrm{eff}{(\beta)}$ reduces to the Hilbert-space dimension $\mathcal{D}$ in the limit of high temperatures ($\beta \rightarrow 0$). This also indicates that the second term in Eq.~\eqref{eq:C_dqt_with_O} disappears exponentially fast as $N$ increases, which makes the first term in Eq.~\eqref{eq:C_dqt_with_O} a good approximation for the current autocorrelation function $C^{(\mu)}{(t; \beta)}$ for a large enough $N$.

To further simplify the time evolution in Eq.~\eqref{eq:C_dqt_with_O}, we introduce two new states
\begin{eqnarray}
    \ket{\Phi_\beta{(t)}} &\coloneqq& e^{-i H t} \ket{\psi_\beta}\,,\label{eq:DQT_Phi_state}\\
    \ket{\varphi_\beta{(t)}} &\coloneqq& e^{-i H t} J^{(\mu)}{(0)} \ket{\psi_\beta}\,.\label{eq:DQT_varphi_state}
\end{eqnarray}
These states represent the time-evolved thermal random states, with and without the application of the total current operator, respectively. Here, by switching from the Heisenberg picture to the Schrödinger picture, we basically express the time evolution in a more convenient form. Using Eqs.~\eqref{eq:DQT_Phi_state} and \eqref{eq:DQT_varphi_state}, we can approximate the current autocorrelation function $C^{(\mu)}{(t; \beta)}$ as
\begin{equation}
    C^{(\mu)}{(t; \beta}) \approx \Re{\frac{\mel{\Phi_\beta{(t)}}{J^{(\mu)}{(0)}}{\varphi_\beta{(t)}}}{N\braket{\psi_\beta}}}\,.
\end{equation}

\begin{figure}
    \centering
    \includegraphics[width=\linewidth]{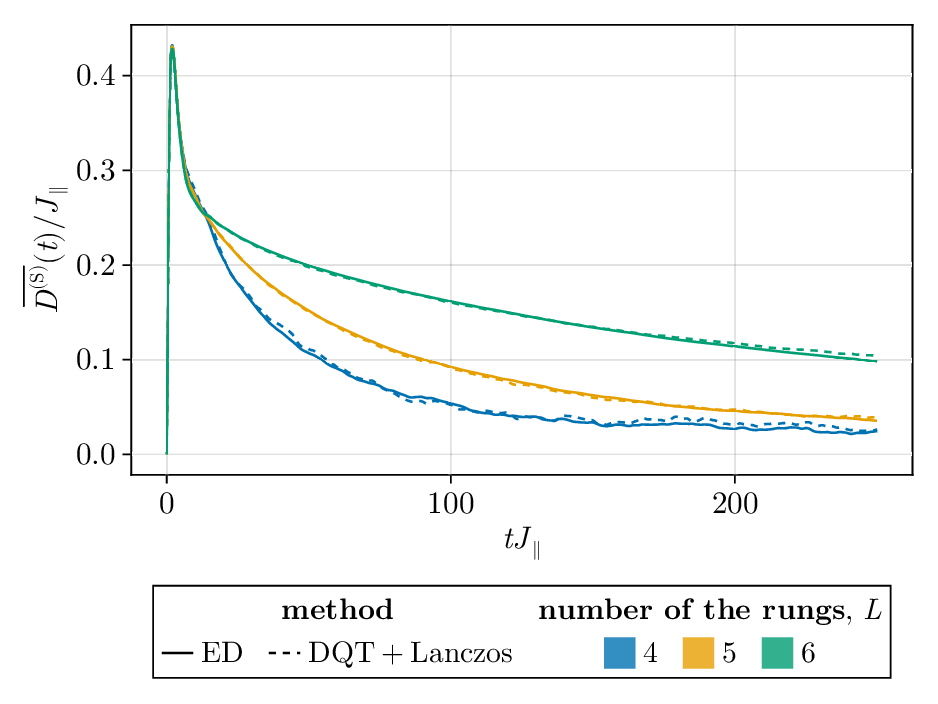}
    \caption{Time-dependent disorder-averaged spin-diffusion constants $\overline{D^{(\mathrm{S})}}{(t)}$ of a spin-$1/2$ XX ladder as a function of time $t$ at disorder strength $W/J_\parallel = 1.5$ and coupling ratio $J_\perp/J_\parallel = 1.0$ for different numbers of rungs $L=4$ (blue), $L=5$ (yellow) and $L=6$ (green). Results obtained with ED are shown as solid lines, while those from the DQT using the Lanczos algorithm are plotted as dashed lines. The parameters employed are identical to those listed in Fig.~\ref{fig:disorder-averaged_D_t_for_Ls}.}
    \label{fig:disorder-averaged_D_t_for_Ls_with_ED}
\end{figure}

To assess the reliability of the diffusion constants obtained with the DQT, we compare them against exact‑diagonalization (ED) results in Fig.~\ref{fig:disorder-averaged_D_t_for_Ls_with_ED}. As $L$ increases, the time-dependent disorder-averaged spin-diffusion constants computed via the DQT with the Lanczos algorithm exhibit a closer correspondence with the ED data. While for small $L\sim 4$, the data quality could be improved by sampling over more independent Haar-random states disorder realization, we find that going to larger $L$ the discrepancy with ED decreases quickly.\footnote{We also verify that the utilization of the same Haar-random state per disorder realization still gives rise to similar results.} In our calculations for diffusion constants, we go up to $L=10$.

\bibliography{references}

\begin{thebibliography}{140}%
\makeatletter
\providecommand \@ifxundefined [1]{%
 \@ifx{#1\undefined}
}%
\providecommand \@ifnum [1]{%
 \ifnum #1\expandafter \@firstoftwo
 \else \expandafter \@secondoftwo
 \fi
}%
\providecommand \@ifx [1]{%
 \ifx #1\expandafter \@firstoftwo
 \else \expandafter \@secondoftwo
 \fi
}%
\providecommand \natexlab [1]{#1}%
\providecommand \enquote  [1]{``#1''}%
\providecommand \bibnamefont  [1]{#1}%
\providecommand \bibfnamefont [1]{#1}%
\providecommand \citenamefont [1]{#1}%
\providecommand \href@noop [0]{\@secondoftwo}%
\providecommand \href [0]{\begingroup \@sanitize@url \@href}%
\providecommand \@href[1]{\@@startlink{#1}\@@href}%
\providecommand \@@href[1]{\endgroup#1\@@endlink}%
\providecommand \@sanitize@url [0]{\catcode `\\12\catcode `\$12\catcode
  `\&12\catcode `\#12\catcode `\^12\catcode `\_12\catcode `\%12\relax}%
\providecommand \@@startlink[1]{}%
\providecommand \@@endlink[0]{}%
\providecommand \url  [0]{\begingroup\@sanitize@url \@url }%
\providecommand \@url [1]{\endgroup\@href {#1}{\urlprefix }}%
\providecommand \urlprefix  [0]{URL }%
\providecommand \Eprint [0]{\href }%
\providecommand \doibase [0]{https://doi.org/}%
\providecommand \selectlanguage [0]{\@gobble}%
\providecommand \bibinfo  [0]{\@secondoftwo}%
\providecommand \bibfield  [0]{\@secondoftwo}%
\providecommand \translation [1]{[#1]}%
\providecommand \BibitemOpen [0]{}%
\providecommand \bibitemStop [0]{}%
\providecommand \bibitemNoStop [0]{.\EOS\space}%
\providecommand \EOS [0]{\spacefactor3000\relax}%
\providecommand \BibitemShut  [1]{\csname bibitem#1\endcsname}%
\let\auto@bib@innerbib\@empty
\bibitem [{\citenamefont {D'Alessio}\ \emph {et~al.}(2016)\citenamefont
  {D'Alessio}, \citenamefont {Kafri}, \citenamefont {Polkovnikov},\ and\
  \citenamefont {Rigol}}]{dalessioQuantumChaosEigenstate2016}%
  \BibitemOpen
  \bibfield  {author} {\bibinfo {author} {\bibfnamefont {L.}~\bibnamefont
  {D'Alessio}}, \bibinfo {author} {\bibfnamefont {Y.}~\bibnamefont {Kafri}},
  \bibinfo {author} {\bibfnamefont {A.}~\bibnamefont {Polkovnikov}},\ and\
  \bibinfo {author} {\bibfnamefont {M.}~\bibnamefont {Rigol}},\ }\bibfield
  {title} {\bibinfo {title} {From quantum chaos and eigenstate thermalization
  to statistical mechanics and thermodynamics},\ }\href
  {https://doi.org/10.1080/00018732.2016.1198134} {\bibfield  {journal}
  {\bibinfo  {journal} {Adv. Phys.}\ }\textbf {\bibinfo {volume} {65}},\
  \bibinfo {pages} {239} (\bibinfo {year} {2016})}\BibitemShut {NoStop}%
\bibitem [{\citenamefont {Short}\ and\ \citenamefont
  {Farrelly}(2012)}]{shortQuantumEquilibrationFinite2012}%
  \BibitemOpen
  \bibfield  {author} {\bibinfo {author} {\bibfnamefont {A.~J.}\ \bibnamefont
  {Short}}\ and\ \bibinfo {author} {\bibfnamefont {T.~C.}\ \bibnamefont
  {Farrelly}},\ }\bibfield  {title} {\bibinfo {title} {Quantum equilibration in
  finite time},\ }\href {https://doi.org/10.1088/1367-2630/14/1/013063}
  {\bibfield  {journal} {\bibinfo  {journal} {New J. Phys.}\ }\textbf {\bibinfo
  {volume} {14}},\ \bibinfo {pages} {013063} (\bibinfo {year}
  {2012})}\BibitemShut {NoStop}%
\bibitem [{\citenamefont {Goldstein}\ \emph {et~al.}(2013)\citenamefont
  {Goldstein}, \citenamefont {Hara},\ and\ \citenamefont
  {Tasaki}}]{goldsteinTimeScalesApproach2013}%
  \BibitemOpen
  \bibfield  {author} {\bibinfo {author} {\bibfnamefont {S.}~\bibnamefont
  {Goldstein}}, \bibinfo {author} {\bibfnamefont {T.}~\bibnamefont {Hara}},\
  and\ \bibinfo {author} {\bibfnamefont {H.}~\bibnamefont {Tasaki}},\
  }\bibfield  {title} {\bibinfo {title} {Time {{Scales}} in the {{Approach}} to
  {{Equilibrium}} of {{Macroscopic Quantum Systems}}},\ }\href
  {https://doi.org/10.1103/PhysRevLett.111.140401} {\bibfield  {journal}
  {\bibinfo  {journal} {Phys. Rev. Lett.}\ }\textbf {\bibinfo {volume} {111}},\
  \bibinfo {pages} {140401} (\bibinfo {year} {2013})}\BibitemShut {NoStop}%
\bibitem [{\citenamefont {Malabarba}\ \emph {et~al.}(2014)\citenamefont
  {Malabarba}, \citenamefont {{Garc{\'i}a-Pintos}}, \citenamefont {Linden},
  \citenamefont {Farrelly},\ and\ \citenamefont
  {Short}}]{malabarbaQuantumSystemsEquilibrate2014}%
  \BibitemOpen
  \bibfield  {author} {\bibinfo {author} {\bibfnamefont {A.~S.~L.}\
  \bibnamefont {Malabarba}}, \bibinfo {author} {\bibfnamefont {L.~P.}\
  \bibnamefont {{Garc{\'i}a-Pintos}}}, \bibinfo {author} {\bibfnamefont
  {N.}~\bibnamefont {Linden}}, \bibinfo {author} {\bibfnamefont {T.~C.}\
  \bibnamefont {Farrelly}},\ and\ \bibinfo {author} {\bibfnamefont {A.~J.}\
  \bibnamefont {Short}},\ }\bibfield  {title} {\bibinfo {title} {Quantum
  systems equilibrate rapidly for most observables},\ }\href
  {https://doi.org/10.1103/PhysRevE.90.012121} {\bibfield  {journal} {\bibinfo
  {journal} {Phys. Rev. E}\ }\textbf {\bibinfo {volume} {90}},\ \bibinfo
  {pages} {012121} (\bibinfo {year} {2014})}\BibitemShut {NoStop}%
\bibitem [{\citenamefont {Goldstein}\ \emph {et~al.}(2015)\citenamefont
  {Goldstein}, \citenamefont {Hara},\ and\ \citenamefont
  {Tasaki}}]{goldsteinExtremelyQuickThermalization2015}%
  \BibitemOpen
  \bibfield  {author} {\bibinfo {author} {\bibfnamefont {S.}~\bibnamefont
  {Goldstein}}, \bibinfo {author} {\bibfnamefont {T.}~\bibnamefont {Hara}},\
  and\ \bibinfo {author} {\bibfnamefont {H.}~\bibnamefont {Tasaki}},\
  }\bibfield  {title} {\bibinfo {title} {Extremely quick thermalization in a
  macroscopic quantum system for a typical nonequilibrium subspace},\ }\href
  {https://doi.org/10.1088/1367-2630/17/4/045002} {\bibfield  {journal}
  {\bibinfo  {journal} {New J. Phys.}\ }\textbf {\bibinfo {volume} {17}},\
  \bibinfo {pages} {045002} (\bibinfo {year} {2015})}\BibitemShut {NoStop}%
\bibitem [{\citenamefont {{Torres-Herrera}}\ \emph {et~al.}(2015)\citenamefont
  {{Torres-Herrera}}, \citenamefont {Kollmar},\ and\ \citenamefont
  {Santos}}]{torres-herreraRelaxationThermalizationIsolated2015}%
  \BibitemOpen
  \bibfield  {author} {\bibinfo {author} {\bibfnamefont {E.~J.}\ \bibnamefont
  {{Torres-Herrera}}}, \bibinfo {author} {\bibfnamefont {D.}~\bibnamefont
  {Kollmar}},\ and\ \bibinfo {author} {\bibfnamefont {L.~F.}\ \bibnamefont
  {Santos}},\ }\bibfield  {title} {\bibinfo {title} {Relaxation and
  thermalization of isolated many-body quantum systems},\ }\href
  {https://doi.org/10.1088/0031-8949/2015/T165/014018} {\bibfield  {journal}
  {\bibinfo  {journal} {Phys. Scr.}\ }\textbf {\bibinfo {volume} {T165}},\
  \bibinfo {pages} {014018} (\bibinfo {year} {2015})}\BibitemShut {NoStop}%
\bibitem [{\citenamefont
  {Reimann}(2016)}]{reimannTypicalFastThermalization2016}%
  \BibitemOpen
  \bibfield  {author} {\bibinfo {author} {\bibfnamefont {P.}~\bibnamefont
  {Reimann}},\ }\bibfield  {title} {\bibinfo {title} {Typical fast
  thermalization processes in closed many-body systems},\ }\href
  {https://doi.org/10.1038/ncomms10821} {\bibfield  {journal} {\bibinfo
  {journal} {Nat. Commun.}\ }\textbf {\bibinfo {volume} {7}},\ \bibinfo {pages}
  {10821} (\bibinfo {year} {2016})}\BibitemShut {NoStop}%
\bibitem [{\citenamefont {Gogolin}\ and\ \citenamefont
  {Eisert}(2016)}]{gogolinEquilibrationThermalisationEmergence2016}%
  \BibitemOpen
  \bibfield  {author} {\bibinfo {author} {\bibfnamefont {C.}~\bibnamefont
  {Gogolin}}\ and\ \bibinfo {author} {\bibfnamefont {J.}~\bibnamefont
  {Eisert}},\ }\bibfield  {title} {\bibinfo {title} {Equilibration,
  thermalisation, and the emergence of statistical mechanics in closed quantum
  systems},\ }\href {https://doi.org/10.1088/0034-4885/79/5/056001} {\bibfield
  {journal} {\bibinfo  {journal} {Rep. Prog. Phys.}\ }\textbf {\bibinfo
  {volume} {79}},\ \bibinfo {pages} {056001} (\bibinfo {year}
  {2016})}\BibitemShut {NoStop}%
\bibitem [{\citenamefont {Friedman}\ \emph {et~al.}(2019)\citenamefont
  {Friedman}, \citenamefont {Chan}, \citenamefont {De~Luca},\ and\
  \citenamefont {Chalker}}]{friedmanSpectralStatisticsManyBody2019}%
  \BibitemOpen
  \bibfield  {author} {\bibinfo {author} {\bibfnamefont {A.~J.}\ \bibnamefont
  {Friedman}}, \bibinfo {author} {\bibfnamefont {A.}~\bibnamefont {Chan}},
  \bibinfo {author} {\bibfnamefont {A.}~\bibnamefont {De~Luca}},\ and\ \bibinfo
  {author} {\bibfnamefont {J.~T.}\ \bibnamefont {Chalker}},\ }\bibfield
  {title} {\bibinfo {title} {Spectral {{Statistics}} and {{Many-Body Quantum
  Chaos}} with {{Conserved Charge}}},\ }\href
  {https://doi.org/10.1103/PhysRevLett.123.210603} {\bibfield  {journal}
  {\bibinfo  {journal} {Phys. Rev. Lett.}\ }\textbf {\bibinfo {volume} {123}},\
  \bibinfo {pages} {210603} (\bibinfo {year} {2019})}\BibitemShut {NoStop}%
\bibitem [{\citenamefont {Edwards}\ and\ \citenamefont
  {Thouless}(1972)}]{edwardsNumericalStudiesLocalization1972}%
  \BibitemOpen
  \bibfield  {author} {\bibinfo {author} {\bibfnamefont {J.~T.}\ \bibnamefont
  {Edwards}}\ and\ \bibinfo {author} {\bibfnamefont {D.~J.}\ \bibnamefont
  {Thouless}},\ }\bibfield  {title} {\bibinfo {title} {Numerical studies of
  localization in disordered systems},\ }\href
  {https://doi.org/10.1088/0022-3719/5/8/007} {\bibfield  {journal} {\bibinfo
  {journal} {J. Phys. C: Solid State Phys.}\ }\textbf {\bibinfo {volume} {5}},\
  \bibinfo {pages} {807} (\bibinfo {year} {1972})}\BibitemShut {NoStop}%
\bibitem [{\citenamefont
  {Thouless}(1974)}]{thoulessElectronsDisorderedSystems1974}%
  \BibitemOpen
  \bibfield  {author} {\bibinfo {author} {\bibfnamefont {D.~J.}\ \bibnamefont
  {Thouless}},\ }\bibfield  {title} {\bibinfo {title} {Electrons in disordered
  systems and the theory of localization},\ }\href
  {https://doi.org/10.1016/0370-1573(74)90029-5} {\bibfield  {journal}
  {\bibinfo  {journal} {Phys. Rep.}\ }\textbf {\bibinfo {volume} {13}},\
  \bibinfo {pages} {93} (\bibinfo {year} {1974})}\BibitemShut {NoStop}%
\bibitem [{\citenamefont
  {Thouless}(1977)}]{thoulessMaximumMetallicResistance1977}%
  \BibitemOpen
  \bibfield  {author} {\bibinfo {author} {\bibfnamefont {D.~J.}\ \bibnamefont
  {Thouless}},\ }\bibfield  {title} {\bibinfo {title} {Maximum {{Metallic
  Resistance}} in {{Thin Wires}}},\ }\href
  {https://doi.org/10.1103/PhysRevLett.39.1167} {\bibfield  {journal} {\bibinfo
   {journal} {Phys. Rev. Lett.}\ }\textbf {\bibinfo {volume} {39}},\ \bibinfo
  {pages} {1167} (\bibinfo {year} {1977})}\BibitemShut {NoStop}%
\bibitem [{\citenamefont
  {Deutsch}(1991)}]{deutschQuantumStatisticalMechanics1991}%
  \BibitemOpen
  \bibfield  {author} {\bibinfo {author} {\bibfnamefont {J.~M.}\ \bibnamefont
  {Deutsch}},\ }\bibfield  {title} {\bibinfo {title} {Quantum statistical
  mechanics in a closed system},\ }\href
  {https://doi.org/10.1103/PhysRevA.43.2046} {\bibfield  {journal} {\bibinfo
  {journal} {Phys. Rev. A}\ }\textbf {\bibinfo {volume} {43}},\ \bibinfo
  {pages} {2046} (\bibinfo {year} {1991})}\BibitemShut {NoStop}%
\bibitem [{\citenamefont
  {Srednicki}(1994)}]{srednickiChaosQuantumThermalization1994}%
  \BibitemOpen
  \bibfield  {author} {\bibinfo {author} {\bibfnamefont {M.}~\bibnamefont
  {Srednicki}},\ }\bibfield  {title} {\bibinfo {title} {Chaos and quantum
  thermalization},\ }\href {https://doi.org/10.1103/PhysRevE.50.888} {\bibfield
   {journal} {\bibinfo  {journal} {Phys. Rev. E}\ }\textbf {\bibinfo {volume}
  {50}},\ \bibinfo {pages} {888} (\bibinfo {year} {1994})}\BibitemShut
  {NoStop}%
\bibitem [{\citenamefont
  {Srednicki}(1999)}]{srednickiApproachThermalEquilibrium1999}%
  \BibitemOpen
  \bibfield  {author} {\bibinfo {author} {\bibfnamefont {M.}~\bibnamefont
  {Srednicki}},\ }\bibfield  {title} {\bibinfo {title} {The approach to thermal
  equilibrium in quantized chaotic systems},\ }\href
  {https://doi.org/10.1088/0305-4470/32/7/007} {\bibfield  {journal} {\bibinfo
  {journal} {J. Phys. A: Math. Gen.}\ }\textbf {\bibinfo {volume} {32}},\
  \bibinfo {pages} {1163} (\bibinfo {year} {1999})}\BibitemShut {NoStop}%
\bibitem [{\citenamefont {Rigol}\ \emph {et~al.}(2008)\citenamefont {Rigol},
  \citenamefont {Dunjko},\ and\ \citenamefont
  {Olshanii}}]{rigolThermalizationItsMechanism2008}%
  \BibitemOpen
  \bibfield  {author} {\bibinfo {author} {\bibfnamefont {M.}~\bibnamefont
  {Rigol}}, \bibinfo {author} {\bibfnamefont {V.}~\bibnamefont {Dunjko}},\ and\
  \bibinfo {author} {\bibfnamefont {M.}~\bibnamefont {Olshanii}},\ }\bibfield
  {title} {\bibinfo {title} {Thermalization and its mechanism for generic
  isolated quantum systems},\ }\href {https://doi.org/10.1038/nature06838}
  {\bibfield  {journal} {\bibinfo  {journal} {Nature}\ }\textbf {\bibinfo
  {volume} {452}},\ \bibinfo {pages} {854} (\bibinfo {year}
  {2008})}\BibitemShut {NoStop}%
\bibitem [{\citenamefont {Bertini}\ \emph {et~al.}(2018)\citenamefont
  {Bertini}, \citenamefont {Kos},\ and\ \citenamefont
  {Prosen}}]{bertiniExactSpectralForm2018}%
  \BibitemOpen
  \bibfield  {author} {\bibinfo {author} {\bibfnamefont {B.}~\bibnamefont
  {Bertini}}, \bibinfo {author} {\bibfnamefont {P.}~\bibnamefont {Kos}},\ and\
  \bibinfo {author} {\bibfnamefont {T.}~\bibnamefont {Prosen}},\ }\bibfield
  {title} {\bibinfo {title} {Exact {{Spectral Form Factor}} in a {{Minimal
  Model}} of {{Many-Body Quantum Chaos}}},\ }\href
  {https://doi.org/10.1103/PhysRevLett.121.264101} {\bibfield  {journal}
  {\bibinfo  {journal} {Phys. Rev. Lett.}\ }\textbf {\bibinfo {volume} {121}},\
  \bibinfo {pages} {264101} (\bibinfo {year} {2018})}\BibitemShut {NoStop}%
\bibitem [{\citenamefont {Chan}\ \emph
  {et~al.}(2018{\natexlab{a}})\citenamefont {Chan}, \citenamefont {De~Luca},\
  and\ \citenamefont {Chalker}}]{chanSolutionMinimalModel2018}%
  \BibitemOpen
  \bibfield  {author} {\bibinfo {author} {\bibfnamefont {A.}~\bibnamefont
  {Chan}}, \bibinfo {author} {\bibfnamefont {A.}~\bibnamefont {De~Luca}},\ and\
  \bibinfo {author} {\bibfnamefont {J.~T.}\ \bibnamefont {Chalker}},\
  }\bibfield  {title} {\bibinfo {title} {Solution of a {{Minimal Model}} for
  {{Many-Body Quantum Chaos}}},\ }\href
  {https://doi.org/10.1103/PhysRevX.8.041019} {\bibfield  {journal} {\bibinfo
  {journal} {Phys. Rev. X}\ }\textbf {\bibinfo {volume} {8}},\ \bibinfo {pages}
  {041019} (\bibinfo {year} {2018}{\natexlab{a}})}\BibitemShut {NoStop}%
\bibitem [{\citenamefont {Chan}\ \emph
  {et~al.}(2018{\natexlab{b}})\citenamefont {Chan}, \citenamefont {De~Luca},\
  and\ \citenamefont {Chalker}}]{chanSpectralStatisticsSpatially2018}%
  \BibitemOpen
  \bibfield  {author} {\bibinfo {author} {\bibfnamefont {A.}~\bibnamefont
  {Chan}}, \bibinfo {author} {\bibfnamefont {A.}~\bibnamefont {De~Luca}},\ and\
  \bibinfo {author} {\bibfnamefont {J.~T.}\ \bibnamefont {Chalker}},\
  }\bibfield  {title} {\bibinfo {title} {Spectral {{Statistics}} in {{Spatially
  Extended Chaotic Quantum Many-Body Systems}}},\ }\href
  {https://doi.org/10.1103/PhysRevLett.121.060601} {\bibfield  {journal}
  {\bibinfo  {journal} {Phys. Rev. Lett.}\ }\textbf {\bibinfo {volume} {121}},\
  \bibinfo {pages} {060601} (\bibinfo {year} {2018}{\natexlab{b}})}\BibitemShut
  {NoStop}%
\bibitem [{\citenamefont {Gharibyan}\ \emph {et~al.}(2018)\citenamefont
  {Gharibyan}, \citenamefont {Hanada}, \citenamefont {Shenker},\ and\
  \citenamefont {Tezuka}}]{gharibyanOnsetRandomMatrix2018}%
  \BibitemOpen
  \bibfield  {author} {\bibinfo {author} {\bibfnamefont {H.}~\bibnamefont
  {Gharibyan}}, \bibinfo {author} {\bibfnamefont {M.}~\bibnamefont {Hanada}},
  \bibinfo {author} {\bibfnamefont {S.~H.}\ \bibnamefont {Shenker}},\ and\
  \bibinfo {author} {\bibfnamefont {M.}~\bibnamefont {Tezuka}},\ }\bibfield
  {title} {\bibinfo {title} {Onset of random matrix behavior in scrambling
  systems},\ }\href {https://doi.org/10.1007/JHEP07(2018)124} {\bibfield
  {journal} {\bibinfo  {journal} {J. High Energy Phys.}\ }\textbf {\bibinfo
  {volume} {2018}}\bibinfo  {number} { (7)},\ \bibinfo {pages}
  {124}}\BibitemShut {NoStop}%
\bibitem [{\citenamefont {Kos}\ \emph {et~al.}(2018)\citenamefont {Kos},
  \citenamefont {Ljubotina},\ and\ \citenamefont
  {Prosen}}]{kosManyBodyQuantumChaos2018}%
  \BibitemOpen
\bibfield  {number} {  }\bibfield  {author} {\bibinfo {author} {\bibfnamefont
  {P.}~\bibnamefont {Kos}}, \bibinfo {author} {\bibfnamefont {M.}~\bibnamefont
  {Ljubotina}},\ and\ \bibinfo {author} {\bibfnamefont {T.}~\bibnamefont
  {Prosen}},\ }\bibfield  {title} {\bibinfo {title} {Many-{{Body Quantum
  Chaos}}: {{Analytic Connection}} to {{Random Matrix Theory}}},\ }\href
  {https://doi.org/10.1103/PhysRevX.8.021062} {\bibfield  {journal} {\bibinfo
  {journal} {Phys. Rev. X}\ }\textbf {\bibinfo {volume} {8}},\ \bibinfo {pages}
  {021062} (\bibinfo {year} {2018})}\BibitemShut {NoStop}%
\bibitem [{\citenamefont {{\v S}untajs}\ \emph {et~al.}(2020)\citenamefont {{\v
  S}untajs}, \citenamefont {Bon{\v c}a}, \citenamefont {Prosen},\ and\
  \citenamefont {Vidmar}}]{suntajsQuantumChaosChallenges2020}%
  \BibitemOpen
  \bibfield  {author} {\bibinfo {author} {\bibfnamefont {J.}~\bibnamefont {{\v
  S}untajs}}, \bibinfo {author} {\bibfnamefont {J.}~\bibnamefont {Bon{\v c}a}},
  \bibinfo {author} {\bibfnamefont {T.}~\bibnamefont {Prosen}},\ and\ \bibinfo
  {author} {\bibfnamefont {L.}~\bibnamefont {Vidmar}},\ }\bibfield  {title}
  {\bibinfo {title} {Quantum chaos challenges many-body localization},\ }\href
  {https://doi.org/10.1103/PhysRevE.102.062144} {\bibfield  {journal} {\bibinfo
   {journal} {Phys. Rev. E}\ }\textbf {\bibinfo {volume} {102}},\ \bibinfo
  {pages} {062144} (\bibinfo {year} {2020})}\BibitemShut {NoStop}%
\bibitem [{\citenamefont {Liao}\ \emph {et~al.}(2020)\citenamefont {Liao},
  \citenamefont {Vikram},\ and\ \citenamefont
  {Galitski}}]{liaoManyBodyLevelStatistics2020}%
  \BibitemOpen
  \bibfield  {author} {\bibinfo {author} {\bibfnamefont {Y.}~\bibnamefont
  {Liao}}, \bibinfo {author} {\bibfnamefont {A.}~\bibnamefont {Vikram}},\ and\
  \bibinfo {author} {\bibfnamefont {V.}~\bibnamefont {Galitski}},\ }\bibfield
  {title} {\bibinfo {title} {Many-{{Body Level Statistics}} of
  {{Single-Particle Quantum Chaos}}},\ }\href
  {https://doi.org/10.1103/PhysRevLett.125.250601} {\bibfield  {journal}
  {\bibinfo  {journal} {Phys. Rev. Lett.}\ }\textbf {\bibinfo {volume} {125}},\
  \bibinfo {pages} {250601} (\bibinfo {year} {2020})}\BibitemShut {NoStop}%
\bibitem [{\citenamefont {Roy}\ and\ \citenamefont
  {Prosen}(2020)}]{royRandomMatrixSpectral2020}%
  \BibitemOpen
  \bibfield  {author} {\bibinfo {author} {\bibfnamefont {D.}~\bibnamefont
  {Roy}}\ and\ \bibinfo {author} {\bibfnamefont {T.}~\bibnamefont {Prosen}},\
  }\bibfield  {title} {\bibinfo {title} {Random matrix spectral form factor in
  kicked interacting fermionic chains},\ }\href
  {https://doi.org/10.1103/PhysRevE.102.060202} {\bibfield  {journal} {\bibinfo
   {journal} {Phys. Rev. E}\ }\textbf {\bibinfo {volume} {102}},\ \bibinfo
  {pages} {060202(R)} (\bibinfo {year} {2020})}\BibitemShut {NoStop}%
\bibitem [{\citenamefont {Sierant}\ \emph {et~al.}(2020)\citenamefont
  {Sierant}, \citenamefont {Delande},\ and\ \citenamefont
  {Zakrzewski}}]{sierantThoulessTimeAnalysis2020}%
  \BibitemOpen
  \bibfield  {author} {\bibinfo {author} {\bibfnamefont {P.}~\bibnamefont
  {Sierant}}, \bibinfo {author} {\bibfnamefont {D.}~\bibnamefont {Delande}},\
  and\ \bibinfo {author} {\bibfnamefont {J.}~\bibnamefont {Zakrzewski}},\
  }\bibfield  {title} {\bibinfo {title} {Thouless {{Time Analysis}} of
  {{Anderson}} and {{Many-Body Localization Transitions}}},\ }\href
  {https://doi.org/10.1103/PhysRevLett.124.186601} {\bibfield  {journal}
  {\bibinfo  {journal} {Phys. Rev. Lett.}\ }\textbf {\bibinfo {volume} {124}},\
  \bibinfo {pages} {186601} (\bibinfo {year} {2020})}\BibitemShut {NoStop}%
\bibitem [{\citenamefont {Bertini}\ \emph
  {et~al.}(2021{\natexlab{a}})\citenamefont {Bertini}, \citenamefont {Kos},\
  and\ \citenamefont {Prosen}}]{bertiniRandomMatrixSpectral2021}%
  \BibitemOpen
  \bibfield  {author} {\bibinfo {author} {\bibfnamefont {B.}~\bibnamefont
  {Bertini}}, \bibinfo {author} {\bibfnamefont {P.}~\bibnamefont {Kos}},\ and\
  \bibinfo {author} {\bibfnamefont {T.}~\bibnamefont {Prosen}},\ }\bibfield
  {title} {\bibinfo {title} {Random {{Matrix Spectral Form Factor}} of
  {{Dual-Unitary Quantum Circuits}}},\ }\href
  {https://doi.org/10.1007/s00220-021-04139-2} {\bibfield  {journal} {\bibinfo
  {journal} {Commun. Math. Phys.}\ }\textbf {\bibinfo {volume} {387}},\
  \bibinfo {pages} {597} (\bibinfo {year} {2021}{\natexlab{a}})}\BibitemShut
  {NoStop}%
\bibitem [{\citenamefont {Moudgalya}\ \emph {et~al.}(2021)\citenamefont
  {Moudgalya}, \citenamefont {Prem}, \citenamefont {Huse},\ and\ \citenamefont
  {Chan}}]{moudgalyaSpectralStatisticsConstrained2021}%
  \BibitemOpen
  \bibfield  {author} {\bibinfo {author} {\bibfnamefont {S.}~\bibnamefont
  {Moudgalya}}, \bibinfo {author} {\bibfnamefont {A.}~\bibnamefont {Prem}},
  \bibinfo {author} {\bibfnamefont {D.~A.}\ \bibnamefont {Huse}},\ and\
  \bibinfo {author} {\bibfnamefont {A.}~\bibnamefont {Chan}},\ }\bibfield
  {title} {\bibinfo {title} {Spectral statistics in constrained many-body
  quantum chaotic systems},\ }\href
  {https://doi.org/10.1103/PhysRevResearch.3.023176} {\bibfield  {journal}
  {\bibinfo  {journal} {Phys. Rev. Res.}\ }\textbf {\bibinfo {volume} {3}},\
  \bibinfo {pages} {023176} (\bibinfo {year} {2021})}\BibitemShut {NoStop}%
\bibitem [{\citenamefont {{\v S}untajs}\ \emph {et~al.}(2021)\citenamefont {{\v
  S}untajs}, \citenamefont {Prosen},\ and\ \citenamefont
  {Vidmar}}]{suntajsSpectralPropertiesThreedimensional2021}%
  \BibitemOpen
  \bibfield  {author} {\bibinfo {author} {\bibfnamefont {J.}~\bibnamefont {{\v
  S}untajs}}, \bibinfo {author} {\bibfnamefont {T.}~\bibnamefont {Prosen}},\
  and\ \bibinfo {author} {\bibfnamefont {L.}~\bibnamefont {Vidmar}},\
  }\bibfield  {title} {\bibinfo {title} {Spectral properties of
  three-dimensional {{Anderson}} model},\ }\href
  {https://doi.org/10.1016/j.aop.2021.168469} {\bibfield  {journal} {\bibinfo
  {journal} {Ann. Phys.}\ }\textbf {\bibinfo {volume} {435}},\ \bibinfo {pages}
  {168469} (\bibinfo {year} {2021})}\BibitemShut {NoStop}%
\bibitem [{\citenamefont {Chan}\ \emph {et~al.}(2021)\citenamefont {Chan},
  \citenamefont {De~Luca},\ and\ \citenamefont
  {Chalker}}]{chanSpectralLyapunovExponents2021}%
  \BibitemOpen
  \bibfield  {author} {\bibinfo {author} {\bibfnamefont {A.}~\bibnamefont
  {Chan}}, \bibinfo {author} {\bibfnamefont {A.}~\bibnamefont {De~Luca}},\ and\
  \bibinfo {author} {\bibfnamefont {J.~T.}\ \bibnamefont {Chalker}},\
  }\bibfield  {title} {\bibinfo {title} {Spectral {{Lyapunov}} exponents in
  chaotic and localized many-body quantum systems},\ }\href
  {https://doi.org/10.1103/PhysRevResearch.3.023118} {\bibfield  {journal}
  {\bibinfo  {journal} {Phys. Rev. Res.}\ }\textbf {\bibinfo {volume} {3}},\
  \bibinfo {pages} {023118} (\bibinfo {year} {2021})}\BibitemShut {NoStop}%
\bibitem [{\citenamefont {{\v S}untajs}\ and\ \citenamefont
  {Vidmar}(2022)}]{suntajsErgodicityBreakingTransition2022}%
  \BibitemOpen
  \bibfield  {author} {\bibinfo {author} {\bibfnamefont {J.}~\bibnamefont {{\v
  S}untajs}}\ and\ \bibinfo {author} {\bibfnamefont {L.}~\bibnamefont
  {Vidmar}},\ }\bibfield  {title} {\bibinfo {title} {Ergodicity {{Breaking
  Transition}} in {{Zero Dimensions}}},\ }\href
  {https://doi.org/10.1103/PhysRevLett.129.060602} {\bibfield  {journal}
  {\bibinfo  {journal} {Phys. Rev. Lett.}\ }\textbf {\bibinfo {volume} {129}},\
  \bibinfo {pages} {060602} (\bibinfo {year} {2022})}\BibitemShut {NoStop}%
\bibitem [{\citenamefont {Chan}\ \emph {et~al.}(2022)\citenamefont {Chan},
  \citenamefont {Shivam}, \citenamefont {Huse},\ and\ \citenamefont
  {De~Luca}}]{chanManybodyQuantumChaos2022}%
  \BibitemOpen
  \bibfield  {author} {\bibinfo {author} {\bibfnamefont {A.}~\bibnamefont
  {Chan}}, \bibinfo {author} {\bibfnamefont {S.}~\bibnamefont {Shivam}},
  \bibinfo {author} {\bibfnamefont {D.~A.}\ \bibnamefont {Huse}},\ and\
  \bibinfo {author} {\bibfnamefont {A.}~\bibnamefont {De~Luca}},\ }\bibfield
  {title} {\bibinfo {title} {Many-body quantum chaos and space-time
  translational invariance},\ }\href
  {https://doi.org/10.1038/s41467-022-34318-1} {\bibfield  {journal} {\bibinfo
  {journal} {Nat. Commun.}\ }\textbf {\bibinfo {volume} {13}},\ \bibinfo
  {pages} {7484} (\bibinfo {year} {2022})}\BibitemShut {NoStop}%
\bibitem [{\citenamefont {Roy}\ \emph {et~al.}(2022)\citenamefont {Roy},
  \citenamefont {Mishra},\ and\ \citenamefont
  {Prosen}}]{roySpectralFormFactor2022}%
  \BibitemOpen
  \bibfield  {author} {\bibinfo {author} {\bibfnamefont {D.}~\bibnamefont
  {Roy}}, \bibinfo {author} {\bibfnamefont {D.}~\bibnamefont {Mishra}},\ and\
  \bibinfo {author} {\bibfnamefont {T.}~\bibnamefont {Prosen}},\ }\bibfield
  {title} {\bibinfo {title} {Spectral form factor in a minimal bosonic model of
  many-body quantum chaos},\ }\href
  {https://doi.org/10.1103/PhysRevE.106.024208} {\bibfield  {journal} {\bibinfo
   {journal} {Phys. Rev. E}\ }\textbf {\bibinfo {volume} {106}},\ \bibinfo
  {pages} {024208} (\bibinfo {year} {2022})}\BibitemShut {NoStop}%
\bibitem [{\citenamefont {Kliczkowski}\ \emph {et~al.}(2024)\citenamefont
  {Kliczkowski}, \citenamefont {\'{S}wi\c{e}tek}, \citenamefont {Hopjan},\ and\
  \citenamefont {Vidmar}}]{kliczkowskiFadingErgodicity2024}%
  \BibitemOpen
  \bibfield  {author} {\bibinfo {author} {\bibfnamefont {M.}~\bibnamefont
  {Kliczkowski}}, \bibinfo {author} {\bibfnamefont {R.}~\bibnamefont
  {\'{S}wi\c{e}tek}}, \bibinfo {author} {\bibfnamefont {M.}~\bibnamefont
  {Hopjan}},\ and\ \bibinfo {author} {\bibfnamefont {L.}~\bibnamefont
  {Vidmar}},\ }\bibfield  {title} {\bibinfo {title} {Fading ergodicity},\
  }\href {https://doi.org/10.1103/PhysRevB.110.134206} {\bibfield  {journal}
  {\bibinfo  {journal} {Phys. Rev. B}\ }\textbf {\bibinfo {volume} {110}},\
  \bibinfo {pages} {134206} (\bibinfo {year} {2024})}\BibitemShut {NoStop}%
\bibitem [{\citenamefont {{Martinez-Azcona}}\ \emph {et~al.}(2025)\citenamefont
  {{Martinez-Azcona}}, \citenamefont {Shir},\ and\ \citenamefont
  {Chenu}}]{martinez-azconaDecomposingSpectralForm2025}%
  \BibitemOpen
  \bibfield  {author} {\bibinfo {author} {\bibfnamefont {P.}~\bibnamefont
  {{Martinez-Azcona}}}, \bibinfo {author} {\bibfnamefont {R.}~\bibnamefont
  {Shir}},\ and\ \bibinfo {author} {\bibfnamefont {A.}~\bibnamefont {Chenu}},\
  }\bibfield  {title} {\bibinfo {title} {Decomposing the spectral form
  factor},\ }\href {https://doi.org/10.1103/PhysRevB.111.165108} {\bibfield
  {journal} {\bibinfo  {journal} {Phys. Rev. B}\ }\textbf {\bibinfo {volume}
  {111}},\ \bibinfo {pages} {165108} (\bibinfo {year} {2025})}\BibitemShut
  {NoStop}%
\bibitem [{\citenamefont {Kumar}\ \emph {et~al.}(2025)\citenamefont {Kumar},
  \citenamefont {Prosen},\ and\ \citenamefont
  {Roy}}]{kumarLeadingLeadingorderSpectral2025}%
  \BibitemOpen
  \bibfield  {author} {\bibinfo {author} {\bibfnamefont {V.}~\bibnamefont
  {Kumar}}, \bibinfo {author} {\bibfnamefont {T.}~\bibnamefont {Prosen}},\ and\
  \bibinfo {author} {\bibfnamefont {D.}~\bibnamefont {Roy}},\ }\bibfield
  {title} {\bibinfo {title} {Leading and beyond leading-order spectral form
  factor in chaotic quantum many-body systems across all {{Dyson}} symmetry
  classes},\ }\Eprint {https://arxiv.org/abs/2502.04152} {arXiv:2502.04152
  [cond-mat]}  (\bibinfo {year} {2025})\BibitemShut {NoStop}%
\bibitem [{\citenamefont {T{\'a}vora}\ \emph {et~al.}(2016)\citenamefont
  {T{\'a}vora}, \citenamefont {{Torres-Herrera}},\ and\ \citenamefont
  {Santos}}]{tavoraInevitablePowerlawBehavior2016}%
  \BibitemOpen
  \bibfield  {author} {\bibinfo {author} {\bibfnamefont {M.}~\bibnamefont
  {T{\'a}vora}}, \bibinfo {author} {\bibfnamefont {E.~J.}\ \bibnamefont
  {{Torres-Herrera}}},\ and\ \bibinfo {author} {\bibfnamefont {L.~F.}\
  \bibnamefont {Santos}},\ }\bibfield  {title} {\bibinfo {title} {Inevitable
  power-law behavior of isolated many-body quantum systems and how it
  anticipates thermalization},\ }\href
  {https://doi.org/10.1103/PhysRevA.94.041603} {\bibfield  {journal} {\bibinfo
  {journal} {Phys. Rev. A}\ }\textbf {\bibinfo {volume} {94}},\ \bibinfo
  {pages} {041603(R)} (\bibinfo {year} {2016})}\BibitemShut {NoStop}%
\bibitem [{\citenamefont {T{\'a}vora}\ \emph {et~al.}(2017)\citenamefont
  {T{\'a}vora}, \citenamefont {{Torres-Herrera}},\ and\ \citenamefont
  {Santos}}]{tavoraPowerlawDecayExponents2017}%
  \BibitemOpen
  \bibfield  {author} {\bibinfo {author} {\bibfnamefont {M.}~\bibnamefont
  {T{\'a}vora}}, \bibinfo {author} {\bibfnamefont {E.~J.}\ \bibnamefont
  {{Torres-Herrera}}},\ and\ \bibinfo {author} {\bibfnamefont {L.~F.}\
  \bibnamefont {Santos}},\ }\bibfield  {title} {\bibinfo {title} {Power-law
  decay exponents: {{A}} dynamical criterion for predicting thermalization},\
  }\href {https://doi.org/10.1103/PhysRevA.95.013604} {\bibfield  {journal}
  {\bibinfo  {journal} {Phys. Rev. A}\ }\textbf {\bibinfo {volume} {95}},\
  \bibinfo {pages} {013604} (\bibinfo {year} {2017})}\BibitemShut {NoStop}%
\bibitem [{\citenamefont {{Torres-Herrera}}\ and\ \citenamefont
  {Santos}(2017)}]{torres-herreraDynamicalManifestationsQuantum2017}%
  \BibitemOpen
  \bibfield  {author} {\bibinfo {author} {\bibfnamefont {E.~J.}\ \bibnamefont
  {{Torres-Herrera}}}\ and\ \bibinfo {author} {\bibfnamefont {L.~F.}\
  \bibnamefont {Santos}},\ }\bibfield  {title} {\bibinfo {title} {Dynamical
  manifestations of quantum chaos: Correlation hole and bulge},\ }\href
  {https://doi.org/10.1098/rsta.2016.0434} {\bibfield  {journal} {\bibinfo
  {journal} {Phil. Trans. R. Soc. A.}\ }\textbf {\bibinfo {volume} {375}},\
  \bibinfo {pages} {20160434} (\bibinfo {year} {2017})}\BibitemShut {NoStop}%
\bibitem [{\citenamefont {{Torres-Herrera}}\ \emph {et~al.}(2018)\citenamefont
  {{Torres-Herrera}}, \citenamefont {{Garc{\'i}a-Garc{\'i}a}},\ and\
  \citenamefont {Santos}}]{torres-herreraGenericDynamicalFeatures2018}%
  \BibitemOpen
  \bibfield  {author} {\bibinfo {author} {\bibfnamefont {E.~J.}\ \bibnamefont
  {{Torres-Herrera}}}, \bibinfo {author} {\bibfnamefont {A.~M.}\ \bibnamefont
  {{Garc{\'i}a-Garc{\'i}a}}},\ and\ \bibinfo {author} {\bibfnamefont {L.~F.}\
  \bibnamefont {Santos}},\ }\bibfield  {title} {\bibinfo {title} {Generic
  dynamical features of quenched interacting quantum systems: {{Survival}}
  probability, density imbalance, and out-of-time-ordered correlator},\ }\href
  {https://doi.org/10.1103/PhysRevB.97.060303} {\bibfield  {journal} {\bibinfo
  {journal} {Phys. Rev. B}\ }\textbf {\bibinfo {volume} {97}},\ \bibinfo
  {pages} {060303(R)} (\bibinfo {year} {2018})}\BibitemShut {NoStop}%
\bibitem [{\citenamefont {Schiulaz}\ \emph {et~al.}(2019)\citenamefont
  {Schiulaz}, \citenamefont {{Torres-Herrera}},\ and\ \citenamefont
  {Santos}}]{schiulazThoulessRelaxationTime2019}%
  \BibitemOpen
  \bibfield  {author} {\bibinfo {author} {\bibfnamefont {M.}~\bibnamefont
  {Schiulaz}}, \bibinfo {author} {\bibfnamefont {E.~J.}\ \bibnamefont
  {{Torres-Herrera}}},\ and\ \bibinfo {author} {\bibfnamefont {L.~F.}\
  \bibnamefont {Santos}},\ }\bibfield  {title} {\bibinfo {title} {Thouless and
  relaxation time scales in many-body quantum systems},\ }\href
  {https://doi.org/10.1103/PhysRevB.99.174313} {\bibfield  {journal} {\bibinfo
  {journal} {Phys. Rev. B}\ }\textbf {\bibinfo {volume} {99}},\ \bibinfo
  {pages} {174313} (\bibinfo {year} {2019})}\BibitemShut {NoStop}%
\bibitem [{\citenamefont {{Lerma-Hern{\'a}ndez}}\ \emph
  {et~al.}(2019)\citenamefont {{Lerma-Hern{\'a}ndez}}, \citenamefont
  {Villase{\~n}or}, \citenamefont {{Bastarrachea-Magnani}}, \citenamefont
  {{Torres-Herrera}}, \citenamefont {Santos},\ and\ \citenamefont
  {Hirsch}}]{lerma-hernandezDynamicalSignaturesQuantum2019}%
  \BibitemOpen
  \bibfield  {author} {\bibinfo {author} {\bibfnamefont {S.}~\bibnamefont
  {{Lerma-Hern{\'a}ndez}}}, \bibinfo {author} {\bibfnamefont {D.}~\bibnamefont
  {Villase{\~n}or}}, \bibinfo {author} {\bibfnamefont {M.~A.}\ \bibnamefont
  {{Bastarrachea-Magnani}}}, \bibinfo {author} {\bibfnamefont {E.~J.}\
  \bibnamefont {{Torres-Herrera}}}, \bibinfo {author} {\bibfnamefont {L.~F.}\
  \bibnamefont {Santos}},\ and\ \bibinfo {author} {\bibfnamefont {J.~G.}\
  \bibnamefont {Hirsch}},\ }\bibfield  {title} {\bibinfo {title} {Dynamical
  signatures of quantum chaos and relaxation time scales in a spin-boson
  system},\ }\href {https://doi.org/10.1103/PhysRevE.100.012218} {\bibfield
  {journal} {\bibinfo  {journal} {Phys. Rev. E}\ }\textbf {\bibinfo {volume}
  {100}},\ \bibinfo {pages} {012218} (\bibinfo {year} {2019})}\BibitemShut
  {NoStop}%
\bibitem [{\citenamefont {Schiulaz}\ \emph {et~al.}(2020)\citenamefont
  {Schiulaz}, \citenamefont {{Torres-Herrera}}, \citenamefont
  {{P{\'e}rez-Bernal}},\ and\ \citenamefont
  {Santos}}]{schiulazSelfaveragingManybodyQuantum2020}%
  \BibitemOpen
  \bibfield  {author} {\bibinfo {author} {\bibfnamefont {M.}~\bibnamefont
  {Schiulaz}}, \bibinfo {author} {\bibfnamefont {E.~J.}\ \bibnamefont
  {{Torres-Herrera}}}, \bibinfo {author} {\bibfnamefont {F.}~\bibnamefont
  {{P{\'e}rez-Bernal}}},\ and\ \bibinfo {author} {\bibfnamefont {L.~F.}\
  \bibnamefont {Santos}},\ }\bibfield  {title} {\bibinfo {title}
  {Self-averaging in many-body quantum systems out of equilibrium: {{Chaotic}}
  systems},\ }\href {https://doi.org/10.1103/PhysRevB.101.174312} {\bibfield
  {journal} {\bibinfo  {journal} {Phys. Rev. B}\ }\textbf {\bibinfo {volume}
  {101}},\ \bibinfo {pages} {174312} (\bibinfo {year} {2020})}\BibitemShut
  {NoStop}%
\bibitem [{\citenamefont {Lezama}\ \emph {et~al.}(2021)\citenamefont {Lezama},
  \citenamefont {{Torres-Herrera}}, \citenamefont {{P{\'e}rez-Bernal}},
  \citenamefont {Bar~Lev},\ and\ \citenamefont
  {Santos}}]{lezamaEquilibrationTimeManybody2021}%
  \BibitemOpen
  \bibfield  {author} {\bibinfo {author} {\bibfnamefont {T.~L.~M.}\
  \bibnamefont {Lezama}}, \bibinfo {author} {\bibfnamefont {E.~J.}\
  \bibnamefont {{Torres-Herrera}}}, \bibinfo {author} {\bibfnamefont
  {F.}~\bibnamefont {{P{\'e}rez-Bernal}}}, \bibinfo {author} {\bibfnamefont
  {Y.}~\bibnamefont {Bar~Lev}},\ and\ \bibinfo {author} {\bibfnamefont {L.~F.}\
  \bibnamefont {Santos}},\ }\bibfield  {title} {\bibinfo {title} {Equilibration
  time in many-body quantum systems},\ }\href
  {https://doi.org/10.1103/PhysRevB.104.085117} {\bibfield  {journal} {\bibinfo
   {journal} {Phys. Rev. B}\ }\textbf {\bibinfo {volume} {104}},\ \bibinfo
  {pages} {085117} (\bibinfo {year} {2021})}\BibitemShut {NoStop}%
\bibitem [{\citenamefont {Khatami}\ \emph {et~al.}(2013)\citenamefont
  {Khatami}, \citenamefont {Pupillo}, \citenamefont {Srednicki},\ and\
  \citenamefont {Rigol}}]{khatamiFluctuationDissipationTheoremIsolated2013}%
  \BibitemOpen
  \bibfield  {author} {\bibinfo {author} {\bibfnamefont {E.}~\bibnamefont
  {Khatami}}, \bibinfo {author} {\bibfnamefont {G.}~\bibnamefont {Pupillo}},
  \bibinfo {author} {\bibfnamefont {M.}~\bibnamefont {Srednicki}},\ and\
  \bibinfo {author} {\bibfnamefont {M.}~\bibnamefont {Rigol}},\ }\bibfield
  {title} {\bibinfo {title} {Fluctuation-{{Dissipation Theorem}} in an
  {{Isolated System}} of {{Quantum Dipolar Bosons}} after a {{Quench}}},\
  }\href {https://doi.org/10.1103/PhysRevLett.111.050403} {\bibfield  {journal}
  {\bibinfo  {journal} {Phys. Rev. Lett.}\ }\textbf {\bibinfo {volume} {111}},\
  \bibinfo {pages} {050403} (\bibinfo {year} {2013})}\BibitemShut {NoStop}%
\bibitem [{\citenamefont {Serbyn}\ \emph {et~al.}(2017)\citenamefont {Serbyn},
  \citenamefont {Papi\ifmmode~\acute{c}\else \'{c}\fi{}},\ and\ \citenamefont
  {Abanin}}]{Serbyn2017}%
  \BibitemOpen
  \bibfield  {author} {\bibinfo {author} {\bibfnamefont {M.}~\bibnamefont
  {Serbyn}}, \bibinfo {author} {\bibfnamefont {Z.}~\bibnamefont
  {Papi\ifmmode~\acute{c}\else \'{c}\fi{}}},\ and\ \bibinfo {author}
  {\bibfnamefont {D.~A.}\ \bibnamefont {Abanin}},\ }\bibfield  {title}
  {\bibinfo {title} {Thouless energy and multifractality across the many-body
  localization transition},\ }\href
  {https://doi.org/10.1103/PhysRevB.96.104201} {\bibfield  {journal} {\bibinfo
  {journal} {Phys. Rev. B}\ }\textbf {\bibinfo {volume} {96}},\ \bibinfo
  {pages} {104201} (\bibinfo {year} {2017})}\BibitemShut {NoStop}%
\bibitem [{\citenamefont {Brenes}\ \emph {et~al.}(2020)\citenamefont {Brenes},
  \citenamefont {Goold},\ and\ \citenamefont
  {Rigol}}]{brenesLowfrequencyBehaviorOffdiagonal2020}%
  \BibitemOpen
  \bibfield  {author} {\bibinfo {author} {\bibfnamefont {M.}~\bibnamefont
  {Brenes}}, \bibinfo {author} {\bibfnamefont {J.}~\bibnamefont {Goold}},\ and\
  \bibinfo {author} {\bibfnamefont {M.}~\bibnamefont {Rigol}},\ }\bibfield
  {title} {\bibinfo {title} {Low-frequency behavior of off-diagonal matrix
  elements in the integrable {{XXZ}} chain and in a locally perturbed
  quantum-chaotic {{XXZ}} chain},\ }\href
  {https://doi.org/10.1103/PhysRevB.102.075127} {\bibfield  {journal} {\bibinfo
   {journal} {Phys. Rev. B}\ }\textbf {\bibinfo {volume} {102}},\ \bibinfo
  {pages} {075127} (\bibinfo {year} {2020})}\BibitemShut {NoStop}%
\bibitem [{\citenamefont {Wang}\ \emph {et~al.}(2022)\citenamefont {Wang},
  \citenamefont {Lamann}, \citenamefont {Richter}, \citenamefont {Steinigeweg},
  \citenamefont {Dymarsky},\ and\ \citenamefont
  {Gemmer}}]{wangEigenstateThermalizationHypothesis2022}%
  \BibitemOpen
  \bibfield  {author} {\bibinfo {author} {\bibfnamefont {J.}~\bibnamefont
  {Wang}}, \bibinfo {author} {\bibfnamefont {M.~H.}\ \bibnamefont {Lamann}},
  \bibinfo {author} {\bibfnamefont {J.}~\bibnamefont {Richter}}, \bibinfo
  {author} {\bibfnamefont {R.}~\bibnamefont {Steinigeweg}}, \bibinfo {author}
  {\bibfnamefont {A.}~\bibnamefont {Dymarsky}},\ and\ \bibinfo {author}
  {\bibfnamefont {J.}~\bibnamefont {Gemmer}},\ }\bibfield  {title} {\bibinfo
  {title} {Eigenstate {{Thermalization Hypothesis}} and {{Its Deviations}} from
  {{Random-Matrix Theory}} beyond the {{Thermalization Time}}},\ }\href
  {https://doi.org/10.1103/PhysRevLett.128.180601} {\bibfield  {journal}
  {\bibinfo  {journal} {Phys. Rev. Lett.}\ }\textbf {\bibinfo {volume} {128}},\
  \bibinfo {pages} {180601} (\bibinfo {year} {2022})}\BibitemShut {NoStop}%
\bibitem [{\citenamefont {Bartsch}\ \emph {et~al.}(2024)\citenamefont
  {Bartsch}, \citenamefont {Dymarsky}, \citenamefont {Lamann}, \citenamefont
  {Wang}, \citenamefont {Steinigeweg},\ and\ \citenamefont
  {Gemmer}}]{bartschEstimationEquilibrationTime2024}%
  \BibitemOpen
  \bibfield  {author} {\bibinfo {author} {\bibfnamefont {C.}~\bibnamefont
  {Bartsch}}, \bibinfo {author} {\bibfnamefont {A.}~\bibnamefont {Dymarsky}},
  \bibinfo {author} {\bibfnamefont {M.~H.}\ \bibnamefont {Lamann}}, \bibinfo
  {author} {\bibfnamefont {J.}~\bibnamefont {Wang}}, \bibinfo {author}
  {\bibfnamefont {R.}~\bibnamefont {Steinigeweg}},\ and\ \bibinfo {author}
  {\bibfnamefont {J.}~\bibnamefont {Gemmer}},\ }\bibfield  {title} {\bibinfo
  {title} {Estimation of equilibration time scales from nested fraction
  approximations},\ }\href {https://doi.org/10.1103/PhysRevE.110.024126}
  {\bibfield  {journal} {\bibinfo  {journal} {Phys. Rev. E}\ }\textbf {\bibinfo
  {volume} {110}},\ \bibinfo {pages} {024126} (\bibinfo {year}
  {2024})}\BibitemShut {NoStop}%
\bibitem [{\citenamefont {Wang}\ \emph {et~al.}(2025)\citenamefont {Wang},
  \citenamefont {F{\"u}llgraf},\ and\ \citenamefont
  {Gemmer}}]{wangEstimateEquilibrationTimes2025}%
  \BibitemOpen
  \bibfield  {author} {\bibinfo {author} {\bibfnamefont {J.}~\bibnamefont
  {Wang}}, \bibinfo {author} {\bibfnamefont {M.}~\bibnamefont {F{\"u}llgraf}},\
  and\ \bibinfo {author} {\bibfnamefont {J.}~\bibnamefont {Gemmer}},\
  }\bibfield  {title} {\bibinfo {title} {Estimate of {{Equilibration Times}} of
  {{Quantum Correlation Functions}} in the {{Thermodynamic Limit Based}} on
  {{Lanczos Coefficients}}},\ }\href {https://doi.org/10.1103/ztrj-2tm6}
  {\bibfield  {journal} {\bibinfo  {journal} {Phys. Rev. Lett.}\ }\textbf
  {\bibinfo {volume} {135}},\ \bibinfo {pages} {010403} (\bibinfo {year}
  {2025})}\BibitemShut {NoStop}%
\bibitem [{\citenamefont {Maceira}\ and\ \citenamefont
  {L{\"a}uchli}(2025)}]{maceiraThermalizationDynamicsClosed2025}%
  \BibitemOpen
  \bibfield  {author} {\bibinfo {author} {\bibfnamefont {I.~A.}\ \bibnamefont
  {Maceira}}\ and\ \bibinfo {author} {\bibfnamefont {A.~M.}\ \bibnamefont
  {L{\"a}uchli}},\ }\bibfield  {title} {\bibinfo {title} {Thermalization
  {{Dynamics}} in {{Closed Quantum Many Body Systems}}: A {{Precision Large
  Scale Exact Diagonalization Study}}},\ }\Eprint
  {https://arxiv.org/abs/2409.18863} {arXiv:2409.18863 [quant-ph]}  (\bibinfo
  {year} {2025})\BibitemShut {NoStop}%
\bibitem [{\citenamefont {{Bouverot-Dupuis}}\ \emph {et~al.}(2025)\citenamefont
  {{Bouverot-Dupuis}}, \citenamefont {Pappalardi}, \citenamefont {Kurchan},
  \citenamefont {Polkovnikov},\ and\ \citenamefont
  {Foini}}]{bouverot-dupuisRandomMatrixUniversality2025}%
  \BibitemOpen
  \bibfield  {author} {\bibinfo {author} {\bibfnamefont {O.}~\bibnamefont
  {{Bouverot-Dupuis}}}, \bibinfo {author} {\bibfnamefont {S.}~\bibnamefont
  {Pappalardi}}, \bibinfo {author} {\bibfnamefont {J.}~\bibnamefont {Kurchan}},
  \bibinfo {author} {\bibfnamefont {A.}~\bibnamefont {Polkovnikov}},\ and\
  \bibinfo {author} {\bibfnamefont {L.}~\bibnamefont {Foini}},\ }\bibfield
  {title} {\bibinfo {title} {Random matrix universality in dynamical
  correlation functions at late times},\ }\href
  {https://doi.org/10.21468/SciPostPhys.19.2.050} {\bibfield  {journal}
  {\bibinfo  {journal} {SciPost Phys.}\ }\textbf {\bibinfo {volume} {19}},\
  \bibinfo {pages} {050} (\bibinfo {year} {2025})}\BibitemShut {NoStop}%
\bibitem [{\citenamefont {Dong}\ \emph {et~al.}(2025)\citenamefont {Dong},
  \citenamefont {Zhang}, \citenamefont {Da{\u g}}, \citenamefont {Gao},
  \citenamefont {Wang}, \citenamefont {Deng}, \citenamefont {Zhang},
  \citenamefont {Chen}, \citenamefont {Xu}, \citenamefont {Wang}, \citenamefont
  {Wu}, \citenamefont {Zhang}, \citenamefont {Jin}, \citenamefont {Zhu},
  \citenamefont {Zhang}, \citenamefont {Zou}, \citenamefont {Tan},
  \citenamefont {Cui}, \citenamefont {Zhu}, \citenamefont {Shen}, \citenamefont
  {Li}, \citenamefont {Zhong}, \citenamefont {Bao}, \citenamefont {Li},
  \citenamefont {Wang}, \citenamefont {Guo}, \citenamefont {Song},
  \citenamefont {Liu}, \citenamefont {Chan}, \citenamefont {Ying},\ and\
  \citenamefont {Wang}}]{dongMeasuringSpectralForm2025}%
  \BibitemOpen
  \bibfield  {author} {\bibinfo {author} {\bibfnamefont {H.}~\bibnamefont
  {Dong}}, \bibinfo {author} {\bibfnamefont {P.}~\bibnamefont {Zhang}},
  \bibinfo {author} {\bibfnamefont {C.~B.}\ \bibnamefont {Da{\u g}}}, \bibinfo
  {author} {\bibfnamefont {Y.}~\bibnamefont {Gao}}, \bibinfo {author}
  {\bibfnamefont {N.}~\bibnamefont {Wang}}, \bibinfo {author} {\bibfnamefont
  {J.}~\bibnamefont {Deng}}, \bibinfo {author} {\bibfnamefont {X.}~\bibnamefont
  {Zhang}}, \bibinfo {author} {\bibfnamefont {J.}~\bibnamefont {Chen}},
  \bibinfo {author} {\bibfnamefont {S.}~\bibnamefont {Xu}}, \bibinfo {author}
  {\bibfnamefont {K.}~\bibnamefont {Wang}}, \bibinfo {author} {\bibfnamefont
  {Y.}~\bibnamefont {Wu}}, \bibinfo {author} {\bibfnamefont {C.}~\bibnamefont
  {Zhang}}, \bibinfo {author} {\bibfnamefont {F.}~\bibnamefont {Jin}}, \bibinfo
  {author} {\bibfnamefont {X.}~\bibnamefont {Zhu}}, \bibinfo {author}
  {\bibfnamefont {A.}~\bibnamefont {Zhang}}, \bibinfo {author} {\bibfnamefont
  {Y.}~\bibnamefont {Zou}}, \bibinfo {author} {\bibfnamefont {Z.}~\bibnamefont
  {Tan}}, \bibinfo {author} {\bibfnamefont {Z.}~\bibnamefont {Cui}}, \bibinfo
  {author} {\bibfnamefont {Z.}~\bibnamefont {Zhu}}, \bibinfo {author}
  {\bibfnamefont {F.}~\bibnamefont {Shen}}, \bibinfo {author} {\bibfnamefont
  {T.}~\bibnamefont {Li}}, \bibinfo {author} {\bibfnamefont {J.}~\bibnamefont
  {Zhong}}, \bibinfo {author} {\bibfnamefont {Z.}~\bibnamefont {Bao}}, \bibinfo
  {author} {\bibfnamefont {H.}~\bibnamefont {Li}}, \bibinfo {author}
  {\bibfnamefont {Z.}~\bibnamefont {Wang}}, \bibinfo {author} {\bibfnamefont
  {Q.}~\bibnamefont {Guo}}, \bibinfo {author} {\bibfnamefont {C.}~\bibnamefont
  {Song}}, \bibinfo {author} {\bibfnamefont {F.}~\bibnamefont {Liu}}, \bibinfo
  {author} {\bibfnamefont {A.}~\bibnamefont {Chan}}, \bibinfo {author}
  {\bibfnamefont {L.}~\bibnamefont {Ying}},\ and\ \bibinfo {author}
  {\bibfnamefont {H.}~\bibnamefont {Wang}},\ }\bibfield  {title} {\bibinfo
  {title} {Measuring the {{Spectral Form Factor}} in {{Many-Body Chaotic}} and
  {{Localized Phases}} of {{Quantum Processors}}},\ }\href
  {https://doi.org/10.1103/PhysRevLett.134.010402} {\bibfield  {journal}
  {\bibinfo  {journal} {Phys. Rev. Lett.}\ }\textbf {\bibinfo {volume} {134}},\
  \bibinfo {pages} {010402} (\bibinfo {year} {2025})}\BibitemShut {NoStop}%
\bibitem [{\citenamefont {Das}\ \emph {et~al.}(2025)\citenamefont {Das},
  \citenamefont {Cianci}, \citenamefont {Cabral}, \citenamefont
  {{Zarate-Herrada}}, \citenamefont {Pinney}, \citenamefont
  {{Pilatowsky-Cameo}}, \citenamefont {{Matsoukas-Roubeas}}, \citenamefont
  {Batista}, \citenamefont {Del~Campo}, \citenamefont {{Torres-Herrera}},\ and\
  \citenamefont {Santos}}]{dasProposalManybodyQuantum2025}%
  \BibitemOpen
  \bibfield  {author} {\bibinfo {author} {\bibfnamefont {A.~K.}\ \bibnamefont
  {Das}}, \bibinfo {author} {\bibfnamefont {C.}~\bibnamefont {Cianci}},
  \bibinfo {author} {\bibfnamefont {D.~G.~A.}\ \bibnamefont {Cabral}}, \bibinfo
  {author} {\bibfnamefont {D.~A.}\ \bibnamefont {{Zarate-Herrada}}}, \bibinfo
  {author} {\bibfnamefont {P.}~\bibnamefont {Pinney}}, \bibinfo {author}
  {\bibfnamefont {S.}~\bibnamefont {{Pilatowsky-Cameo}}}, \bibinfo {author}
  {\bibfnamefont {A.~S.}\ \bibnamefont {{Matsoukas-Roubeas}}}, \bibinfo
  {author} {\bibfnamefont {V.~S.}\ \bibnamefont {Batista}}, \bibinfo {author}
  {\bibfnamefont {A.}~\bibnamefont {Del~Campo}}, \bibinfo {author}
  {\bibfnamefont {E.~J.}\ \bibnamefont {{Torres-Herrera}}},\ and\ \bibinfo
  {author} {\bibfnamefont {L.~F.}\ \bibnamefont {Santos}},\ }\bibfield  {title}
  {\bibinfo {title} {Proposal for many-body quantum chaos detection},\ }\href
  {https://doi.org/10.1103/PhysRevResearch.7.013181} {\bibfield  {journal}
  {\bibinfo  {journal} {Phys. Rev. Res.}\ }\textbf {\bibinfo {volume} {7}},\
  \bibinfo {pages} {013181} (\bibinfo {year} {2025})}\BibitemShut {NoStop}%
\bibitem [{\citenamefont {Karch}\ \emph {et~al.}(2025)\citenamefont {Karch},
  \citenamefont {Bandyopadhyay}, \citenamefont {Sun}, \citenamefont {Impertro},
  \citenamefont {Huh}, \citenamefont {Rodr{\'i}guez}, \citenamefont {Wienand},
  \citenamefont {Ketterle}, \citenamefont {Heyl}, \citenamefont {Polkovnikov},
  \citenamefont {Bloch},\ and\ \citenamefont
  {Aidelsburger}}]{karchProbingQuantumManybody2025}%
  \BibitemOpen
  \bibfield  {author} {\bibinfo {author} {\bibfnamefont {S.}~\bibnamefont
  {Karch}}, \bibinfo {author} {\bibfnamefont {S.}~\bibnamefont
  {Bandyopadhyay}}, \bibinfo {author} {\bibfnamefont {Z.-H.}\ \bibnamefont
  {Sun}}, \bibinfo {author} {\bibfnamefont {A.}~\bibnamefont {Impertro}},
  \bibinfo {author} {\bibfnamefont {S.}~\bibnamefont {Huh}}, \bibinfo {author}
  {\bibfnamefont {I.~P.}\ \bibnamefont {Rodr{\'i}guez}}, \bibinfo {author}
  {\bibfnamefont {J.~F.}\ \bibnamefont {Wienand}}, \bibinfo {author}
  {\bibfnamefont {W.}~\bibnamefont {Ketterle}}, \bibinfo {author}
  {\bibfnamefont {M.}~\bibnamefont {Heyl}}, \bibinfo {author} {\bibfnamefont
  {A.}~\bibnamefont {Polkovnikov}}, \bibinfo {author} {\bibfnamefont
  {I.}~\bibnamefont {Bloch}},\ and\ \bibinfo {author} {\bibfnamefont
  {M.}~\bibnamefont {Aidelsburger}},\ }\bibfield  {title} {\bibinfo {title}
  {Probing quantum many-body dynamics using subsystem {{Loschmidt}} echos},\
  }\Eprint {https://arxiv.org/abs/2501.16995} {arXiv:2501.16995 [cond-mat]}
  (\bibinfo {year} {2025})\BibitemShut {NoStop}%
\bibitem [{\citenamefont {Bohigas}\ \emph {et~al.}(1984)\citenamefont
  {Bohigas}, \citenamefont {Giannoni},\ and\ \citenamefont
  {Schmit}}]{bohigasCharacterizationChaoticQuantum1984}%
  \BibitemOpen
  \bibfield  {author} {\bibinfo {author} {\bibfnamefont {O.}~\bibnamefont
  {Bohigas}}, \bibinfo {author} {\bibfnamefont {M.~J.}\ \bibnamefont
  {Giannoni}},\ and\ \bibinfo {author} {\bibfnamefont {C.}~\bibnamefont
  {Schmit}},\ }\bibfield  {title} {\bibinfo {title} {Characterization of
  {{Chaotic Quantum Spectra}} and {{Universality}} of {{Level Fluctuation
  Laws}}},\ }\href {https://doi.org/10.1103/PhysRevLett.52.1} {\bibfield
  {journal} {\bibinfo  {journal} {Phys. Rev. Lett.}\ }\textbf {\bibinfo
  {volume} {52}},\ \bibinfo {pages} {1} (\bibinfo {year} {1984})}\BibitemShut
  {NoStop}%
\bibitem [{\citenamefont {Berry}(1985)}]{berrySemiclassicalTheorySpectral1985}%
  \BibitemOpen
  \bibfield  {author} {\bibinfo {author} {\bibfnamefont {M.~V.}\ \bibnamefont
  {Berry}},\ }\bibfield  {title} {\bibinfo {title} {Semiclassical theory of
  spectral rigidity},\ }\href {https://doi.org/10.1098/rspa.1985.0078}
  {\bibfield  {journal} {\bibinfo  {journal} {Proc. R. Soc. Lond. A}\ }\textbf
  {\bibinfo {volume} {400}},\ \bibinfo {pages} {229} (\bibinfo {year}
  {1985})}\BibitemShut {NoStop}%
\bibitem [{\citenamefont {Haake}(1991)}]{haakeQuantumSignaturesChaos1991}%
  \BibitemOpen
  \bibfield  {author} {\bibinfo {author} {\bibfnamefont {F.}~\bibnamefont
  {Haake}},\ }\href@noop {} {\emph {\bibinfo {title} {Quantum {{Signatures}} of
  {{Chaos}}}}}\ (\bibinfo  {publisher} {Springer US},\ \bibinfo {address}
  {Boston},\ \bibinfo {year} {1991})\ pp.\ \bibinfo {pages}
  {583--595}\BibitemShut {NoStop}%
\bibitem [{\citenamefont
  {St{\"o}ckmann}(1999)}]{stockmannQuantumChaosIntroduction1999}%
  \BibitemOpen
  \bibfield  {author} {\bibinfo {author} {\bibfnamefont {H.-J.}\ \bibnamefont
  {St{\"o}ckmann}},\ }\href@noop {} {\emph {\bibinfo {title} {Quantum Chaos:
  {{An}} Introduction}}}\ (\bibinfo  {publisher} {Cambridge University Press},\
  \bibinfo {address} {Cambridge},\ \bibinfo {year} {1999})\BibitemShut
  {NoStop}%
\bibitem [{\citenamefont {Mehta}(2004)}]{mehtaRandomMatrices2004}%
  \BibitemOpen
  \bibfield  {author} {\bibinfo {author} {\bibfnamefont {M.~L.}\ \bibnamefont
  {Mehta}},\ }\href@noop {} {\emph {\bibinfo {title} {Random {{Matrices}}}}},\
  \bibinfo {edition} {3rd}\ ed.,\ Vol.\ \bibinfo {volume} {142}\ (\bibinfo
  {publisher} {Elsevier},\ \bibinfo {address} {New York},\ \bibinfo {year}
  {2004})\BibitemShut {NoStop}%
\bibitem [{\citenamefont {Guhr}\ \emph {et~al.}(1998)\citenamefont {Guhr},
  \citenamefont {{M{\"u}ller--Groeling}},\ and\ \citenamefont
  {Weidenm{\"u}ller}}]{guhrRandommatrixTheoriesQuantum1998}%
  \BibitemOpen
  \bibfield  {author} {\bibinfo {author} {\bibfnamefont {T.}~\bibnamefont
  {Guhr}}, \bibinfo {author} {\bibfnamefont {A.}~\bibnamefont
  {{M{\"u}ller--Groeling}}},\ and\ \bibinfo {author} {\bibfnamefont {H.~A.}\
  \bibnamefont {Weidenm{\"u}ller}},\ }\bibfield  {title} {\bibinfo {title}
  {Random-matrix theories in quantum physics: Common concepts},\ }\href
  {https://doi.org/10.1016/S0370-1573(97)00088-4} {\bibfield  {journal}
  {\bibinfo  {journal} {Phys. Rep.}\ }\textbf {\bibinfo {volume} {299}},\
  \bibinfo {pages} {189} (\bibinfo {year} {1998})}\BibitemShut {NoStop}%
\bibitem [{\citenamefont {M{\"u}ller}\ \emph {et~al.}(2004)\citenamefont
  {M{\"u}ller}, \citenamefont {Heusler}, \citenamefont {Braun}, \citenamefont
  {Haake},\ and\ \citenamefont
  {Altland}}]{mullerSemiclassicalFoundationUniversality2004}%
  \BibitemOpen
  \bibfield  {author} {\bibinfo {author} {\bibfnamefont {S.}~\bibnamefont
  {M{\"u}ller}}, \bibinfo {author} {\bibfnamefont {S.}~\bibnamefont {Heusler}},
  \bibinfo {author} {\bibfnamefont {P.}~\bibnamefont {Braun}}, \bibinfo
  {author} {\bibfnamefont {F.}~\bibnamefont {Haake}},\ and\ \bibinfo {author}
  {\bibfnamefont {A.}~\bibnamefont {Altland}},\ }\bibfield  {title} {\bibinfo
  {title} {Semiclassical {{Foundation}} of {{Universality}} in {{Quantum
  Chaos}}},\ }\href {https://doi.org/10.1103/PhysRevLett.93.014103} {\bibfield
  {journal} {\bibinfo  {journal} {Phys. Rev. Lett.}\ }\textbf {\bibinfo
  {volume} {93}},\ \bibinfo {pages} {014103} (\bibinfo {year}
  {2004})}\BibitemShut {NoStop}%
\bibitem [{\citenamefont {M{\"u}ller}\ \emph {et~al.}(2005)\citenamefont
  {M{\"u}ller}, \citenamefont {Heusler}, \citenamefont {Braun}, \citenamefont
  {Haake},\ and\ \citenamefont
  {Altland}}]{mullerPeriodicorbitTheoryUniversality2005}%
  \BibitemOpen
  \bibfield  {author} {\bibinfo {author} {\bibfnamefont {S.}~\bibnamefont
  {M{\"u}ller}}, \bibinfo {author} {\bibfnamefont {S.}~\bibnamefont {Heusler}},
  \bibinfo {author} {\bibfnamefont {P.}~\bibnamefont {Braun}}, \bibinfo
  {author} {\bibfnamefont {F.}~\bibnamefont {Haake}},\ and\ \bibinfo {author}
  {\bibfnamefont {A.}~\bibnamefont {Altland}},\ }\bibfield  {title} {\bibinfo
  {title} {Periodic-orbit theory of universality in quantum chaos},\ }\href
  {https://doi.org/10.1103/PhysRevE.72.046207} {\bibfield  {journal} {\bibinfo
  {journal} {Phys. Rev. E}\ }\textbf {\bibinfo {volume} {72}},\ \bibinfo
  {pages} {046207} (\bibinfo {year} {2005})}\BibitemShut {NoStop}%
\bibitem [{\citenamefont {Sierant}\ \emph {et~al.}(2025)\citenamefont
  {Sierant}, \citenamefont {Lewenstein}, \citenamefont {Scardicchio},
  \citenamefont {Vidmar},\ and\ \citenamefont
  {Zakrzewski}}]{sierantManybodyLocalizationAge2025}%
  \BibitemOpen
  \bibfield  {author} {\bibinfo {author} {\bibfnamefont {P.}~\bibnamefont
  {Sierant}}, \bibinfo {author} {\bibfnamefont {M.}~\bibnamefont {Lewenstein}},
  \bibinfo {author} {\bibfnamefont {A.}~\bibnamefont {Scardicchio}}, \bibinfo
  {author} {\bibfnamefont {L.}~\bibnamefont {Vidmar}},\ and\ \bibinfo {author}
  {\bibfnamefont {J.}~\bibnamefont {Zakrzewski}},\ }\bibfield  {title}
  {\bibinfo {title} {Many-body localization in the age of classical
  computing},\ }\href {https://doi.org/10.1088/1361-6633/ad9756} {\bibfield
  {journal} {\bibinfo  {journal} {Rep. Prog. Phys.}\ }\textbf {\bibinfo
  {volume} {88}},\ \bibinfo {pages} {026502} (\bibinfo {year}
  {2025})}\BibitemShut {NoStop}%
\bibitem [{\citenamefont {Kumar}\ and\ \citenamefont
  {Roy}(2024)}]{kumarManybodyQuantumChaos2024}%
  \BibitemOpen
  \bibfield  {author} {\bibinfo {author} {\bibfnamefont {V.}~\bibnamefont
  {Kumar}}\ and\ \bibinfo {author} {\bibfnamefont {D.}~\bibnamefont {Roy}},\
  }\bibfield  {title} {\bibinfo {title} {Many-body quantum chaos in mixtures of
  multiple species},\ }\href {https://doi.org/10.1103/PhysRevE.109.L032201}
  {\bibfield  {journal} {\bibinfo  {journal} {Phys. Rev. E}\ }\textbf {\bibinfo
  {volume} {109}},\ \bibinfo {pages} {L032201} (\bibinfo {year}
  {2024})}\BibitemShut {NoStop}%
\bibitem [{\citenamefont {Colmenarez}\ \emph {et~al.}(2022)\citenamefont
  {Colmenarez}, \citenamefont {Luitz}, \citenamefont {Khaymovich},\ and\
  \citenamefont {De~Tomasi}}]{colmenarezSubdiffusiveThoulessTime2022}%
  \BibitemOpen
  \bibfield  {author} {\bibinfo {author} {\bibfnamefont {L.}~\bibnamefont
  {Colmenarez}}, \bibinfo {author} {\bibfnamefont {D.~J.}\ \bibnamefont
  {Luitz}}, \bibinfo {author} {\bibfnamefont {I.~M.}\ \bibnamefont
  {Khaymovich}},\ and\ \bibinfo {author} {\bibfnamefont {G.}~\bibnamefont
  {De~Tomasi}},\ }\bibfield  {title} {\bibinfo {title} {Subdiffusive
  {{Thouless}} time scaling in the {{Anderson}} model on random regular
  graphs},\ }\href {https://doi.org/10.1103/PhysRevB.105.174207} {\bibfield
  {journal} {\bibinfo  {journal} {Phys. Rev. B}\ }\textbf {\bibinfo {volume}
  {105}},\ \bibinfo {pages} {174207} (\bibinfo {year} {2022})}\BibitemShut
  {NoStop}%
\bibitem [{\citenamefont {Prelov{\v s}ek}\ \emph {et~al.}(2023)\citenamefont
  {Prelov{\v s}ek}, \citenamefont {Herbrych},\ and\ \citenamefont
  {Mierzejewski}}]{prelovsekSlowDiffusionThouless2023}%
  \BibitemOpen
  \bibfield  {author} {\bibinfo {author} {\bibfnamefont {P.}~\bibnamefont
  {Prelov{\v s}ek}}, \bibinfo {author} {\bibfnamefont {J.}~\bibnamefont
  {Herbrych}},\ and\ \bibinfo {author} {\bibfnamefont {M.}~\bibnamefont
  {Mierzejewski}},\ }\bibfield  {title} {\bibinfo {title} {Slow diffusion and
  {{Thouless}} localization criterion in modulated spin chains},\ }\href
  {https://doi.org/10.1103/PhysRevB.108.035106} {\bibfield  {journal} {\bibinfo
   {journal} {Phys. Rev. B}\ }\textbf {\bibinfo {volume} {108}},\ \bibinfo
  {pages} {035106} (\bibinfo {year} {2023})}\BibitemShut {NoStop}%
\bibitem [{\citenamefont {Zotos}\ \emph {et~al.}(1997)\citenamefont {Zotos},
  \citenamefont {Naef},\ and\ \citenamefont
  {Prelovsek}}]{zotosTransportConservationLaws1997}%
  \BibitemOpen
  \bibfield  {author} {\bibinfo {author} {\bibfnamefont {X.}~\bibnamefont
  {Zotos}}, \bibinfo {author} {\bibfnamefont {F.}~\bibnamefont {Naef}},\ and\
  \bibinfo {author} {\bibfnamefont {P.}~\bibnamefont {Prelovsek}},\ }\bibfield
  {title} {\bibinfo {title} {Transport and conservation laws},\ }\href
  {https://doi.org/10.1103/physrevb.55.11029} {\bibfield  {journal} {\bibinfo
  {journal} {Phys. Rev. B}\ }\textbf {\bibinfo {volume} {55}},\ \bibinfo
  {pages} {11029} (\bibinfo {year} {1997})}\BibitemShut {NoStop}%
\bibitem [{\citenamefont {Zotos}\ and\ \citenamefont {Prelov{\v
  s}ek}(1996)}]{zotosEvidenceIdealInsulating1996}%
  \BibitemOpen
  \bibfield  {author} {\bibinfo {author} {\bibfnamefont {X.}~\bibnamefont
  {Zotos}}\ and\ \bibinfo {author} {\bibfnamefont {P.}~\bibnamefont {Prelov{\v
  s}ek}},\ }\bibfield  {title} {\bibinfo {title} {Evidence for ideal insulating
  or conducting state in a one-dimensional integrable system},\ }\href
  {https://doi.org/10.1103/PhysRevB.53.983} {\bibfield  {journal} {\bibinfo
  {journal} {Phys. Rev. B}\ }\textbf {\bibinfo {volume} {53}},\ \bibinfo
  {pages} {983} (\bibinfo {year} {1996})}\BibitemShut {NoStop}%
\bibitem [{\citenamefont {Bertini}\ \emph
  {et~al.}(2021{\natexlab{b}})\citenamefont {Bertini}, \citenamefont
  {{Heidrich-Meisner}}, \citenamefont {Karrasch}, \citenamefont {Prosen},
  \citenamefont {Steinigeweg},\ and\ \citenamefont {{\v Z}nidari{\v
  c}}}]{bertiniFinitetemperatureTransportOnedimensional2021}%
  \BibitemOpen
  \bibfield  {author} {\bibinfo {author} {\bibfnamefont {B.}~\bibnamefont
  {Bertini}}, \bibinfo {author} {\bibfnamefont {F.}~\bibnamefont
  {{Heidrich-Meisner}}}, \bibinfo {author} {\bibfnamefont {C.}~\bibnamefont
  {Karrasch}}, \bibinfo {author} {\bibfnamefont {T.}~\bibnamefont {Prosen}},
  \bibinfo {author} {\bibfnamefont {R.}~\bibnamefont {Steinigeweg}},\ and\
  \bibinfo {author} {\bibfnamefont {M.}~\bibnamefont {{\v Z}nidari{\v c}}},\
  }\bibfield  {title} {\bibinfo {title} {Finite-temperature transport in
  one-dimensional quantum lattice models},\ }\href
  {https://doi.org/10.1103/RevModPhys.93.025003} {\bibfield  {journal}
  {\bibinfo  {journal} {Rev. Mod. Phys.}\ }\textbf {\bibinfo {volume} {93}},\
  \bibinfo {pages} {025003} (\bibinfo {year} {2021}{\natexlab{b}})}\BibitemShut
  {NoStop}%
\bibitem [{\citenamefont {Hess}(2007)}]{hessHeatConductionLowdimensional2007}%
  \BibitemOpen
  \bibfield  {author} {\bibinfo {author} {\bibfnamefont {C.}~\bibnamefont
  {Hess}},\ }\bibfield  {title} {\bibinfo {title} {Heat conduction in
  low-dimensional quantum magnets},\ }\href
  {https://doi.org/10.1140/epjst/e2007-00363-8} {\bibfield  {journal} {\bibinfo
   {journal} {Eur. Phys. J. Spec. Top.}\ }\textbf {\bibinfo {volume} {151}},\
  \bibinfo {pages} {73} (\bibinfo {year} {2007})}\BibitemShut {NoStop}%
\bibitem [{\citenamefont {Bulchandani}\ \emph {et~al.}(2021)\citenamefont
  {Bulchandani}, \citenamefont {Gopalakrishnan},\ and\ \citenamefont
  {Ilievski}}]{Bulchandani2021}%
  \BibitemOpen
  \bibfield  {author} {\bibinfo {author} {\bibfnamefont {V.~B.}\ \bibnamefont
  {Bulchandani}}, \bibinfo {author} {\bibfnamefont {S.}~\bibnamefont
  {Gopalakrishnan}},\ and\ \bibinfo {author} {\bibfnamefont {E.}~\bibnamefont
  {Ilievski}},\ }\bibfield  {title} {\bibinfo {title} {Superdiffusion in spin
  chains},\ }\href {https://doi.org/10.1088/1742-5468/ac12c7} {\bibfield
  {journal} {\bibinfo  {journal} {J. Stat. Mech.: Theory Exp.}\ }\textbf
  {\bibinfo {volume} {2021}}\bibinfo  {number} { (8)},\ \bibinfo {pages}
  {084001}}\BibitemShut {NoStop}%
\bibitem [{\citenamefont {Vidmar}\ and\ \citenamefont
  {Rigol}(2016)}]{Vidmar2016}%
  \BibitemOpen
\bibfield  {number} {  }\bibfield  {author} {\bibinfo {author} {\bibfnamefont
  {L.}~\bibnamefont {Vidmar}}\ and\ \bibinfo {author} {\bibfnamefont
  {M.}~\bibnamefont {Rigol}},\ }\bibfield  {title} {\bibinfo {title}
  {Generalized {{Gibbs}} ensemble in integrable lattice models},\ }\href
  {https://doi.org/10.1088/1742-5468/2016/06/064007} {\bibfield  {journal}
  {\bibinfo  {journal} {J. Stat. Mech.: Theory Exp.}\ }\textbf {\bibinfo
  {volume} {2016}}\bibinfo  {number} { (6)},\ \bibinfo {pages}
  {064007}}\BibitemShut {NoStop}%
\bibitem [{\citenamefont {Ljubotina}\ \emph {et~al.}(2023)\citenamefont
  {Ljubotina}, \citenamefont {Desaules}, \citenamefont {Serbyn},\ and\
  \citenamefont {Papi\ifmmode~\acute{c}\else \'{c}\fi{}}}]{Ljubotina2023}%
  \BibitemOpen
\bibfield  {number} {  }\bibfield  {author} {\bibinfo {author} {\bibfnamefont
  {M.}~\bibnamefont {Ljubotina}}, \bibinfo {author} {\bibfnamefont {J.-Y.}\
  \bibnamefont {Desaules}}, \bibinfo {author} {\bibfnamefont {M.}~\bibnamefont
  {Serbyn}},\ and\ \bibinfo {author} {\bibfnamefont {Z.}~\bibnamefont
  {Papi\ifmmode~\acute{c}\else \'{c}\fi{}}},\ }\bibfield  {title} {\bibinfo
  {title} {Superdiffusive energy transport in kinetically constrained models},\
  }\href {https://doi.org/10.1103/PhysRevX.13.011033} {\bibfield  {journal}
  {\bibinfo  {journal} {Phys. Rev. X}\ }\textbf {\bibinfo {volume} {13}},\
  \bibinfo {pages} {011033} (\bibinfo {year} {2023})}\BibitemShut {NoStop}%
\bibitem [{\citenamefont {Jepsen}\ \emph {et~al.}(2020)\citenamefont {Jepsen},
  \citenamefont {{Amato-Grill}}, \citenamefont {Dimitrova}, \citenamefont {Ho},
  \citenamefont {Demler},\ and\ \citenamefont
  {Ketterle}}]{jepsenSpinTransportTunable2020}%
  \BibitemOpen
  \bibfield  {author} {\bibinfo {author} {\bibfnamefont {P.~N.}\ \bibnamefont
  {Jepsen}}, \bibinfo {author} {\bibfnamefont {J.}~\bibnamefont
  {{Amato-Grill}}}, \bibinfo {author} {\bibfnamefont {I.}~\bibnamefont
  {Dimitrova}}, \bibinfo {author} {\bibfnamefont {W.~W.}\ \bibnamefont {Ho}},
  \bibinfo {author} {\bibfnamefont {E.}~\bibnamefont {Demler}},\ and\ \bibinfo
  {author} {\bibfnamefont {W.}~\bibnamefont {Ketterle}},\ }\bibfield  {title}
  {\bibinfo {title} {Spin transport in a tunable {{Heisenberg}} model realized
  with ultracold atoms},\ }\href {https://doi.org/10.1038/s41586-020-3033-y}
  {\bibfield  {journal} {\bibinfo  {journal} {Nature}\ }\textbf {\bibinfo
  {volume} {588}},\ \bibinfo {pages} {403} (\bibinfo {year}
  {2020})}\BibitemShut {NoStop}%
\bibitem [{\citenamefont {Scheie}\ \emph {et~al.}(2021)\citenamefont {Scheie},
  \citenamefont {Sherman}, \citenamefont {Dupont}, \citenamefont {Nagler},
  \citenamefont {Stone}, \citenamefont {Granroth}, \citenamefont {Moore},\ and\
  \citenamefont {Tennant}}]{Scheie2021}%
  \BibitemOpen
  \bibfield  {author} {\bibinfo {author} {\bibfnamefont {A.}~\bibnamefont
  {Scheie}}, \bibinfo {author} {\bibfnamefont {N.~E.}\ \bibnamefont {Sherman}},
  \bibinfo {author} {\bibfnamefont {M.}~\bibnamefont {Dupont}}, \bibinfo
  {author} {\bibfnamefont {S.~E.}\ \bibnamefont {Nagler}}, \bibinfo {author}
  {\bibfnamefont {M.~B.}\ \bibnamefont {Stone}}, \bibinfo {author}
  {\bibfnamefont {G.~E.}\ \bibnamefont {Granroth}}, \bibinfo {author}
  {\bibfnamefont {J.~E.}\ \bibnamefont {Moore}},\ and\ \bibinfo {author}
  {\bibfnamefont {D.~A.}\ \bibnamefont {Tennant}},\ }\bibfield  {title}
  {\bibinfo {title} {Detection of {{Kardar}}--{{Parisi}}--{{Zhang}}
  hydrodynamics in a quantum {{Heisenberg}} spin-1/2 chain},\ }\href
  {https://doi.org/10.1038/s41567-021-01191-6} {\bibfield  {journal} {\bibinfo
  {journal} {Nat. Phys.}\ }\textbf {\bibinfo {volume} {17}},\ \bibinfo {pages}
  {726} (\bibinfo {year} {2021})}\BibitemShut {NoStop}%
\bibitem [{\citenamefont {Wei}\ \emph {et~al.}(2022)\citenamefont {Wei},
  \citenamefont {{Rubio-Abadal}}, \citenamefont {Ye}, \citenamefont {Machado},
  \citenamefont {Kemp}, \citenamefont {Srakaew}, \citenamefont {Hollerith},
  \citenamefont {Rui}, \citenamefont {Gopalakrishnan}, \citenamefont {Yao},
  \citenamefont {Bloch},\ and\ \citenamefont {Zeiher}}]{Wei2022}%
  \BibitemOpen
  \bibfield  {author} {\bibinfo {author} {\bibfnamefont {D.}~\bibnamefont
  {Wei}}, \bibinfo {author} {\bibfnamefont {A.}~\bibnamefont {{Rubio-Abadal}}},
  \bibinfo {author} {\bibfnamefont {B.}~\bibnamefont {Ye}}, \bibinfo {author}
  {\bibfnamefont {F.}~\bibnamefont {Machado}}, \bibinfo {author} {\bibfnamefont
  {J.}~\bibnamefont {Kemp}}, \bibinfo {author} {\bibfnamefont {K.}~\bibnamefont
  {Srakaew}}, \bibinfo {author} {\bibfnamefont {S.}~\bibnamefont {Hollerith}},
  \bibinfo {author} {\bibfnamefont {J.}~\bibnamefont {Rui}}, \bibinfo {author}
  {\bibfnamefont {S.}~\bibnamefont {Gopalakrishnan}}, \bibinfo {author}
  {\bibfnamefont {N.~Y.}\ \bibnamefont {Yao}}, \bibinfo {author} {\bibfnamefont
  {I.}~\bibnamefont {Bloch}},\ and\ \bibinfo {author} {\bibfnamefont
  {J.}~\bibnamefont {Zeiher}},\ }\bibfield  {title} {\bibinfo {title} {Quantum
  gas microscopy of {{Kardar-Parisi-Zhang}} superdiffusion},\ }\href
  {https://doi.org/10.1126/science.abk2397} {\bibfield  {journal} {\bibinfo
  {journal} {Science}\ }\textbf {\bibinfo {volume} {376}},\ \bibinfo {pages}
  {716} (\bibinfo {year} {2022})}\BibitemShut {NoStop}%
\bibitem [{\citenamefont {Feldmeier}\ \emph {et~al.}(2020)\citenamefont
  {Feldmeier}, \citenamefont {Sala}, \citenamefont {De~Tomasi}, \citenamefont
  {Pollmann},\ and\ \citenamefont {Knap}}]{Feldmeier2020}%
  \BibitemOpen
  \bibfield  {author} {\bibinfo {author} {\bibfnamefont {J.}~\bibnamefont
  {Feldmeier}}, \bibinfo {author} {\bibfnamefont {P.}~\bibnamefont {Sala}},
  \bibinfo {author} {\bibfnamefont {G.}~\bibnamefont {De~Tomasi}}, \bibinfo
  {author} {\bibfnamefont {F.}~\bibnamefont {Pollmann}},\ and\ \bibinfo
  {author} {\bibfnamefont {M.}~\bibnamefont {Knap}},\ }\bibfield  {title}
  {\bibinfo {title} {{Anomalous Diffusion in Dipole- and
  Higher-Moment-Conserving Systems}},\ }\href
  {https://doi.org/10.1103/PhysRevLett.125.245303} {\bibfield  {journal}
  {\bibinfo  {journal} {Phys. Rev. Lett.}\ }\textbf {\bibinfo {volume} {125}},\
  \bibinfo {pages} {245303} (\bibinfo {year} {2020})}\BibitemShut {NoStop}%
\bibitem [{\citenamefont {{\v Z}nidari{\v
  c}}(2013)}]{znidaricMagnetizationTransportSpin2013}%
  \BibitemOpen
  \bibfield  {author} {\bibinfo {author} {\bibfnamefont {M.}~\bibnamefont {{\v
  Z}nidari{\v c}}},\ }\bibfield  {title} {\bibinfo {title} {Magnetization
  transport in spin ladders and next-nearest-neighbor chains},\ }\href
  {https://doi.org/10.1103/PhysRevB.88.205135} {\bibfield  {journal} {\bibinfo
  {journal} {Phys. Rev. B}\ }\textbf {\bibinfo {volume} {88}},\ \bibinfo
  {pages} {205135} (\bibinfo {year} {2013})}\BibitemShut {NoStop}%
\bibitem [{\citenamefont {Steinigeweg}\ \emph
  {et~al.}(2014{\natexlab{a}})\citenamefont {Steinigeweg}, \citenamefont
  {{Heidrich-Meisner}}, \citenamefont {Gemmer}, \citenamefont {Michielsen},\
  and\ \citenamefont {De~Raedt}}]{steinigewegScalingDiffusionConstants2014}%
  \BibitemOpen
  \bibfield  {author} {\bibinfo {author} {\bibfnamefont {R.}~\bibnamefont
  {Steinigeweg}}, \bibinfo {author} {\bibfnamefont {F.}~\bibnamefont
  {{Heidrich-Meisner}}}, \bibinfo {author} {\bibfnamefont {J.}~\bibnamefont
  {Gemmer}}, \bibinfo {author} {\bibfnamefont {K.}~\bibnamefont {Michielsen}},\
  and\ \bibinfo {author} {\bibfnamefont {H.}~\bibnamefont {De~Raedt}},\
  }\bibfield  {title} {\bibinfo {title} {Scaling of diffusion constants in the
  spin-$\frac{1}{2}$ {{XX}} ladder},\ }\href
  {https://doi.org/10.1103/PhysRevB.90.094417} {\bibfield  {journal} {\bibinfo
  {journal} {Phys. Rev. B}\ }\textbf {\bibinfo {volume} {90}},\ \bibinfo
  {pages} {094417} (\bibinfo {year} {2014}{\natexlab{a}})}\BibitemShut
  {NoStop}%
\bibitem [{\citenamefont {Kloss}\ \emph {et~al.}(2018)\citenamefont {Kloss},
  \citenamefont {Lev},\ and\ \citenamefont
  {Reichman}}]{klossTimedependentVariationalPrinciple2018}%
  \BibitemOpen
  \bibfield  {author} {\bibinfo {author} {\bibfnamefont {B.}~\bibnamefont
  {Kloss}}, \bibinfo {author} {\bibfnamefont {Y.~B.}\ \bibnamefont {Lev}},\
  and\ \bibinfo {author} {\bibfnamefont {D.}~\bibnamefont {Reichman}},\
  }\bibfield  {title} {\bibinfo {title} {Time-dependent variational principle
  in matrix-product state manifolds: {{Pitfalls}} and potential},\ }\href
  {https://doi.org/10.1103/PhysRevB.97.024307} {\bibfield  {journal} {\bibinfo
  {journal} {Phys. Rev. B}\ }\textbf {\bibinfo {volume} {97}},\ \bibinfo
  {pages} {024307} (\bibinfo {year} {2018})}\BibitemShut {NoStop}%
\bibitem [{\citenamefont {Rakovszky}\ \emph {et~al.}(2022)\citenamefont
  {Rakovszky}, \citenamefont {von Keyserlingk},\ and\ \citenamefont
  {Pollmann}}]{rakovszkyDissipationassistedOperatorEvolution2022}%
  \BibitemOpen
  \bibfield  {author} {\bibinfo {author} {\bibfnamefont {T.}~\bibnamefont
  {Rakovszky}}, \bibinfo {author} {\bibfnamefont {C.~W.}\ \bibnamefont {von
  Keyserlingk}},\ and\ \bibinfo {author} {\bibfnamefont {F.}~\bibnamefont
  {Pollmann}},\ }\bibfield  {title} {\bibinfo {title} {Dissipation-assisted
  operator evolution method for capturing hydrodynamic transport},\ }\href
  {https://doi.org/10.1103/PhysRevB.105.075131} {\bibfield  {journal} {\bibinfo
   {journal} {Phys. Rev. B}\ }\textbf {\bibinfo {volume} {105}},\ \bibinfo
  {pages} {075131} (\bibinfo {year} {2022})}\BibitemShut {NoStop}%
\bibitem [{\citenamefont {Uskov}\ and\ \citenamefont
  {Lychkovskiy}(2024)}]{uskovQuantumDynamicsOne2024}%
  \BibitemOpen
  \bibfield  {author} {\bibinfo {author} {\bibfnamefont {F.}~\bibnamefont
  {Uskov}}\ and\ \bibinfo {author} {\bibfnamefont {O.}~\bibnamefont
  {Lychkovskiy}},\ }\bibfield  {title} {\bibinfo {title} {Quantum dynamics in
  one and two dimensions via the recursion method},\ }\href
  {https://doi.org/10.1103/PhysRevB.109.L140301} {\bibfield  {journal}
  {\bibinfo  {journal} {Phys. Rev. B}\ }\textbf {\bibinfo {volume} {109}},\
  \bibinfo {pages} {L140301} (\bibinfo {year} {2024})}\BibitemShut {NoStop}%
\bibitem [{\citenamefont {Artiaco}\ \emph {et~al.}(2024)\citenamefont
  {Artiaco}, \citenamefont {Fleckenstein}, \citenamefont {Aceituno~Ch{\'a}vez},
  \citenamefont {Kvorning},\ and\ \citenamefont
  {Bardarson}}]{artiacoEfficientLargeScaleManyBody2024}%
  \BibitemOpen
  \bibfield  {author} {\bibinfo {author} {\bibfnamefont {C.}~\bibnamefont
  {Artiaco}}, \bibinfo {author} {\bibfnamefont {C.}~\bibnamefont
  {Fleckenstein}}, \bibinfo {author} {\bibfnamefont {D.}~\bibnamefont
  {Aceituno~Ch{\'a}vez}}, \bibinfo {author} {\bibfnamefont {T.~K.}\
  \bibnamefont {Kvorning}},\ and\ \bibinfo {author} {\bibfnamefont {J.~H.}\
  \bibnamefont {Bardarson}},\ }\bibfield  {title} {\bibinfo {title} {Efficient
  {{Large-Scale Many-Body Quantum Dynamics}} via {{Local-Information Time
  Evolution}}},\ }\href {https://doi.org/10.1103/PRXQuantum.5.020352}
  {\bibfield  {journal} {\bibinfo  {journal} {PRX Quantum}\ }\textbf {\bibinfo
  {volume} {5}},\ \bibinfo {pages} {020352} (\bibinfo {year}
  {2024})}\BibitemShut {NoStop}%
\bibitem [{\citenamefont {Wang}\ \emph {et~al.}(2024)\citenamefont {Wang},
  \citenamefont {Lamann}, \citenamefont {Steinigeweg},\ and\ \citenamefont
  {Gemmer}}]{wangDiffusionConstantsRecursion2024}%
  \BibitemOpen
  \bibfield  {author} {\bibinfo {author} {\bibfnamefont {J.}~\bibnamefont
  {Wang}}, \bibinfo {author} {\bibfnamefont {M.~H.}\ \bibnamefont {Lamann}},
  \bibinfo {author} {\bibfnamefont {R.}~\bibnamefont {Steinigeweg}},\ and\
  \bibinfo {author} {\bibfnamefont {J.}~\bibnamefont {Gemmer}},\ }\bibfield
  {title} {\bibinfo {title} {Diffusion constants from the recursion method},\
  }\href {https://doi.org/10.1103/PhysRevB.110.104413} {\bibfield  {journal}
  {\bibinfo  {journal} {Phys. Rev. B}\ }\textbf {\bibinfo {volume} {110}},\
  \bibinfo {pages} {104413} (\bibinfo {year} {2024})}\BibitemShut {NoStop}%
\bibitem [{\citenamefont {F{\"u}llgraf}\ \emph {et~al.}(2025)\citenamefont
  {F{\"u}llgraf}, \citenamefont {Wang},\ and\ \citenamefont
  {Gemmer}}]{fullgrafLanczosPascalApproachCorrelation2025}%
  \BibitemOpen
  \bibfield  {author} {\bibinfo {author} {\bibfnamefont {M.}~\bibnamefont
  {F{\"u}llgraf}}, \bibinfo {author} {\bibfnamefont {J.}~\bibnamefont {Wang}},\
  and\ \bibinfo {author} {\bibfnamefont {J.}~\bibnamefont {Gemmer}},\
  }\bibfield  {title} {\bibinfo {title} {Lanczos-{{Pascal}} approach to
  correlation functions in chaotic quantum systems},\ }\href
  {https://doi.org/10.1103/zvgy-hzpl} {\bibfield  {journal} {\bibinfo
  {journal} {Phys. Rev. Lett.}\ }\textbf {\bibinfo {volume} {135}},\ \bibinfo
  {pages} {250401} (\bibinfo {year} {2025})}\BibitemShut {NoStop}%
\bibitem [{\citenamefont {Long}\ \emph {et~al.}(2003)\citenamefont {Long},
  \citenamefont {Prelov{\v s}ek}, \citenamefont {El~Shawish}, \citenamefont
  {Karadamoglou},\ and\ \citenamefont
  {Zotos}}]{longFinitetemperatureDynamicalCorrelations2003}%
  \BibitemOpen
  \bibfield  {author} {\bibinfo {author} {\bibfnamefont {M.~W.}\ \bibnamefont
  {Long}}, \bibinfo {author} {\bibfnamefont {P.}~\bibnamefont {Prelov{\v
  s}ek}}, \bibinfo {author} {\bibfnamefont {S.}~\bibnamefont {El~Shawish}},
  \bibinfo {author} {\bibfnamefont {J.}~\bibnamefont {Karadamoglou}},\ and\
  \bibinfo {author} {\bibfnamefont {X.}~\bibnamefont {Zotos}},\ }\bibfield
  {title} {\bibinfo {title} {Finite-temperature dynamical correlations using
  the microcanonical ensemble and the {{Lanczos}} algorithm},\ }\href
  {https://doi.org/10.1103/PhysRevB.68.235106} {\bibfield  {journal} {\bibinfo
  {journal} {Phys. Rev. B}\ }\textbf {\bibinfo {volume} {68}},\ \bibinfo
  {pages} {235106} (\bibinfo {year} {2003})}\BibitemShut {NoStop}%
\bibitem [{\citenamefont {Paw{\l}owski}\ \emph {et~al.}(2025)\citenamefont
  {Paw{\l}owski}, \citenamefont {Mierzejewski},\ and\ \citenamefont {Prelov{\v
  s}ek}}]{pawlowskiThoulessApproachTransport2025}%
  \BibitemOpen
  \bibfield  {author} {\bibinfo {author} {\bibfnamefont {J.}~\bibnamefont
  {Paw{\l}owski}}, \bibinfo {author} {\bibfnamefont {M.}~\bibnamefont
  {Mierzejewski}},\ and\ \bibinfo {author} {\bibfnamefont {P.}~\bibnamefont
  {Prelov{\v s}ek}},\ }\bibfield  {title} {\bibinfo {title} {Thouless approach
  in transport in integrable and perturbed easy-axis {{Heisenberg}} chains},\
  }\href {https://doi.org/10.1103/physrevb.111.l201113} {\bibfield  {journal}
  {\bibinfo  {journal} {Phys. Rev. B}\ }\textbf {\bibinfo {volume} {111}},\
  \bibinfo {pages} {L201113} (\bibinfo {year} {2025})}\BibitemShut {NoStop}%
\bibitem [{\citenamefont {Schneider}\ \emph {et~al.}(2012)\citenamefont
  {Schneider}, \citenamefont {Hackerm{\"u}ller}, \citenamefont {Ronzheimer},
  \citenamefont {Will}, \citenamefont {Braun}, \citenamefont {Best},
  \citenamefont {Bloch}, \citenamefont {Demler}, \citenamefont {Mandt},
  \citenamefont {Rasch},\ and\ \citenamefont
  {Rosch}}]{schneiderFermionicTransportOutofequilibrium2012}%
  \BibitemOpen
  \bibfield  {author} {\bibinfo {author} {\bibfnamefont {U.}~\bibnamefont
  {Schneider}}, \bibinfo {author} {\bibfnamefont {L.}~\bibnamefont
  {Hackerm{\"u}ller}}, \bibinfo {author} {\bibfnamefont {J.~P.}\ \bibnamefont
  {Ronzheimer}}, \bibinfo {author} {\bibfnamefont {S.}~\bibnamefont {Will}},
  \bibinfo {author} {\bibfnamefont {S.}~\bibnamefont {Braun}}, \bibinfo
  {author} {\bibfnamefont {T.}~\bibnamefont {Best}}, \bibinfo {author}
  {\bibfnamefont {I.}~\bibnamefont {Bloch}}, \bibinfo {author} {\bibfnamefont
  {E.}~\bibnamefont {Demler}}, \bibinfo {author} {\bibfnamefont
  {S.}~\bibnamefont {Mandt}}, \bibinfo {author} {\bibfnamefont
  {D.}~\bibnamefont {Rasch}},\ and\ \bibinfo {author} {\bibfnamefont
  {A.}~\bibnamefont {Rosch}},\ }\bibfield  {title} {\bibinfo {title} {Fermionic
  transport and out-of-equilibrium dynamics in a homogeneous {{Hubbard}} model
  with ultracold atoms},\ }\href {https://doi.org/10.1038/nphys2205} {\bibfield
   {journal} {\bibinfo  {journal} {Nat. Phys.}\ }\textbf {\bibinfo {volume}
  {8}},\ \bibinfo {pages} {213} (\bibinfo {year} {2012})}\BibitemShut {NoStop}%
\bibitem [{\citenamefont {Ronzheimer}\ \emph {et~al.}(2013)\citenamefont
  {Ronzheimer}, \citenamefont {Schreiber}, \citenamefont {Braun}, \citenamefont
  {Hodgman}, \citenamefont {Langer}, \citenamefont {McCulloch}, \citenamefont
  {{Heidrich-Meisner}}, \citenamefont {Bloch},\ and\ \citenamefont
  {Schneider}}]{ronzheimerExpansionDynamicsInteracting2013}%
  \BibitemOpen
  \bibfield  {author} {\bibinfo {author} {\bibfnamefont {J.~P.}\ \bibnamefont
  {Ronzheimer}}, \bibinfo {author} {\bibfnamefont {M.}~\bibnamefont
  {Schreiber}}, \bibinfo {author} {\bibfnamefont {S.}~\bibnamefont {Braun}},
  \bibinfo {author} {\bibfnamefont {S.~S.}\ \bibnamefont {Hodgman}}, \bibinfo
  {author} {\bibfnamefont {S.}~\bibnamefont {Langer}}, \bibinfo {author}
  {\bibfnamefont {I.~P.}\ \bibnamefont {McCulloch}}, \bibinfo {author}
  {\bibfnamefont {F.}~\bibnamefont {{Heidrich-Meisner}}}, \bibinfo {author}
  {\bibfnamefont {I.}~\bibnamefont {Bloch}},\ and\ \bibinfo {author}
  {\bibfnamefont {U.}~\bibnamefont {Schneider}},\ }\bibfield  {title} {\bibinfo
  {title} {Expansion {{Dynamics}} of {{Interacting Bosons}} in {{Homogeneous
  Lattices}} in {{One}} and {{Two Dimensions}}},\ }\href
  {https://doi.org/10.1103/PhysRevLett.110.205301} {\bibfield  {journal}
  {\bibinfo  {journal} {Phys. Rev. Lett.}\ }\textbf {\bibinfo {volume} {110}},\
  \bibinfo {pages} {205301} (\bibinfo {year} {2013})}\BibitemShut {NoStop}%
\bibitem [{\citenamefont {Hild}\ \emph {et~al.}(2014)\citenamefont {Hild},
  \citenamefont {Fukuhara}, \citenamefont {Schau\ss{}}, \citenamefont {Zeiher},
  \citenamefont {Knap}, \citenamefont {Demler}, \citenamefont {Bloch},\ and\
  \citenamefont {Gross}}]{Hild2014}%
  \BibitemOpen
  \bibfield  {author} {\bibinfo {author} {\bibfnamefont {S.}~\bibnamefont
  {Hild}}, \bibinfo {author} {\bibfnamefont {T.}~\bibnamefont {Fukuhara}},
  \bibinfo {author} {\bibfnamefont {P.}~\bibnamefont {Schau\ss{}}}, \bibinfo
  {author} {\bibfnamefont {J.}~\bibnamefont {Zeiher}}, \bibinfo {author}
  {\bibfnamefont {M.}~\bibnamefont {Knap}}, \bibinfo {author} {\bibfnamefont
  {E.}~\bibnamefont {Demler}}, \bibinfo {author} {\bibfnamefont
  {I.}~\bibnamefont {Bloch}},\ and\ \bibinfo {author} {\bibfnamefont
  {C.}~\bibnamefont {Gross}},\ }\bibfield  {title} {\bibinfo {title}
  {{Far-from-Equilibrium Spin Transport in Heisenberg Quantum Magnets}},\
  }\href {https://doi.org/10.1103/PhysRevLett.113.147205} {\bibfield  {journal}
  {\bibinfo  {journal} {Phys. Rev. Lett.}\ }\textbf {\bibinfo {volume} {113}},\
  \bibinfo {pages} {147205} (\bibinfo {year} {2014})}\BibitemShut {NoStop}%
\bibitem [{\citenamefont {Wienand}\ \emph {et~al.}(2024)\citenamefont
  {Wienand}, \citenamefont {Karch}, \citenamefont {Impertro}, \citenamefont
  {Schweizer}, \citenamefont {McCulloch}, \citenamefont {Vasseur},
  \citenamefont {Gopalakrishnan}, \citenamefont {Aidelsburger},\ and\
  \citenamefont {Bloch}}]{wienandEmergenceFluctuatingHydrodynamics2024}%
  \BibitemOpen
  \bibfield  {author} {\bibinfo {author} {\bibfnamefont {J.~F.}\ \bibnamefont
  {Wienand}}, \bibinfo {author} {\bibfnamefont {S.}~\bibnamefont {Karch}},
  \bibinfo {author} {\bibfnamefont {A.}~\bibnamefont {Impertro}}, \bibinfo
  {author} {\bibfnamefont {C.}~\bibnamefont {Schweizer}}, \bibinfo {author}
  {\bibfnamefont {E.}~\bibnamefont {McCulloch}}, \bibinfo {author}
  {\bibfnamefont {R.}~\bibnamefont {Vasseur}}, \bibinfo {author} {\bibfnamefont
  {S.}~\bibnamefont {Gopalakrishnan}}, \bibinfo {author} {\bibfnamefont
  {M.}~\bibnamefont {Aidelsburger}},\ and\ \bibinfo {author} {\bibfnamefont
  {I.}~\bibnamefont {Bloch}},\ }\bibfield  {title} {\bibinfo {title} {Emergence
  of fluctuating hydrodynamics in chaotic quantum systems},\ }\href
  {https://doi.org/10.1038/s41567-024-02611-z} {\bibfield  {journal} {\bibinfo
  {journal} {Nat. Phys.}\ }\textbf {\bibinfo {volume} {20}},\ \bibinfo {pages}
  {1732} (\bibinfo {year} {2024})}\BibitemShut {NoStop}%
\bibitem [{\citenamefont {Hess}(2019)}]{hessHeatTransportCupratebased2019}%
  \BibitemOpen
  \bibfield  {author} {\bibinfo {author} {\bibfnamefont {C.}~\bibnamefont
  {Hess}},\ }\bibfield  {title} {\bibinfo {title} {Heat transport of
  cuprate-based low-dimensional quantum magnets with strong exchange
  coupling},\ }\href {https://doi.org/10.1016/j.physrep.2019.02.004} {\bibfield
   {journal} {\bibinfo  {journal} {Phys. Rep.}\ }\textbf {\bibinfo {volume}
  {811}},\ \bibinfo {pages} {1} (\bibinfo {year} {2019})}\BibitemShut {NoStop}%
\bibitem [{\citenamefont {Hopjan}\ and\ \citenamefont
  {Vidmar}(2025)}]{hopjanCriticalDynamicsShortRange2025}%
  \BibitemOpen
  \bibfield  {author} {\bibinfo {author} {\bibfnamefont {M.}~\bibnamefont
  {Hopjan}}\ and\ \bibinfo {author} {\bibfnamefont {L.}~\bibnamefont
  {Vidmar}},\ }\bibfield  {title} {\bibinfo {title} {Critical {{Dynamics}} in
  {{Short-Range Quadratic Hamiltonians}}},\ }\href
  {https://doi.org/10.1103/5vyx-k877} {\bibfield  {journal} {\bibinfo
  {journal} {Phys. Rev. Lett.}\ }\textbf {\bibinfo {volume} {135}},\ \bibinfo
  {pages} {060401} (\bibinfo {year} {2025})}\BibitemShut {NoStop}%
\bibitem [{\citenamefont {Capizzi}\ \emph {et~al.}(2025)\citenamefont
  {Capizzi}, \citenamefont {Wang}, \citenamefont {Xu}, \citenamefont {Mazza},\
  and\ \citenamefont {Poletti}}]{Capizzi2025}%
  \BibitemOpen
  \bibfield  {author} {\bibinfo {author} {\bibfnamefont {L.}~\bibnamefont
  {Capizzi}}, \bibinfo {author} {\bibfnamefont {J.}~\bibnamefont {Wang}},
  \bibinfo {author} {\bibfnamefont {X.}~\bibnamefont {Xu}}, \bibinfo {author}
  {\bibfnamefont {L.}~\bibnamefont {Mazza}},\ and\ \bibinfo {author}
  {\bibfnamefont {D.}~\bibnamefont {Poletti}},\ }\bibfield  {title} {\bibinfo
  {title} {Hydrodynamics and the eigenstate thermalization hypothesis},\ }\href
  {https://doi.org/10.1103/PhysRevX.15.011059} {\bibfield  {journal} {\bibinfo
  {journal} {Phys. Rev. X}\ }\textbf {\bibinfo {volume} {15}},\ \bibinfo
  {pages} {011059} (\bibinfo {year} {2025})}\BibitemShut {NoStop}%
\bibitem [{\citenamefont {Mierzejewski}\ \emph {et~al.}(2022)\citenamefont
  {Mierzejewski}, \citenamefont {Paw{\l}owski}, \citenamefont {Prelovsek},\
  and\ \citenamefont {Herbrych}}]{mierzejewskiMultipleRelaxationTimes2022}%
  \BibitemOpen
  \bibfield  {author} {\bibinfo {author} {\bibfnamefont {M.}~\bibnamefont
  {Mierzejewski}}, \bibinfo {author} {\bibfnamefont {J.}~\bibnamefont
  {Paw{\l}owski}}, \bibinfo {author} {\bibfnamefont {P.}~\bibnamefont
  {Prelovsek}},\ and\ \bibinfo {author} {\bibfnamefont {J.}~\bibnamefont
  {Herbrych}},\ }\bibfield  {title} {\bibinfo {title} {Multiple relaxation
  times in perturbed {{XXZ}} chain},\ }\href
  {https://doi.org/10.21468/SciPostPhys.13.2.013} {\bibfield  {journal}
  {\bibinfo  {journal} {SciPost Phys.}\ }\textbf {\bibinfo {volume} {13}},\
  \bibinfo {pages} {013} (\bibinfo {year} {2022})}\BibitemShut {NoStop}%
\bibitem [{\citenamefont {Bulchandani}\ \emph {et~al.}(2020)\citenamefont
  {Bulchandani}, \citenamefont {Karrasch},\ and\ \citenamefont
  {Moore}}]{bulchandaniSuperdiffusiveTransportEnergy2020}%
  \BibitemOpen
  \bibfield  {author} {\bibinfo {author} {\bibfnamefont {V.~B.}\ \bibnamefont
  {Bulchandani}}, \bibinfo {author} {\bibfnamefont {C.}~\bibnamefont
  {Karrasch}},\ and\ \bibinfo {author} {\bibfnamefont {J.~E.}\ \bibnamefont
  {Moore}},\ }\bibfield  {title} {\bibinfo {title} {Superdiffusive transport of
  energy in one-dimensional metals},\ }\href
  {https://doi.org/10.1073/pnas.1916213117} {\bibfield  {journal} {\bibinfo
  {journal} {Proc. Natl. Acad. Sci. U.S.A.}\ }\textbf {\bibinfo {volume}
  {117}},\ \bibinfo {pages} {12713} (\bibinfo {year} {2020})}\BibitemShut
  {NoStop}%
\bibitem [{\citenamefont {{Chen-Lin}}\ \emph {et~al.}(2019)\citenamefont
  {{Chen-Lin}}, \citenamefont {Delacr{\'e}taz},\ and\ \citenamefont
  {Hartnoll}}]{chen-linTheoryDiffusiveFluctuations2019}%
  \BibitemOpen
  \bibfield  {author} {\bibinfo {author} {\bibfnamefont {X.}~\bibnamefont
  {{Chen-Lin}}}, \bibinfo {author} {\bibfnamefont {L.~V.}\ \bibnamefont
  {Delacr{\'e}taz}},\ and\ \bibinfo {author} {\bibfnamefont {S.~A.}\
  \bibnamefont {Hartnoll}},\ }\bibfield  {title} {\bibinfo {title} {Theory of
  {{Diffusive Fluctuations}}},\ }\href
  {https://doi.org/10.1103/PhysRevLett.122.091602} {\bibfield  {journal}
  {\bibinfo  {journal} {Phys. Rev. Lett.}\ }\textbf {\bibinfo {volume} {122}},\
  \bibinfo {pages} {091602} (\bibinfo {year} {2019})}\BibitemShut {NoStop}%
\bibitem [{\citenamefont
  {Delacr{\'e}taz}(2025)}]{delacretazBoundThermalizationDiffusive2025}%
  \BibitemOpen
  \bibfield  {author} {\bibinfo {author} {\bibfnamefont {L.~V.}\ \bibnamefont
  {Delacr{\'e}taz}},\ }\bibfield  {title} {\bibinfo {title} {A bound on
  thermalization from diffusive fluctuations},\ }\href
  {https://doi.org/10.1038/s41567-024-02774-9} {\bibfield  {journal} {\bibinfo
  {journal} {Nat. Phy.}\ }\textbf {\bibinfo {volume} {21}},\ \bibinfo {pages}
  {669} (\bibinfo {year} {2025})}\BibitemShut {NoStop}%
\bibitem [{\citenamefont {Kiendl}\ and\ \citenamefont
  {Marquardt}(2017)}]{Kiendl2017}%
  \BibitemOpen
  \bibfield  {author} {\bibinfo {author} {\bibfnamefont {T.}~\bibnamefont
  {Kiendl}}\ and\ \bibinfo {author} {\bibfnamefont {F.}~\bibnamefont
  {Marquardt}},\ }\bibfield  {title} {\bibinfo {title} {Many-particle dephasing
  after a quench},\ }\href {https://doi.org/10.1103/PhysRevLett.118.130601}
  {\bibfield  {journal} {\bibinfo  {journal} {Phys. Rev. Lett.}\ }\textbf
  {\bibinfo {volume} {118}},\ \bibinfo {pages} {130601} (\bibinfo {year}
  {2017})}\BibitemShut {NoStop}%
\bibitem [{\citenamefont {{Matsoukas-Roubeas}}\ \emph
  {et~al.}(2023)\citenamefont {{Matsoukas-Roubeas}}, \citenamefont {Beau},
  \citenamefont {Santos},\ and\ \citenamefont {del
  Campo}}]{matsoukas-roubeasUnitarityBreakingSelfaveraging2023}%
  \BibitemOpen
  \bibfield  {author} {\bibinfo {author} {\bibfnamefont {A.~S.}\ \bibnamefont
  {{Matsoukas-Roubeas}}}, \bibinfo {author} {\bibfnamefont {M.}~\bibnamefont
  {Beau}}, \bibinfo {author} {\bibfnamefont {L.~F.}\ \bibnamefont {Santos}},\
  and\ \bibinfo {author} {\bibfnamefont {A.}~\bibnamefont {del Campo}},\
  }\bibfield  {title} {\bibinfo {title} {Unitarity breaking in self-averaging
  spectral form factors},\ }\href {https://doi.org/10.1103/PhysRevA.108.062201}
  {\bibfield  {journal} {\bibinfo  {journal} {Phys. Rev. A}\ }\textbf {\bibinfo
  {volume} {108}},\ \bibinfo {pages} {062201} (\bibinfo {year}
  {2023})}\BibitemShut {NoStop}%
\bibitem [{\citenamefont {Vidmar}\ \emph {et~al.}(2013)\citenamefont {Vidmar},
  \citenamefont {Langer}, \citenamefont {McCulloch}, \citenamefont {Schneider},
  \citenamefont {Schollw{\"o}ck},\ and\ \citenamefont
  {{Heidrich-Meisner}}}]{vidmarSuddenExpansionMott2013}%
  \BibitemOpen
  \bibfield  {author} {\bibinfo {author} {\bibfnamefont {L.}~\bibnamefont
  {Vidmar}}, \bibinfo {author} {\bibfnamefont {S.}~\bibnamefont {Langer}},
  \bibinfo {author} {\bibfnamefont {I.~P.}\ \bibnamefont {McCulloch}}, \bibinfo
  {author} {\bibfnamefont {U.}~\bibnamefont {Schneider}}, \bibinfo {author}
  {\bibfnamefont {U.}~\bibnamefont {Schollw{\"o}ck}},\ and\ \bibinfo {author}
  {\bibfnamefont {F.}~\bibnamefont {{Heidrich-Meisner}}},\ }\bibfield  {title}
  {\bibinfo {title} {Sudden expansion of {{Mott}} insulators in one
  dimension},\ }\href {https://doi.org/10.1103/PhysRevB.88.235117} {\bibfield
  {journal} {\bibinfo  {journal} {Phys. Rev. B}\ }\textbf {\bibinfo {volume}
  {88}},\ \bibinfo {pages} {235117} (\bibinfo {year} {2013})}\BibitemShut
  {NoStop}%
\bibitem [{\citenamefont {Karrasch}\ \emph {et~al.}(2014)\citenamefont
  {Karrasch}, \citenamefont {Moore},\ and\ \citenamefont
  {{Heidrich-Meisner}}}]{karraschRealtimeRealspaceSpin2014}%
  \BibitemOpen
  \bibfield  {author} {\bibinfo {author} {\bibfnamefont {C.}~\bibnamefont
  {Karrasch}}, \bibinfo {author} {\bibfnamefont {J.~E.}\ \bibnamefont
  {Moore}},\ and\ \bibinfo {author} {\bibfnamefont {F.}~\bibnamefont
  {{Heidrich-Meisner}}},\ }\bibfield  {title} {\bibinfo {title} {Real-time and
  real-space spin and energy dynamics in one-dimensional spin-$\frac{1}{2}$
  systems induced by local quantum quenches at finite temperatures},\ }\href
  {https://doi.org/10.1103/PhysRevB.89.075139} {\bibfield  {journal} {\bibinfo
  {journal} {Phys. Rev. B}\ }\textbf {\bibinfo {volume} {89}},\ \bibinfo
  {pages} {075139} (\bibinfo {year} {2014})}\BibitemShut {NoStop}%
\bibitem [{\citenamefont {Karrasch}\ \emph {et~al.}(2015)\citenamefont
  {Karrasch}, \citenamefont {Kennes},\ and\ \citenamefont
  {{Heidrich-Meisner}}}]{karraschSpinThermalConductivity2015}%
  \BibitemOpen
  \bibfield  {author} {\bibinfo {author} {\bibfnamefont {C.}~\bibnamefont
  {Karrasch}}, \bibinfo {author} {\bibfnamefont {D.~M.}\ \bibnamefont
  {Kennes}},\ and\ \bibinfo {author} {\bibfnamefont {F.}~\bibnamefont
  {{Heidrich-Meisner}}},\ }\bibfield  {title} {\bibinfo {title} {Spin and
  thermal conductivity of quantum spin chains and ladders},\ }\href
  {https://doi.org/10.1103/PhysRevB.91.115130} {\bibfield  {journal} {\bibinfo
  {journal} {Phys. Rev. B}\ }\textbf {\bibinfo {volume} {91}},\ \bibinfo
  {pages} {115130} (\bibinfo {year} {2015})}\BibitemShut {NoStop}%
\bibitem [{\citenamefont {Iadecola}\ and\ \citenamefont {\ifmmode
  \check{Z}\else \v{Z}\fi{}nidari\ifmmode~\check{c}\else
  \v{c}\fi{}}(2019)}]{Iadecola2019}%
  \BibitemOpen
  \bibfield  {author} {\bibinfo {author} {\bibfnamefont {T.}~\bibnamefont
  {Iadecola}}\ and\ \bibinfo {author} {\bibfnamefont {M.}~\bibnamefont
  {\ifmmode \check{Z}\else \v{Z}\fi{}nidari\ifmmode~\check{c}\else
  \v{c}\fi{}}},\ }\bibfield  {title} {\bibinfo {title} {{Exact Localized and
  Ballistic Eigenstates in Disordered Chaotic Spin Ladders and the
  Fermi-Hubbard Model}},\ }\href
  {https://doi.org/10.1103/PhysRevLett.123.036403} {\bibfield  {journal}
  {\bibinfo  {journal} {Phys. Rev. Lett.}\ }\textbf {\bibinfo {volume} {123}},\
  \bibinfo {pages} {036403} (\bibinfo {year} {2019})}\BibitemShut {NoStop}%
\bibitem [{\citenamefont {Luitz}\ and\ \citenamefont
  {Lev}(2017)}]{luitzErgodicSideManybody2017}%
  \BibitemOpen
  \bibfield  {author} {\bibinfo {author} {\bibfnamefont {D.~J.}\ \bibnamefont
  {Luitz}}\ and\ \bibinfo {author} {\bibfnamefont {Y.~B.}\ \bibnamefont
  {Lev}},\ }\bibfield  {title} {\bibinfo {title} {The ergodic side of the
  many-body localization transition},\ }\href
  {https://doi.org/10.1002/andp.201600350} {\bibfield  {journal} {\bibinfo
  {journal} {Ann. Phys.}\ }\textbf {\bibinfo {volume} {529}},\ \bibinfo {pages}
  {1600350} (\bibinfo {year} {2017})}\BibitemShut {NoStop}%
\bibitem [{\citenamefont {Richter}\ \emph {et~al.}(2019)\citenamefont
  {Richter}, \citenamefont {Jin}, \citenamefont {Knipschild}, \citenamefont
  {Herbrych}, \citenamefont {De~Raedt}, \citenamefont {Michielsen},
  \citenamefont {Gemmer},\ and\ \citenamefont
  {Steinigeweg}}]{richterMagnetizationEnergyDynamics2019}%
  \BibitemOpen
  \bibfield  {author} {\bibinfo {author} {\bibfnamefont {J.}~\bibnamefont
  {Richter}}, \bibinfo {author} {\bibfnamefont {F.}~\bibnamefont {Jin}},
  \bibinfo {author} {\bibfnamefont {L.}~\bibnamefont {Knipschild}}, \bibinfo
  {author} {\bibfnamefont {J.}~\bibnamefont {Herbrych}}, \bibinfo {author}
  {\bibfnamefont {H.}~\bibnamefont {De~Raedt}}, \bibinfo {author}
  {\bibfnamefont {K.}~\bibnamefont {Michielsen}}, \bibinfo {author}
  {\bibfnamefont {J.}~\bibnamefont {Gemmer}},\ and\ \bibinfo {author}
  {\bibfnamefont {R.}~\bibnamefont {Steinigeweg}},\ }\bibfield  {title}
  {\bibinfo {title} {Magnetization and energy dynamics in spin ladders:
  {{Evidence}} of diffusion in time, frequency, position, and momentum},\
  }\href {https://doi.org/10.1103/PhysRevB.99.144422} {\bibfield  {journal}
  {\bibinfo  {journal} {Phys. Rev. B}\ }\textbf {\bibinfo {volume} {99}},\
  \bibinfo {pages} {144422} (\bibinfo {year} {2019})}\BibitemShut {NoStop}%
\bibitem [{\citenamefont {Herbrych}\ and\ \citenamefont {Prelov{\v
  s}ek}(2025)}]{herbrychSpinEnergyDiffusion2025}%
  \BibitemOpen
  \bibfield  {author} {\bibinfo {author} {\bibfnamefont {J.}~\bibnamefont
  {Herbrych}}\ and\ \bibinfo {author} {\bibfnamefont {P.}~\bibnamefont
  {Prelov{\v s}ek}},\ }\bibfield  {title} {\bibinfo {title} {Spin and energy
  diffusion versus subdiffusion in disordered spin chains},\ }\href
  {https://doi.org/10.1103/kv2f-m8vk} {\bibfield  {journal} {\bibinfo
  {journal} {Phys. Rev. B}\ }\textbf {\bibinfo {volume} {112}},\ \bibinfo
  {pages} {045108} (\bibinfo {year} {2025})}\BibitemShut {NoStop}%
\bibitem [{\citenamefont {Narozhny}\ \emph {et~al.}(1998)\citenamefont
  {Narozhny}, \citenamefont {Millis},\ and\ \citenamefont
  {Andrei}}]{narozhnyTransportXXZModel1998}%
  \BibitemOpen
  \bibfield  {author} {\bibinfo {author} {\bibfnamefont {B.~N.}\ \bibnamefont
  {Narozhny}}, \bibinfo {author} {\bibfnamefont {A.~J.}\ \bibnamefont
  {Millis}},\ and\ \bibinfo {author} {\bibfnamefont {N.}~\bibnamefont
  {Andrei}},\ }\bibfield  {title} {\bibinfo {title} {Transport in the {{XXZ}}
  model},\ }\href {https://doi.org/10.1103/PhysRevB.58.R2921} {\bibfield
  {journal} {\bibinfo  {journal} {Phys. Rev. B}\ }\textbf {\bibinfo {volume}
  {58}},\ \bibinfo {pages} {R2921} (\bibinfo {year} {1998})}\BibitemShut
  {NoStop}%
\bibitem [{\citenamefont
  {Rigol}(2009{\natexlab{a}})}]{rigolBreakdownThermalizationFinite2009}%
  \BibitemOpen
  \bibfield  {author} {\bibinfo {author} {\bibfnamefont {M.}~\bibnamefont
  {Rigol}},\ }\bibfield  {title} {\bibinfo {title} {Breakdown of
  {{Thermalization}} in {{Finite One-Dimensional Systems}}},\ }\href
  {https://doi.org/10.1103/PhysRevLett.103.100403} {\bibfield  {journal}
  {\bibinfo  {journal} {Phys. Rev. Lett.}\ }\textbf {\bibinfo {volume} {103}},\
  \bibinfo {pages} {100403} (\bibinfo {year} {2009}{\natexlab{a}})}\BibitemShut
  {NoStop}%
\bibitem [{\citenamefont
  {Rigol}(2009{\natexlab{b}})}]{rigolQuantumQuenchesThermalization2009}%
  \BibitemOpen
  \bibfield  {author} {\bibinfo {author} {\bibfnamefont {M.}~\bibnamefont
  {Rigol}},\ }\bibfield  {title} {\bibinfo {title} {Quantum quenches and
  thermalization in one-dimensional fermionic systems},\ }\href
  {https://doi.org/10.1103/PhysRevA.80.053607} {\bibfield  {journal} {\bibinfo
  {journal} {Phys. Rev. A}\ }\textbf {\bibinfo {volume} {80}},\ \bibinfo
  {pages} {053607} (\bibinfo {year} {2009}{\natexlab{b}})}\BibitemShut
  {NoStop}%
\bibitem [{\citenamefont
  {Steinigeweg}(2011)}]{steinigewegDecayCurrentsStrong2011}%
  \BibitemOpen
  \bibfield  {author} {\bibinfo {author} {\bibfnamefont {R.}~\bibnamefont
  {Steinigeweg}},\ }\bibfield  {title} {\bibinfo {title} {Decay of currents for
  strong interactions},\ }\href {https://doi.org/10.1103/PhysRevE.84.011136}
  {\bibfield  {journal} {\bibinfo  {journal} {Phys. Rev. E}\ }\textbf {\bibinfo
  {volume} {84}},\ \bibinfo {pages} {011136} (\bibinfo {year}
  {2011})}\BibitemShut {NoStop}%
\bibitem [{\citenamefont {Oganesyan}\ and\ \citenamefont
  {Huse}(2007)}]{oganesyanLocalizationInteractingFermions2007}%
  \BibitemOpen
  \bibfield  {author} {\bibinfo {author} {\bibfnamefont {V.}~\bibnamefont
  {Oganesyan}}\ and\ \bibinfo {author} {\bibfnamefont {D.~A.}\ \bibnamefont
  {Huse}},\ }\bibfield  {title} {\bibinfo {title} {Localization of interacting
  fermions at high temperature},\ }\href
  {https://doi.org/10.1103/PhysRevB.75.155111} {\bibfield  {journal} {\bibinfo
  {journal} {Phys. Rev. B}\ }\textbf {\bibinfo {volume} {75}},\ \bibinfo
  {pages} {155111} (\bibinfo {year} {2007})}\BibitemShut {NoStop}%
\bibitem [{\citenamefont {Santos}\ and\ \citenamefont
  {Rigol}(2010)}]{santosOnsetQuantumChaos2010}%
  \BibitemOpen
  \bibfield  {author} {\bibinfo {author} {\bibfnamefont {L.~F.}\ \bibnamefont
  {Santos}}\ and\ \bibinfo {author} {\bibfnamefont {M.}~\bibnamefont {Rigol}},\
  }\bibfield  {title} {\bibinfo {title} {Onset of quantum chaos in
  one-dimensional bosonic and fermionic systems and its relation to
  thermalization},\ }\href {https://doi.org/10.1103/physreve.81.036206}
  {\bibfield  {journal} {\bibinfo  {journal} {Phys. Rev. E}\ }\textbf {\bibinfo
  {volume} {81}},\ \bibinfo {pages} {036206} (\bibinfo {year}
  {2010})}\BibitemShut {NoStop}%
\bibitem [{\citenamefont {Pal}\ and\ \citenamefont
  {Huse}(2010)}]{palManybodyLocalizationPhase2010}%
  \BibitemOpen
  \bibfield  {author} {\bibinfo {author} {\bibfnamefont {A.}~\bibnamefont
  {Pal}}\ and\ \bibinfo {author} {\bibfnamefont {D.~A.}\ \bibnamefont {Huse}},\
  }\bibfield  {title} {\bibinfo {title} {Many-body localization phase
  transition},\ }\href {https://doi.org/10.1103/PhysRevB.82.174411} {\bibfield
  {journal} {\bibinfo  {journal} {Phys. Rev. B}\ }\textbf {\bibinfo {volume}
  {82}},\ \bibinfo {pages} {174411} (\bibinfo {year} {2010})}\BibitemShut
  {NoStop}%
\bibitem [{\citenamefont {Atas}\ \emph {et~al.}(2013)\citenamefont {Atas},
  \citenamefont {Bogomolny}, \citenamefont {Giraud},\ and\ \citenamefont
  {Roux}}]{atasDistributionRatioConsecutive2013}%
  \BibitemOpen
  \bibfield  {author} {\bibinfo {author} {\bibfnamefont {Y.~Y.}\ \bibnamefont
  {Atas}}, \bibinfo {author} {\bibfnamefont {E.}~\bibnamefont {Bogomolny}},
  \bibinfo {author} {\bibfnamefont {O.}~\bibnamefont {Giraud}},\ and\ \bibinfo
  {author} {\bibfnamefont {G.}~\bibnamefont {Roux}},\ }\bibfield  {title}
  {\bibinfo {title} {Distribution of the {{Ratio}} of {{Consecutive Level
  Spacings}} in {{Random Matrix Ensembles}}},\ }\href
  {https://doi.org/10.1103/PhysRevLett.110.084101} {\bibfield  {journal}
  {\bibinfo  {journal} {Phys. Rev. Lett.}\ }\textbf {\bibinfo {volume} {110}},\
  \bibinfo {pages} {084101} (\bibinfo {year} {2013})}\BibitemShut {NoStop}%
\bibitem [{\citenamefont {Song}\ \emph {et~al.}(2012)\citenamefont {Song},
  \citenamefont {Rachel}, \citenamefont {Flindt}, \citenamefont {Klich},
  \citenamefont {Laflorencie},\ and\ \citenamefont
  {Le~Hur}}]{songBipartiteFluctuationsProbe2012}%
  \BibitemOpen
  \bibfield  {author} {\bibinfo {author} {\bibfnamefont {H.~F.}\ \bibnamefont
  {Song}}, \bibinfo {author} {\bibfnamefont {S.}~\bibnamefont {Rachel}},
  \bibinfo {author} {\bibfnamefont {C.}~\bibnamefont {Flindt}}, \bibinfo
  {author} {\bibfnamefont {I.}~\bibnamefont {Klich}}, \bibinfo {author}
  {\bibfnamefont {N.}~\bibnamefont {Laflorencie}},\ and\ \bibinfo {author}
  {\bibfnamefont {K.}~\bibnamefont {Le~Hur}},\ }\bibfield  {title} {\bibinfo
  {title} {Bipartite fluctuations as a probe of many-body entanglement},\
  }\href {https://doi.org/10.1103/PhysRevB.85.035409} {\bibfield  {journal}
  {\bibinfo  {journal} {Phys. Rev. B}\ }\textbf {\bibinfo {volume} {85}},\
  \bibinfo {pages} {035409} (\bibinfo {year} {2012})}\BibitemShut {NoStop}%
\bibitem [{\citenamefont {Bauer}\ and\ \citenamefont
  {Nayak}(2013)}]{bauerAreaLawsManybody2013}%
  \BibitemOpen
  \bibfield  {author} {\bibinfo {author} {\bibfnamefont {B.}~\bibnamefont
  {Bauer}}\ and\ \bibinfo {author} {\bibfnamefont {C.}~\bibnamefont {Nayak}},\
  }\bibfield  {title} {\bibinfo {title} {Area laws in a many-body localized
  state and its implications for topological order},\ }\href
  {https://doi.org/10.1088/1742-5468/2013/09/P09005} {\bibfield  {journal}
  {\bibinfo  {journal} {J. Stat. Mech.: Theory Exp.}\ }\textbf {\bibinfo
  {volume} {2013}}\bibinfo  {number} { (09)},\ \bibinfo {pages}
  {P09005}}\BibitemShut {NoStop}%
\bibitem [{\citenamefont {Kj{\"a}ll}\ \emph {et~al.}(2014)\citenamefont
  {Kj{\"a}ll}, \citenamefont {Bardarson},\ and\ \citenamefont
  {Pollmann}}]{kjallManyBodyLocalizationDisordered2014}%
  \BibitemOpen
\bibfield  {number} {  }\bibfield  {author} {\bibinfo {author} {\bibfnamefont
  {J.~A.}\ \bibnamefont {Kj{\"a}ll}}, \bibinfo {author} {\bibfnamefont {J.~H.}\
  \bibnamefont {Bardarson}},\ and\ \bibinfo {author} {\bibfnamefont
  {F.}~\bibnamefont {Pollmann}},\ }\bibfield  {title} {\bibinfo {title}
  {Many-{{Body Localization}} in a {{Disordered Quantum Ising Chain}}},\ }\href
  {https://doi.org/10.1103/physrevlett.113.107204} {\bibfield  {journal}
  {\bibinfo  {journal} {Phys. Rev. Lett.}\ }\textbf {\bibinfo {volume} {113}},\
  \bibinfo {pages} {107204} (\bibinfo {year} {2014})}\BibitemShut {NoStop}%
\bibitem [{\citenamefont {Luitz}\ \emph {et~al.}(2015)\citenamefont {Luitz},
  \citenamefont {Laflorencie},\ and\ \citenamefont
  {Alet}}]{luitzManybodyLocalizationEdge2015}%
  \BibitemOpen
  \bibfield  {author} {\bibinfo {author} {\bibfnamefont {D.~J.}\ \bibnamefont
  {Luitz}}, \bibinfo {author} {\bibfnamefont {N.}~\bibnamefont {Laflorencie}},\
  and\ \bibinfo {author} {\bibfnamefont {F.}~\bibnamefont {Alet}},\ }\bibfield
  {title} {\bibinfo {title} {Many-body localization edge in the random-field
  {{Heisenberg}} chain},\ }\href {https://doi.org/10.1103/PhysRevB.91.081103}
  {\bibfield  {journal} {\bibinfo  {journal} {Phys. Rev. B}\ }\textbf {\bibinfo
  {volume} {91}},\ \bibinfo {pages} {081103(R)} (\bibinfo {year}
  {2015})}\BibitemShut {NoStop}%
\bibitem [{\citenamefont {Nandkishore}\ and\ \citenamefont
  {Huse}(2015)}]{nandkishoreManyBodyLocalizationThermalization2015}%
  \BibitemOpen
  \bibfield  {author} {\bibinfo {author} {\bibfnamefont {R.}~\bibnamefont
  {Nandkishore}}\ and\ \bibinfo {author} {\bibfnamefont {D.~A.}\ \bibnamefont
  {Huse}},\ }\bibfield  {title} {\bibinfo {title} {Many-{{Body Localization}}
  and {{Thermalization}} in {{Quantum Statistical Mechanics}}},\ }\href
  {https://doi.org/10.1146/annurev-conmatphys-031214-014726} {\bibfield
  {journal} {\bibinfo  {journal} {Annu. Rev. Condens. Matter Phys.}\ }\textbf
  {\bibinfo {volume} {6}},\ \bibinfo {pages} {15} (\bibinfo {year}
  {2015})}\BibitemShut {NoStop}%
\bibitem [{\citenamefont {Lim}\ and\ \citenamefont
  {Sheng}(2016)}]{limManybodyLocalizationTransition2016}%
  \BibitemOpen
  \bibfield  {author} {\bibinfo {author} {\bibfnamefont {S.~P.}\ \bibnamefont
  {Lim}}\ and\ \bibinfo {author} {\bibfnamefont {D.~N.}\ \bibnamefont
  {Sheng}},\ }\bibfield  {title} {\bibinfo {title} {Many-body localization and
  transition by density matrix renormalization group and exact diagonalization
  studies},\ }\href {https://doi.org/10.1103/PhysRevB.94.045111} {\bibfield
  {journal} {\bibinfo  {journal} {Phys. Rev. B}\ }\textbf {\bibinfo {volume}
  {94}},\ \bibinfo {pages} {045111} (\bibinfo {year} {2016})}\BibitemShut
  {NoStop}%
\bibitem [{\citenamefont {Prange}(1997)}]{prangeSpectralFormFactor1997}%
  \BibitemOpen
  \bibfield  {author} {\bibinfo {author} {\bibfnamefont {R.~E.}\ \bibnamefont
  {Prange}},\ }\bibfield  {title} {\bibinfo {title} {The {{Spectral Form Factor
  Is Not Self-Averaging}}},\ }\href
  {https://doi.org/10.1103/PhysRevLett.78.2280} {\bibfield  {journal} {\bibinfo
   {journal} {Phys. Rev. Lett.}\ }\textbf {\bibinfo {volume} {78}},\ \bibinfo
  {pages} {2280} (\bibinfo {year} {1997})}\BibitemShut {NoStop}%
\bibitem [{\citenamefont {Brody}\ \emph {et~al.}(1981)\citenamefont {Brody},
  \citenamefont {Flores}, \citenamefont {French}, \citenamefont {Mello},
  \citenamefont {Pandey},\ and\ \citenamefont
  {Wong}}]{brodyRandommatrixPhysicsSpectrum1981}%
  \BibitemOpen
  \bibfield  {author} {\bibinfo {author} {\bibfnamefont {T.~A.}\ \bibnamefont
  {Brody}}, \bibinfo {author} {\bibfnamefont {J.}~\bibnamefont {Flores}},
  \bibinfo {author} {\bibfnamefont {J.~B.}\ \bibnamefont {French}}, \bibinfo
  {author} {\bibfnamefont {P.~A.}\ \bibnamefont {Mello}}, \bibinfo {author}
  {\bibfnamefont {A.}~\bibnamefont {Pandey}},\ and\ \bibinfo {author}
  {\bibfnamefont {S.~S.~M.}\ \bibnamefont {Wong}},\ }\bibfield  {title}
  {\bibinfo {title} {Random-matrix physics: Spectrum and strength
  fluctuations},\ }\href {https://doi.org/10.1103/RevModPhys.53.385} {\bibfield
   {journal} {\bibinfo  {journal} {Rev. Mod. Phys.}\ }\textbf {\bibinfo
  {volume} {53}},\ \bibinfo {pages} {385} (\bibinfo {year} {1981})}\BibitemShut
  {NoStop}%
\bibitem [{\citenamefont {G{\'o}mez}\ \emph {et~al.}(2002)\citenamefont
  {G{\'o}mez}, \citenamefont {Molina}, \citenamefont {Rela{\~n}o},\ and\
  \citenamefont {Retamosa}}]{gomezMisleadingSignaturesQuantum2002}%
  \BibitemOpen
  \bibfield  {author} {\bibinfo {author} {\bibfnamefont {J.~M.~G.}\
  \bibnamefont {G{\'o}mez}}, \bibinfo {author} {\bibfnamefont {R.~A.}\
  \bibnamefont {Molina}}, \bibinfo {author} {\bibfnamefont {A.}~\bibnamefont
  {Rela{\~n}o}},\ and\ \bibinfo {author} {\bibfnamefont {J.}~\bibnamefont
  {Retamosa}},\ }\bibfield  {title} {\bibinfo {title} {Misleading signatures of
  quantum chaos},\ }\href {https://doi.org/10.1103/PhysRevE.66.036209}
  {\bibfield  {journal} {\bibinfo  {journal} {Phys. Rev. E}\ }\textbf {\bibinfo
  {volume} {66}},\ \bibinfo {pages} {036209} (\bibinfo {year}
  {2002})}\BibitemShut {NoStop}%
\bibitem [{\citenamefont {{Abul-Magd}}\ and\ \citenamefont
  {{Abul-Magd}}(2014)}]{abul-magdUnfoldingSpectrumChaotic2014}%
  \BibitemOpen
  \bibfield  {author} {\bibinfo {author} {\bibfnamefont {A.~A.}\ \bibnamefont
  {{Abul-Magd}}}\ and\ \bibinfo {author} {\bibfnamefont {A.~Y.}\ \bibnamefont
  {{Abul-Magd}}},\ }\bibfield  {title} {\bibinfo {title} {Unfolding of the
  spectrum for chaotic and mixed systems},\ }\href
  {https://doi.org/10.1016/j.physa.2013.11.012} {\bibfield  {journal} {\bibinfo
   {journal} {Physica A}\ }\textbf {\bibinfo {volume} {396}},\ \bibinfo {pages}
  {185} (\bibinfo {year} {2014})}\BibitemShut {NoStop}%
\bibitem [{\citenamefont {Li}\ \emph {et~al.}(2024)\citenamefont {Li},
  \citenamefont {Yan}, \citenamefont {Prosen},\ and\ \citenamefont
  {Chan}}]{liSpectralFormFactor2024}%
  \BibitemOpen
  \bibfield  {author} {\bibinfo {author} {\bibfnamefont {J.}~\bibnamefont
  {Li}}, \bibinfo {author} {\bibfnamefont {S.}~\bibnamefont {Yan}}, \bibinfo
  {author} {\bibfnamefont {T.}~\bibnamefont {Prosen}},\ and\ \bibinfo {author}
  {\bibfnamefont {A.}~\bibnamefont {Chan}},\ }\bibfield  {title} {\bibinfo
  {title} {Spectral form factor in chaotic, localized, and integrable open
  quantum many-body systems},\ }\Eprint {https://arxiv.org/abs/2405.01641}
  {arXiv:2405.01641 [cond-mat.stat-mech]}  (\bibinfo {year} {2024})\BibitemShut
  {NoStop}%
\bibitem [{\citenamefont {Steinigeweg}\ and\ \citenamefont
  {Brenig}(2011)}]{steinigewegSpinTransportChain2011}%
  \BibitemOpen
  \bibfield  {author} {\bibinfo {author} {\bibfnamefont {R.}~\bibnamefont
  {Steinigeweg}}\ and\ \bibinfo {author} {\bibfnamefont {W.}~\bibnamefont
  {Brenig}},\ }\bibfield  {title} {\bibinfo {title} {Spin {{Transport}} in the
  {{$XXZ$}} {{Chain}} at {{Finite Temperature}} and {{Momentum}}},\ }\href
  {https://doi.org/10.1103/PhysRevLett.107.250602} {\bibfield  {journal}
  {\bibinfo  {journal} {Phys. Rev. Lett.}\ }\textbf {\bibinfo {volume} {107}},\
  \bibinfo {pages} {250602} (\bibinfo {year} {2011})}\BibitemShut {NoStop}%
\bibitem [{\citenamefont {Steinigeweg}\ and\ \citenamefont
  {Gemmer}(2009)}]{steinigewegDensityDynamicsTranslationally2009}%
  \BibitemOpen
  \bibfield  {author} {\bibinfo {author} {\bibfnamefont {R.}~\bibnamefont
  {Steinigeweg}}\ and\ \bibinfo {author} {\bibfnamefont {J.}~\bibnamefont
  {Gemmer}},\ }\bibfield  {title} {\bibinfo {title} {Density dynamics in
  translationally invariant spin-$\frac{1}{2}$ chains at high temperatures:
  {{A}} current-autocorrelation approach to finite time and length scales},\
  }\href {https://doi.org/10.1103/PhysRevB.80.184402} {\bibfield  {journal}
  {\bibinfo  {journal} {Phys. Rev. B}\ }\textbf {\bibinfo {volume} {80}},\
  \bibinfo {pages} {184402} (\bibinfo {year} {2009})}\BibitemShut {NoStop}%
\bibitem [{\citenamefont {Hams}\ and\ \citenamefont
  {De~Raedt}(2000)}]{hamsFastAlgorithmFinding2000}%
  \BibitemOpen
  \bibfield  {author} {\bibinfo {author} {\bibfnamefont {A.}~\bibnamefont
  {Hams}}\ and\ \bibinfo {author} {\bibfnamefont {H.}~\bibnamefont
  {De~Raedt}},\ }\bibfield  {title} {\bibinfo {title} {Fast algorithm for
  finding the eigenvalue distribution of very large matrices},\ }\href
  {https://doi.org/10.1103/PhysRevE.62.4365} {\bibfield  {journal} {\bibinfo
  {journal} {Phys. Rev. E}\ }\textbf {\bibinfo {volume} {62}},\ \bibinfo
  {pages} {4365} (\bibinfo {year} {2000})}\BibitemShut {NoStop}%
\bibitem [{\citenamefont {Bartsch}\ and\ \citenamefont
  {Gemmer}(2009)}]{bartschDynamicalTypicalityQuantum2009}%
  \BibitemOpen
  \bibfield  {author} {\bibinfo {author} {\bibfnamefont {C.}~\bibnamefont
  {Bartsch}}\ and\ \bibinfo {author} {\bibfnamefont {J.}~\bibnamefont
  {Gemmer}},\ }\bibfield  {title} {\bibinfo {title} {Dynamical {{Typicality}}
  of {{Quantum Expectation Values}}},\ }\href
  {https://doi.org/10.1103/PhysRevLett.102.110403} {\bibfield  {journal}
  {\bibinfo  {journal} {Phys. Rev. Lett.}\ }\textbf {\bibinfo {volume} {102}},\
  \bibinfo {pages} {110403} (\bibinfo {year} {2009})}\BibitemShut {NoStop}%
\bibitem [{\citenamefont {Elsayed}\ and\ \citenamefont
  {Fine}(2013)}]{elsayedRegressionRelationPure2013}%
  \BibitemOpen
  \bibfield  {author} {\bibinfo {author} {\bibfnamefont {T.~A.}\ \bibnamefont
  {Elsayed}}\ and\ \bibinfo {author} {\bibfnamefont {B.~V.}\ \bibnamefont
  {Fine}},\ }\bibfield  {title} {\bibinfo {title} {Regression {{Relation}} for
  {{Pure Quantum States}} and {{Its Implications}} for {{Efficient
  Computing}}},\ }\href {https://doi.org/10.1103/PhysRevLett.110.070404}
  {\bibfield  {journal} {\bibinfo  {journal} {Phys. Rev. Lett.}\ }\textbf
  {\bibinfo {volume} {110}},\ \bibinfo {pages} {070404} (\bibinfo {year}
  {2013})}\BibitemShut {NoStop}%
\bibitem [{\citenamefont {Steinigeweg}\ \emph
  {et~al.}(2014{\natexlab{b}})\citenamefont {Steinigeweg}, \citenamefont
  {Gemmer},\ and\ \citenamefont
  {Brenig}}]{steinigewegSpinCurrentAutocorrelationsSingle2014}%
  \BibitemOpen
  \bibfield  {author} {\bibinfo {author} {\bibfnamefont {R.}~\bibnamefont
  {Steinigeweg}}, \bibinfo {author} {\bibfnamefont {J.}~\bibnamefont
  {Gemmer}},\ and\ \bibinfo {author} {\bibfnamefont {W.}~\bibnamefont
  {Brenig}},\ }\bibfield  {title} {\bibinfo {title} {Spin-{{Current
  Autocorrelations}} from {{Single Pure-State Propagation}}},\ }\href
  {https://doi.org/10.1103/PhysRevLett.112.120601} {\bibfield  {journal}
  {\bibinfo  {journal} {Phys. Rev. Lett.}\ }\textbf {\bibinfo {volume} {112}},\
  \bibinfo {pages} {120601} (\bibinfo {year} {2014}{\natexlab{b}})}\BibitemShut
  {NoStop}%
\bibitem [{\citenamefont {Steinigeweg}\ \emph
  {et~al.}(2014{\natexlab{c}})\citenamefont {Steinigeweg}, \citenamefont
  {Khodja}, \citenamefont {Niemeyer}, \citenamefont {Gogolin},\ and\
  \citenamefont {Gemmer}}]{steinigewegPushingLimitsEigenstate2014}%
  \BibitemOpen
  \bibfield  {author} {\bibinfo {author} {\bibfnamefont {R.}~\bibnamefont
  {Steinigeweg}}, \bibinfo {author} {\bibfnamefont {A.}~\bibnamefont {Khodja}},
  \bibinfo {author} {\bibfnamefont {H.}~\bibnamefont {Niemeyer}}, \bibinfo
  {author} {\bibfnamefont {C.}~\bibnamefont {Gogolin}},\ and\ \bibinfo {author}
  {\bibfnamefont {J.}~\bibnamefont {Gemmer}},\ }\bibfield  {title} {\bibinfo
  {title} {Pushing the {{Limits}} of the {{Eigenstate Thermalization
  Hypothesis}} towards {{Mesoscopic Quantum Systems}}},\ }\href
  {https://doi.org/10.1103/PhysRevLett.112.130403} {\bibfield  {journal}
  {\bibinfo  {journal} {Phys. Rev. Lett.}\ }\textbf {\bibinfo {volume} {112}},\
  \bibinfo {pages} {130403} (\bibinfo {year} {2014}{\natexlab{c}})}\BibitemShut
  {NoStop}%
\bibitem [{\citenamefont {Steinigeweg}\ \emph {et~al.}(2015)\citenamefont
  {Steinigeweg}, \citenamefont {Gemmer},\ and\ \citenamefont
  {Brenig}}]{steinigewegSpinEnergyCurrents2015}%
  \BibitemOpen
  \bibfield  {author} {\bibinfo {author} {\bibfnamefont {R.}~\bibnamefont
  {Steinigeweg}}, \bibinfo {author} {\bibfnamefont {J.}~\bibnamefont
  {Gemmer}},\ and\ \bibinfo {author} {\bibfnamefont {W.}~\bibnamefont
  {Brenig}},\ }\bibfield  {title} {\bibinfo {title} {Spin and energy currents
  in integrable and nonintegrable spin-$\frac{1}{2}$ chains: {{A}} typicality
  approach to real-time autocorrelations},\ }\href
  {https://doi.org/10.1103/PhysRevB.91.104404} {\bibfield  {journal} {\bibinfo
  {journal} {Phys. Rev. B}\ }\textbf {\bibinfo {volume} {91}},\ \bibinfo
  {pages} {104404} (\bibinfo {year} {2015})}\BibitemShut {NoStop}%
\bibitem [{\citenamefont {Steinigeweg}\ \emph {et~al.}(2016)\citenamefont
  {Steinigeweg}, \citenamefont {Herbrych}, \citenamefont {Pollmann},\ and\
  \citenamefont {Brenig}}]{steinigewegTypicalityApproachOptical2016}%
  \BibitemOpen
  \bibfield  {author} {\bibinfo {author} {\bibfnamefont {R.}~\bibnamefont
  {Steinigeweg}}, \bibinfo {author} {\bibfnamefont {J.}~\bibnamefont
  {Herbrych}}, \bibinfo {author} {\bibfnamefont {F.}~\bibnamefont {Pollmann}},\
  and\ \bibinfo {author} {\bibfnamefont {W.}~\bibnamefont {Brenig}},\
  }\bibfield  {title} {\bibinfo {title} {Typicality approach to the optical
  conductivity in thermal and many-body localized phases},\ }\href
  {https://doi.org/10.1103/PhysRevB.94.180401} {\bibfield  {journal} {\bibinfo
  {journal} {Phys. Rev. B}\ }\textbf {\bibinfo {volume} {94}},\ \bibinfo
  {pages} {180401(R)} (\bibinfo {year} {2016})}\BibitemShut {NoStop}%
\bibitem [{\citenamefont {Karahalios}\ \emph {et~al.}(2009)\citenamefont
  {Karahalios}, \citenamefont {Metavitsiadis}, \citenamefont {Zotos},
  \citenamefont {Gorczyca},\ and\ \citenamefont {Prelov{\v
  s}ek}}]{karahaliosFinitetemperatureTransportDisordered2009}%
  \BibitemOpen
  \bibfield  {author} {\bibinfo {author} {\bibfnamefont {A.}~\bibnamefont
  {Karahalios}}, \bibinfo {author} {\bibfnamefont {A.}~\bibnamefont
  {Metavitsiadis}}, \bibinfo {author} {\bibfnamefont {X.}~\bibnamefont
  {Zotos}}, \bibinfo {author} {\bibfnamefont {A.}~\bibnamefont {Gorczyca}},\
  and\ \bibinfo {author} {\bibfnamefont {P.}~\bibnamefont {Prelov{\v s}ek}},\
  }\bibfield  {title} {\bibinfo {title} {Finite-temperature transport in
  disordered {{Heisenberg}} chains},\ }\href
  {https://doi.org/10.1103/PhysRevB.79.024425} {\bibfield  {journal} {\bibinfo
  {journal} {Phys. Rev. B}\ }\textbf {\bibinfo {volume} {79}},\ \bibinfo
  {pages} {024425} (\bibinfo {year} {2009})}\BibitemShut {NoStop}%
\bibitem [{\citenamefont {Bari{\v s}i{\'c}}\ and\ \citenamefont {Prelov{\v
  s}ek}(2010)}]{barisicConductivityDisorderedOnedimensional2010}%
  \BibitemOpen
  \bibfield  {author} {\bibinfo {author} {\bibfnamefont {O.~S.}\ \bibnamefont
  {Bari{\v s}i{\'c}}}\ and\ \bibinfo {author} {\bibfnamefont {P.}~\bibnamefont
  {Prelov{\v s}ek}},\ }\bibfield  {title} {\bibinfo {title} {Conductivity in a
  disordered one-dimensional system of interacting fermions},\ }\href
  {https://doi.org/10.1103/PhysRevB.82.161106} {\bibfield  {journal} {\bibinfo
  {journal} {Phys. Rev. B}\ }\textbf {\bibinfo {volume} {82}},\ \bibinfo
  {pages} {161106(R)} (\bibinfo {year} {2010})}\BibitemShut {NoStop}%
\bibitem [{\citenamefont {Bari{\v s}i{\'c}}\ \emph {et~al.}(2016)\citenamefont
  {Bari{\v s}i{\'c}}, \citenamefont {Kokalj}, \citenamefont {Balog},\ and\
  \citenamefont {Prelov{\v s}ek}}]{barisicDynamicalConductivityIts2016}%
  \BibitemOpen
  \bibfield  {author} {\bibinfo {author} {\bibfnamefont {O.~S.}\ \bibnamefont
  {Bari{\v s}i{\'c}}}, \bibinfo {author} {\bibfnamefont {J.}~\bibnamefont
  {Kokalj}}, \bibinfo {author} {\bibfnamefont {I.}~\bibnamefont {Balog}},\ and\
  \bibinfo {author} {\bibfnamefont {P.}~\bibnamefont {Prelov{\v s}ek}},\
  }\bibfield  {title} {\bibinfo {title} {Dynamical conductivity and its
  fluctuations along the crossover to many-body localization},\ }\href
  {https://doi.org/10.1103/PhysRevB.94.045126} {\bibfield  {journal} {\bibinfo
  {journal} {Phys. Rev. B}\ }\textbf {\bibinfo {volume} {94}},\ \bibinfo
  {pages} {045126} (\bibinfo {year} {2016})}\BibitemShut {NoStop}%
\bibitem [{\citenamefont {Fritzsch}\ and\ \citenamefont
  {Prosen}(2021)}]{fritzschEigenstateThermalizationDualunitary2021}%
  \BibitemOpen
  \bibfield  {author} {\bibinfo {author} {\bibfnamefont {F.}~\bibnamefont
  {Fritzsch}}\ and\ \bibinfo {author} {\bibfnamefont {T.}~\bibnamefont
  {Prosen}},\ }\bibfield  {title} {\bibinfo {title} {Eigenstate thermalization
  in dual-unitary quantum circuits: {{Asymptotics}} of spectral functions},\
  }\href {https://doi.org/10.1103/PhysRevE.103.062133} {\bibfield  {journal}
  {\bibinfo  {journal} {Phys. Rev. E}\ }\textbf {\bibinfo {volume} {103}},\
  \bibinfo {pages} {062133} (\bibinfo {year} {2021})}\BibitemShut {NoStop}%
\bibitem [{\citenamefont {{\c C}even}\ \emph {et~al.}(2025)\citenamefont {{\c
  C}even}, \citenamefont {Peinemann},\ and\ \citenamefont
  {{Heidrich-Meisner}}}]{zenodo}%
  \BibitemOpen
  \bibfield  {author} {\bibinfo {author} {\bibfnamefont {K.}~\bibnamefont {{\c
  C}even}}, \bibinfo {author} {\bibfnamefont {L.}~\bibnamefont {Peinemann}},\
  and\ \bibinfo {author} {\bibfnamefont {F.}~\bibnamefont
  {{Heidrich-Meisner}}},\ }\bibfield  {title} {\bibinfo {title} {Data for
  ``{{Hierarchy of timescales in a disordered spin-{{$1/2$}} {{XX}}
  ladder}}''},\ }\href {https://doi.org/10.5281/zenodo.16838209}
  {10.5281/zenodo.16838209} (\bibinfo {year} {2025}),\ \bibinfo {note}
  {{Zenodo}}\BibitemShut {NoStop}%
\end{thebibliography}%

\end{document}